\documentclass[11pt,letterpaper]{article}

\usepackage[T1]{fontenc}
\usepackage[utf8]{inputenc}
\usepackage{lmodern}
\usepackage{microtype}
\usepackage{indentfirst}
\usepackage{titlesec}
\usepackage{tocloft}
\usepackage{float}
\usepackage{tabularx}
\usepackage{booktabs}
\usepackage{placeins}

\usepackage[margin=1in]{geometry} 
\usepackage{amsmath,amssymb,amsthm,mathtools}
\usepackage{mathrsfs}
\usepackage{bm}
\numberwithin{equation}{section}
\allowdisplaybreaks[3]
\usepackage{graphicx}
\usepackage{caption}
\usepackage{subcaption}

\usepackage[colorlinks=true,linkcolor=blue,citecolor=blue,urlcolor=blue]{hyperref}

\usepackage[numbers,sort&compress]{natbib}

\theoremstyle{definition}

\theoremstyle{remark}

\newcommand{\regtag}[2]{\overset{\scriptscriptstyle\eqref{#2}}{#1}}

\titleformat{\section}
  {\normalfont\Large\bfseries}
  {\thesection}{0.75em}{}
\titleformat{\subsection}
  {\normalfont\large\bfseries}
  {\thesubsection}{0.65em}{}
\titlespacing*{\section}{0pt}{2.4ex plus 0.8ex minus 0.2ex}{1.2ex}
\titlespacing*{\subsection}{0pt}{1.8ex plus 0.6ex minus 0.2ex}{0.8ex}

\newcommand{\setupappendixstyle}{%
  \titleformat{\section}{\normalfont\Large\bfseries}{Appendix \thesection}{0.75em}{}%
  \renewcommand{\cftsecpresnum}{Appendix\ }%
  \renewcommand{\cftsecaftersnum}{\quad}%
  \settowidth{\cftsecnumwidth}{Appendix Z\quad}%
}

\title{\textbf{2PN point-mass \(N\)-body equations of motion in harmonic gauge: A computable formulation}}
\author{
  Hongkun Huang\textsuperscript{1,2}%
  \thanks{Contact author: \href{mailto:hhongkun2024@lzu.edu.cn}{hhongkun2024@lzu.edu.cn}}
  \qquad
  Jie Yang\textsuperscript{1,2}%
  \thanks{Contact author: \href{mailto:yangjiev@lzu.edu.cn}{yangjiev@lzu.edu.cn}}
  \qquad
  Wei-Tou Ni\textsuperscript{1,3,4}%
  \thanks{Contact author: \href{mailto:niweidou@lzu.edu.cn}{niweidou@lzu.edu.cn}}
  \\[0.8em]
  \parbox{0.92\textwidth}{\centering\small
    \textsuperscript{1}Lanzhou Center for Theoretical Physics, Key Laboratory of Theoretical Physics of Gansu Province,
    Key Laboratory of Quantum Theory and Applications of MoE, Gansu Provincial Research Center for
    Basic Disciplines of Quantum Physics, Lanzhou University, Lanzhou 730000, China \\
    \textsuperscript{2}Institute of Theoretical Physics and Research Center of Gravitation, School of Physical Science and
    Technology, Lanzhou University, Lanzhou 730000, China \\
    \textsuperscript{3}Wuhan Institute of Physics and Mathematics, Innovation Academy for Precision Measurement Science
    and Technology (APM), Chinese Academy of Sciences, Wuhan 430071, China \\
    \textsuperscript{4}International Centre for Theoretical Physics Asia-Pacific, University of Chinese Academy of Sciences,
    Beijing 100190, China
  }
}
\date{} 

\renewcommand{\thesection}{\Roman{section}}
\renewcommand{\thesubsection}{\thesection.\arabic{subsection}}

\begin{document}

\maketitle

\begin{abstract}

We develop a semianalytic and seminumerical formulation of the
harmonic-gauge second post-Newtonian (2PN) equations of motion for a general
point-mass \(N\)-body system within Hadamard regularization. The equations of
motion are separated into a closed analytic contribution and a nonclosed
integral contribution. We analyze the singular structure of the latter and
further regularize it into a numerically evaluable representation. We apply
the formulation to the Sun--Jupiter--Saturn and Sun--Mercury--Venus systems,
evaluating the instantaneous nonclosed 2PN acceleration along Newtonian
trajectories and its leading finite-time relative-distance response through
the corresponding perturbation equations. In both benchmarks, the nonclosed
acceleration remains a small fraction of the complete 2PN acceleration, while
the induced relative-distance perturbation remains oscillatory and can reach
larger oscillation amplitudes at later times.

\end{abstract}

\setcounter{tocdepth}{2}
\tableofcontents
\section{Introduction}

In the weak-field and slow-motion regime, relativistic gravitational
\(N\)-body dynamics can be treated systematically within the
post-Newtonian (PN) approximation, with
\begin{align}
    \frac{v^2}{c^2}
    \sim
    \frac{GM}{rc^2}.
\end{align}
Solar System applications require such a formulation for general
many-body configurations. Modern Solar System ephemerides are based on numerical integrations of
many-body dynamical models, with relativistic effects generally included
through 1PN order (see, e.g.,
\cite{Fienga2024,doi:10.1142/S0218271816300032} and references therein).
The parametrized post-Newtonian framework likewise underlies Solar System
tests of relativistic gravity. As observational and ephemeris accuracies
improve, higher-order many-body effects become increasingly relevant.
Here we consider general point-mass \(N\)-body dynamics at 2PN order in
harmonic coordinates.

The relativistic many-body problem has a long history. Lorentz and
Droste~\cite{lorentz1937motion} studied extended gravitating bodies whose
sizes were small compared with their mutual separations, neglecting tidal
effects and restricting each body to overall translational motion, with
matter modeled as an incompressible fluid. They showed that the overall motion of the bodies
could be described by Lagrange equations derived from a Lagrangian depending
on the instantaneous separations and velocities. Although not written as
explicit point-particle acceleration equations, their result already
contained the 1PN \(N\)-body Lagrangian structure.

Einstein, Infeld, and Hoffmann
\cite{7c44806c-bf75-3a8d-b622-7761a1e00cd9,%
9c298a78-4f65-3b62-b231-cb66fa9d86b7,einstein1949motion}
later developed a field-based approach. Rather than assuming an
a priori model for the stress-energy tensor, they derived the 1PN
motion of gravitating particles directly from the vacuum Einstein field
equations. The resulting Lorentz--Droste--Einstein--Infeld--Hoffmann
(LDEIH) equations became the standard 1PN description of relativistic
\(N\)-body dynamics.

In parallel, post-Newtonian methods were developed for continuous matter
and extended self-gravitating bodies. Chandrasekhar
~\cite{1965ApJ...142.1488C} derived the general-relativistic hydrodynamic
equations and conservation laws through 1PN order, and Chandrasekhar and
Nutku~\cite{1969ApJ...158...55C} extended them to 2PN order. These works
established a systematic PN framework for self-gravitating fluids.

Another line of development concerned relativistic reference systems for
celestial mechanics. Brumberg and Kopeikin developed a relativistic
global--local reference-system framework, including barycentric and
geocentric harmonic reference systems and their relation through
asymptotic matching
~\cite{BrumbergKopeikin1989Earth,BrumbergKopeikin1989Frames}.
Damour, Soffel, and Xu subsequently developed the DSX formalism
~\cite{PhysRevD.43.3273,PhysRevD.45.1017,PhysRevD.47.3124,PhysRevD.49.618},
which provided a systematic 1PN framework for weakly self-gravitating
extended bodies using global and body-centered local reference systems,
with internal structure described in terms of relativistic mass and spin
multipole moments. The Brumberg--Kopeikin and DSX approaches together
contributed to the relativistic reference-system framework later adopted
in the IAU 2000 resolutions
~\cite{Transactions,Soffel2003IAU2000}.

The treatment of extended-body dynamics with general multipole structure
was developed further in subsequent work. Kopeikin derived covariant
translational and rotational equations of motion for \(N\) extended bodies
with arbitrary mass and spin multipoles
~\cite{Kopeikin2019Multipoles}, and later obtained the corresponding
post-Newtonian \(N\)-body Lagrangian including arbitrary mass and spin
multipoles~\cite{Kopeikin2020Lagrangian}.

For compact binaries, the post-Newtonian dynamics has been developed in considerable detail,
and the conservative equations of motion through 2PN order are well established in complementary
harmonic-coordinate formulations. Blanchet, Faye, Ponsot, and collaborators developed a point-
particle formulation in which compact objects are represented by singular sources and the associated
self-field singularities are treated by Hadamard regularization~\cite{blanchet1998gravitational,Blanchet_2000,blanchet2024postnewtoniantheorygravitationalwaves}.
A complementary harmonic-coordinate approach is the Direct Integration of the relaxed Einstein equations
(DIRE) framework developed by Pati and Will, with applications to 2PN binary dynamics using extended fluid sources
by Pati and Will and by Mitchell and Will~\cite{PhysRevD.62.124015,PhysRevD.65.104008,PhysRevD.75.124025}
(see also Will's review~\cite{Will2014} for a broader post-Newtonian context).
Although the point-particle and extended-matter approaches differ in their treatment of the
sources, their harmonic-coordinate 2PN potentials share related logarithmic triangle-kernel structures~\cite{PhysRevD.65.104008}.

Will also studied systems with a dominant central mass and hierarchical triples, emphasizing
PN cross terms that couple two-body relativistic effects to Newtonian perturbations from other
bodies~\cite{PhysRevD.89.044043,Will2014Triple}.

For general point-mass \(N\)-body dynamics at 2PN order, obtaining comparably explicit formulations has been more challenging.
Ohta et al.~\cite{10.1143/PTP.50.492,10.1143/PTP.51.1220} carried out an
early systematic study of the many-body structure at this order, deriving
the metric in a class of asymptotically Minkowskian coordinate systems and
obtaining important potential, Lagrangian, and Hamiltonian contributions.
Their work showed that genuinely nonlinear many-body interactions arise at
2PN order.

Damour and Sch\"afer~\cite{1985Lagrangians} later clarified the
coordinate dependence of the 2PN point-mass Lagrangian, showing that it
generally depends on accelerations in harmonic coordinates but can be made
acceleration-free in special coordinates such as the Arnowitt--Deser--Misner (ADM) gauge. Within the ADM
framework, Sch\"afer~\cite{SCHAFER1987336} derived the 2PN
three-body Hamiltonian, including genuine three-body interactions. Combinatorial shortcomings
in its reduction from the general many-body Hamiltonian were subsequently identified and corrected
by Lousto and Nakano~\cite{Lousto_2008} and by Galaviz and Br\"ugmann~\cite{PhysRevD.83.084013}.
Chu~\cite{Chu2009NBody2PN} derived the general \(N\)-body
effective Lagrangian through 2PN order using a perturbative field-theory
approach with a de Donder gauge choice, leaving certain integrals
unevaluated.

More recently, Heinze, Sch\"afer, and
Br\"ugmann~\cite{heinze2026nbody2pnhamiltoniannumerical} extended the ADM
Hamiltonian formulation to general \(N\), including the genuinely
four-body 2PN contribution and numerically evaluating the remaining
integral term with high accuracy. Their work provides a semianalytic and
numerical formulation of the general \(N\)-body 2PN Hamiltonian in the ADM
gauge.

To formulate general \(N\)-body dynamics directly in harmonic coordinates
at 2PN order, we adopt the point-mass post-Newtonian formalism and Hadamard
regularization framework developed by Blanchet, Faye, and collaborators
~\cite{blanchet1998gravitational,Blanchet_2000,%
blanchet2024postnewtoniantheorygravitationalwaves}.
Starting from the harmonic-gauge metric potentials, we derive the
corresponding equations of motion (EOM) and decompose them into a closed analytic contribution
and a nonclosed integral contribution, denoted by \((I_k)_A\), where
\(A\) labels the particle under consideration. The latter is termed
nonclosed because it has not yet been reduced to a finite algebraic
expression in the particle variables; this terminology does not imply
that such a reduction is mathematically impossible.

Our main focus is the nonclosed contribution. For fixed \(A\), its terms
are separated into the \(D\neq A\) and \(D=A\) sectors according to the
center \(D\) of the singular kernel. Their distributional and self-singular
structures are treated separately, yielding a semianalytic representation
consisting of explicit analytic terms and numerically evaluable spatial
integrals.

We apply the formulation to two representative three-body systems,
Sun--Jupiter--Saturn (SJS) and Sun--Mercury--Venus (SMV). In both cases,
we evaluate the instantaneous nonclosed 2PN acceleration along Newtonian
trajectories, characterize its absolute magnitude and its fraction of the
complete 2PN acceleration, and solve the corresponding perturbation
equations for the leading finite-time relative-distance perturbation. The
two systems provide outer- and inner-planet benchmarks for the same 2PN
contribution.

The remainder of the paper is organized as follows.
Section~\ref{sec:notation-conventions} introduces the notation and
conventions. Section~\ref{sec:harmonic-gauge-2pn-eom} derives the
harmonic-gauge 2PN EOM within Hadamard regularization.
Section~\ref{sec:singular-integral-reformulation} analyzes the singular
structure of the nonclosed contribution and develops its integral
reformulation. Section~\ref{sec:numerical-evaluation-nonclosed} describes
the numerical evaluation, and
Sec.~\ref{sec:numerical-results-benchmarks} presents the instantaneous
nonclosed 2PN acceleration and leading finite-time relative-distance
perturbation for the two benchmark systems. Finally,
Sec.~\ref{sec:conclusion} summarizes the main results.

\section{Notation and Conventions}
\label{sec:notation-conventions}

Throughout this paper, we consider a system of \(N\) point particles with
constant masses \(m_A\). Capital Latin letters \(A,B,C,\ldots\) denote
particle labels in \(\{1,\ldots,N\}\). Repeated particle labels are not
summed unless an explicit summation sign is present. When a quantity
carries a free particle label \(A\), this label identifies the particle
whose EOM or particle value is being evaluated.

We adopt the metric signature \((-+++)\), with
\begin{align}
    \eta_{\mu\nu}
    =
    \operatorname{diag}(-1,1,1,1).
\end{align}
Greek indices \(\mu,\nu,\rho,\sigma,\ldots\) run over spacetime components
\(0,1,2,3\), while lowercase Latin indices \(i,j,k,\ldots\) run over
spatial components \(1,2,3\). Repeated spacetime and spatial indices are
summed. Spatial indices are raised and lowered with the Euclidean metric
\(\delta_{ij}\).

The position of particle \(A\) is denoted by
\(\mathbf z_A(t)=(z_A^i(t))\), with velocity and acceleration
\begin{align}
    v_A^i
    &\equiv
    \frac{d z_A^i}{dt},
    &
    a_A^i
    &\equiv
    \frac{d v_A^i}{dt}.
\end{align}
For a field point \(\mathbf x\), we define
\begin{align}
    r_A
    &\equiv
    |\mathbf x-\mathbf z_A|,
    &
    n_A^i
    &\equiv
    \frac{x^i-z_A^i}{r_A}.
\end{align}
For an auxiliary integration point \(\mathbf x'\), we similarly define
\begin{align}
    r'_A
    &\equiv
    |\mathbf x'-\mathbf z_A|,
    &
    n_A'{}^i
    &\equiv
    \frac{x'{}^i-z_A^i}{r'_A}.
\end{align}
For two distinct particles \(A\ne B\),
\begin{align}
    r_{AB}
    &\equiv
    |\mathbf z_A-\mathbf z_B|,
    &
    n_{AB}^i
    &\equiv
    \frac{z_A^i-z_B^i}{r_{AB}},
\end{align}
so that
\begin{align}
    r_{BA}
    &=
    r_{AB},
    &
    n_{BA}^i
    &=
    -n_{AB}^i.
\end{align}
All spatial dot products are Euclidean; for example,
\begin{align}
    \mathbf v_A\!\cdot\!\mathbf v_B
    =
    v_A^i v_B^i.
\end{align}

Derivatives with respect to the field point, auxiliary integration point,
and particle positions are denoted by
\begin{align}
    \partial_i
    &\equiv
    \frac{\partial}{\partial x^i},
    &
    \partial_i'
    &\equiv
    \frac{\partial}{\partial x'{}^i},
    &
    (\partial_A)_i
    &\equiv
    \frac{\partial}{\partial z_A^i}.
\end{align}
In \((\partial_A)_i\), the field or integration point and all other
particle positions are held fixed. In particular,
\begin{align}
    \partial_i r_A
    &=
    n_A^i,
    &
    \partial_i' r'_A
    &=
    n_A'{}^i,
    &
    (\partial_A)_i r_A
    &=
    -n_A^i,
    &
    (\partial_A)_i r'_A
    &=
    -n_A'{}^i.
\end{align}

For a singular quantity \(P\), the notation
\begin{align}
    (P)_A
\end{align}
denotes its Hadamard finite value at \(\mathbf z_A\). This operation acts
on the complete expression; thus, \((V\partial_iV)_A\) is not in general
equal to \((V)_A(\partial_iV)_A\) unless this relation is established
separately. Likewise, in \((\partial_iP)_A\), the derivative is taken
before the finite value is assigned.

Post-Newtonian orders are counted in powers of \(1/c\). The notation
\(O(c^{-n})\) denotes terms proportional to \(c^{-n}\) and higher inverse
powers of \(c\).

\section{Harmonic-Gauge 2PN Equations of Motion}
\label{sec:harmonic-gauge-2pn-eom}

The derivation in this section follows the harmonic-coordinate point-mass
post-Newtonian framework developed by Blanchet, Faye, and
Ponsot~\cite{blanchet1998gravitational,Blanchet_2000,blanchet2024postnewtoniantheorygravitationalwaves}.
We use their metric potentials, point-mass source prescription, and
Hadamard finite-part treatment as the starting point, and then organize the
resulting 2PN equations into the working form used in the later
analysis.

\subsection{Harmonic-Gauge Metric, Near-Zone Expansion, and Point-Mass Source}
\label{sec:metric-point-mass-source}

We use the standard harmonic-gauge 2PN metric written in terms of the
potentials \(V\), \(V_i\), \(\hat W_{ij}\), \(\hat R_i\), and \(\hat X\):
\begin{align}
    g_{00}
    &=
    -1
    +\frac{2V}{c^2}
    -\frac{2V^2}{c^4}
    +\frac{8}{c^6}
    \left(
        \hat X
        +V_iV_i
        +\frac{1}{6}V^3
    \right)
    +O(c^{-8}),
    \nonumber\\
    g_{0i}
    &=
    -\frac{4V_i}{c^3}
    -\frac{8\hat R_i}{c^5}
    +O(c^{-7}),
    \nonumber\\
    g_{ij}
    &=
    \delta_{ij}
    \left(
        1+\frac{2V}{c^2}+\frac{2V^2}{c^4}
    \right)
    +\frac{4\hat W_{ij}}{c^4}
    +O(c^{-6}).
    \label{eq:2pn-harmonic-metric}
\end{align}

These potentials are defined as retarded solutions of flat-space wave
equations. With our conventions,
\begin{align}
    \Box
    &\equiv
    \Delta-\frac{1}{c^2}\partial_t^2,
    &
    \Delta
    &\equiv
    \partial_i\partial_i .
\end{align}
For a source \(F\), the retarded inverse of the wave operator is
\begin{align}
    (\Box_R^{-1}F)(\mathbf x,t)
    &=
    -\frac{1}{4\pi}
    \int
    \frac{
        F\!\left(
            \mathbf x',
            t-\dfrac{|\mathbf x-\mathbf x'|}{c}
        \right)
    }
    {|\mathbf x-\mathbf x'|}
    \,d^3x' .
    \label{eq:retarded-inverse-definition}
\end{align}

In the post-Newtonian near zone, the retarded solution is expanded in
powers of \(1/c\). For the conservative dynamics through 2PN order, only
the instantaneous even-power part is required:
\begin{align}
    \Box_{\rm inst}^{-1}
    &\equiv
    \sum_{q\geq0}
    \frac{1}{c^{2q}}
    \partial_t^{2q}\Delta^{-q-1}.
    \label{eq:box-inst-definition}
\end{align}
Here, the Poisson inverse is defined by
\begin{align}
    (\Delta^{-1}F)(\mathbf x,t)
    &=
    -\frac{1}{4\pi}
    \int
    \frac{F(\mathbf x',t)}
    {|\mathbf x-\mathbf x'|}
    \,d^3x',
    \qquad
    \Delta(\Delta^{-1}F)=F,
    \label{eq:poisson-inverse-definition}
\end{align}
and its iterates are denoted by
\begin{align}
    \Delta^{-q-1}
    &\equiv
    (\Delta^{-1})^{q+1},
    \qquad q\geq0 .
    \label{eq:iterated-poisson-inverse-definition}
\end{align}
The omitted time-antisymmetric part of the retarded solution gives rise to
radiation-reaction effects, whose leading contribution to the equations of
motion enters at 2.5PN order and is therefore not included here.

The matter source is modeled as a system of point particles. For each
particle \(A\), we introduce the coordinate velocity
\begin{align}
    v_A^\mu
    &=
    (c,\mathbf v_A).
\end{align}
The corresponding stress-energy tensor is
\begin{align}
    T^{\mu\nu}(\mathbf x,t)
    &=
    \sum_A
    \mu_A(t)\,
    v_A^\mu v_A^\nu\,
    \delta^{(3)}
    \bigl(\mathbf x-\mathbf z_A(t)\bigr),
    \label{eq:point-mass-stress-energy}
\end{align}
where
\begin{align}
    \mu_A(t)
    &=
    \left[
        \frac{m_A}{
            \sqrt{-g}\,
            \sqrt{
                -g_{\rho\sigma}
                \dfrac{v_A^\rho v_A^\sigma}{c^2}
            }
        }
    \right]_{\mathbf x=\mathbf z_A(t)} .
    \label{eq:point-mass-mu}
\end{align}

Because the point-particle sources are singular, the metric potentials
and their derivatives are treated within the Hadamard partie-finie
pseudofunction framework of Blanchet and Faye~\cite{Blanchet_2000}.
In this formalism, singular functions and delta sources are extended as
pseudofunctions, allowing particle values, derivatives, and singular
spatial integrals to be defined consistently. The specific prescriptions
used below are summarized in what follows.

For point-particle dynamics, Hadamard regularization is sufficient through 2PN order. At higher PN orders, beginning at 3PN, dimensional regularization is needed to resolve ambiguities in the Hadamard treatment~\cite{blanchet2024postnewtoniantheorygravitationalwaves,SchaeferJaranowski2024}.

Substituting the point-mass source into the wave equations for the
potentials and applying the instantaneous post-Newtonian expansion yields
the explicit point-mass potentials used throughout this work. Their
expressions are collected in Appendix~\ref{app:2pn-metric-potentials}.

\subsection{Momentum--Force Form of the 2PN EOM}
\label{sec:momentum-force-eom}

The EOM follow from the geodesic equation for each point
particle.  Singular metric quantities at the position of particle \(A\)
are evaluated using the Hadamard partie finie prescription
\cite{Blanchet_2000}.  It is convenient to write the equations in the
Newtonian-like momentum--force form
\begin{align}
    \frac{d\mathcal P_A^i}{dt}
    =
    \mathcal F_A^i ,
    \label{eq:eom-momentum-force-form}
\end{align}
where
\begin{align}
    \mathcal P_A^i
    &\equiv
    \left(
        \frac{
            v_A^\mu g_{i\mu}
        }{
            \sqrt{
                -g_{\rho\sigma}
                \dfrac{v_A^\rho v_A^\sigma}{c^2}
            }
        }
    \right)_A,
    &
    \mathcal F_A^i
    &\equiv
    \frac{1}{2}
    \left(
        \frac{
            v_A^\mu v_A^\nu \partial_i g_{\mu\nu}
        }{
            \sqrt{
                -g_{\rho\sigma}
                \dfrac{v_A^\rho v_A^\sigma}{c^2}
            }
        }
    \right)_A .
    \label{eq:momentum-force-definitions}
\end{align}

Here and below, \((\cdots)_A\) denotes the Hadamard finite value of the
complete singular expression enclosed by the parentheses.  This operation
is not generally distributive:
\begin{align}
    (FG)_A
    \neq
    (F)_A(G)_A .
    \label{eq:hadamard-nondistributive}
\end{align}
Thus, quantities such as \((V^2)_A\), \((V V_i)_A\), and
\((V\,\partial_i V_j)_A\) must be regularized as complete expressions
rather than formed from separately regularized factors.  The total time
derivatives below act on the corresponding regularized quantities along
the trajectory \(\mathbf z_A(t)\).

Substituting the 2PN metric into
Eq.~\eqref{eq:momentum-force-definitions} and differentiating the
generalized momentum gives
\begin{align}
    \frac{d\mathcal P_A^i}{dt}
    &=
    a_A^i
    \nonumber\\
    &\quad
    +\frac{1}{c^2}
    \Bigg[
        -4\,\frac{d}{dt}(V_i)_A
        +3\,\frac{d}{dt}(V)_A\,v_A^i
        +3(V)_A a_A^i
        +(\mathbf v_A\!\cdot\!\mathbf a_A)v_A^i
        +\frac{1}{2}v_A^2a_A^i
    \Bigg]
    \nonumber\\
    &\quad
    +\frac{1}{c^4}
    \Bigg[
        -8\,\frac{d}{dt}(\hat R_i)_A
        +\frac{9}{2}\frac{d}{dt}(V^2)_A\,v_A^i
        +\frac{9}{2}(V^2)_A a_A^i
    \nonumber\\
    &\qquad\qquad
        +4\,\frac{d}{dt}(\hat W_{ij})_A\,v_A^j
        +4(\hat W_{ij})_A a_A^j
        -4\,\frac{d}{dt}(V V_i)_A
    \nonumber\\
    &\qquad\qquad
        +\frac{7}{2}\frac{d}{dt}(V)_A\,v_A^2v_A^i
        +7(V)_A
            (\mathbf v_A\!\cdot\!\mathbf a_A)v_A^i
        +\frac{7}{2}(V)_A v_A^2a_A^i
    \nonumber\\
    &\qquad\qquad
        -2v_A^2\frac{d}{dt}(V_i)_A
        -4(\mathbf v_A\!\cdot\!\mathbf a_A)(V_i)_A
    \nonumber\\
    &\qquad\qquad
        -4a_A^i v_A^j(V_j)_A
        -4v_A^i a_A^j(V_j)_A
        -4v_A^i v_A^j\frac{d}{dt}(V_j)_A
    \nonumber\\
    &\qquad\qquad
        +\frac{3}{2}v_A^2
            (\mathbf v_A\!\cdot\!\mathbf a_A)v_A^i
        +\frac{3}{8}v_A^4a_A^i
    \Bigg]
    +O(c^{-6}).
    \label{eq:dPdt-2pn}
\end{align}

The corresponding force is
\begin{align}
    \mathcal F_A^i
    &=
    (\partial_iV)_A
    +\frac{1}{c^2}
    \left[
        -(V\partial_iV)_A
        +\frac{3}{2}v_A^2(\partial_iV)_A
        -4v_A^j(\partial_iV_j)_A
    \right]
    \nonumber\\
    &\quad
    +\frac{1}{c^4}
    \left[
        4(\partial_i\hat X)_A
        +8(V_j\partial_iV_j)_A
        -8v_A^j(\partial_i\hat R_j)_A
        +\frac{9}{2}v_A^2(V\partial_iV)_A
    \right.
    \nonumber\\
    &\qquad\qquad
        +2v_A^jv_A^k(\partial_i\hat W_{jk})_A
        -2v_A^2v_A^j(\partial_iV_j)_A
        +\frac{7}{8}v_A^4(\partial_iV)_A
        +\frac{1}{2}(V^2\partial_iV)_A
    \nonumber\\
    &\qquad\qquad\left.
        -4v_A^j(V_j\partial_iV)_A
        -4v_A^j(V\partial_iV_j)_A
    \right]
    +O(c^{-6}).
    \label{eq:force-term-2pn}
\end{align}

Equating Eqs.~\eqref{eq:dPdt-2pn} and
\eqref{eq:force-term-2pn} gives an implicit equation for
\(\mathbf a_A\), because accelerations remain inside the post-Newtonian
corrections.  We next rearrange this equation and eliminate these
accelerations by order reduction.

\subsection{Working Form of the 2PN Acceleration}
\label{sec:working-acceleration-order-reduction}

Solving Eqs.~\eqref{eq:dPdt-2pn} and
\eqref{eq:force-term-2pn} for \(a_A^i\) gives the following unreduced
working form:
\begin{align}
a_A^i
&=
\regtag{(\partial_iV)_A}{eq:reg-dV-A}
\nonumber\\
&\quad
+\frac{1}{c^2}
\Bigg[
    -\regtag{(V\partial_iV)_A}{eq:reg-VdV-A}
    +\frac{3}{2}v_A^2\regtag{(\partial_iV)_A}{eq:reg-dV-A}
    -4v_A^j\regtag{(\partial_iV_j)_A}{eq:reg-dVi-A}
\nonumber\\
&\qquad\qquad
    +4\,\regtag{\frac{d}{dt}(V_i)_A}{eq:reg-dt-Vi-A}
    -3\,\regtag{\frac{d}{dt}(V)_A}{eq:reg-dt-V-A}\,v_A^i
    -3\regtag{(V)_A}{eq:reg-V-A} a_A^i
    -(\mathbf v_A\!\cdot\!\mathbf a_A)v_A^i
    -\frac{1}{2}v_A^2a_A^i
\Bigg]
\nonumber\\
&\quad
+\frac{1}{c^4}
\Bigg[
    4\regtag{(\partial_i\hat X)_A}{eq:reg-dXhat-A}
    +8\regtag{(V_j\partial_iV_j)_A}{eq:reg-VjdVj-A}
    -8v_A^j\regtag{(\partial_i\hat R_j)_A}{eq:reg-dRhat-A}
    +\frac{9}{2}v_A^2\regtag{(V\partial_iV)_A}{eq:reg-VdV-A}
\nonumber\\
&\qquad\qquad
    +2v_A^jv_A^k\regtag{(\partial_i\hat W_{jk})_A}{eq:reg-dWhat-A}
    -2v_A^2v_A^j\regtag{(\partial_iV_j)_A}{eq:reg-dVi-A}
    +\frac{7}{8}v_A^4\regtag{(\partial_iV)_A}{eq:reg-dV-A}
    +\frac{1}{2}\regtag{(V^2\partial_iV)_A}{eq:reg-V2dV-A}
\nonumber\\
&\qquad\qquad
    -4v_A^j\regtag{(V_j\partial_iV)_A}{eq:reg-VjdV-A}
    -4v_A^j\regtag{(V\partial_iV_j)_A}{eq:reg-VdVj-A}
\nonumber\\
&\qquad\qquad
    +8\,\regtag{\frac{d}{dt}(\hat R_i)_A}{eq:reg-dt-Rhat-A}
    -\frac{9}{2}\regtag{\frac{d}{dt}(V^2)_A}{eq:reg-dt-V2-A}\,v_A^i
    -\frac{9}{2}\regtag{(V^2)_A}{eq:reg-V2-A} a_A^i
\nonumber\\
&\qquad\qquad
    -4\,\regtag{\frac{d}{dt}(\hat W_{ij})_A}{eq:reg-dt-What-A}\,v_A^j
    -4\regtag{(\hat W_{ij})_A}{eq:reg-What-A} a_A^j
    +4\,\regtag{\frac{d}{dt}(V V_i)_A}{eq:reg-dt-VVi-A}
\nonumber\\
&\qquad\qquad
    -\frac{7}{2}\regtag{\frac{d}{dt}(V)_A}{eq:reg-dt-V-A}\,v_A^2v_A^i
    -7\regtag{(V)_A}{eq:reg-V-A}(\mathbf v_A\!\cdot\!\mathbf a_A)v_A^i
    -\frac{7}{2}\regtag{(V)_A}{eq:reg-V-A} v_A^2a_A^i
\nonumber\\
&\qquad\qquad
    +2v_A^2\regtag{\frac{d}{dt}(V_i)_A}{eq:reg-dt-Vi-A}
    +4(\mathbf v_A\!\cdot\!\mathbf a_A)\regtag{(V_i)_A}{eq:reg-Vi-A}
\nonumber\\
&\qquad\qquad
    +4a_A^i v_A^j\regtag{(V_j)_A}{eq:reg-Vi-A}
    +4v_A^i a_A^j\regtag{(V_j)_A}{eq:reg-Vi-A}
    +4v_A^i v_A^j\regtag{\frac{d}{dt}(V_j)_A}{eq:reg-dt-Vi-A}
\nonumber\\
&\qquad\qquad
    -\frac{3}{2}v_A^2
        (\mathbf v_A\!\cdot\!\mathbf a_A)v_A^i
    -\frac{3}{8}v_A^4a_A^i
\Bigg]
+O(c^{-6}) .
\label{eq:eom-2pn}
\end{align}

The small equation numbers placed above the regularized quantities and
their total time derivatives in Eq.~\eqref{eq:eom-2pn} are cross-reference
tags to Appendix~\ref{app:regularized-quantities-2pn-eom}. 

The accelerations on the right-hand side are removed order by order.  In
the \(c^{-2}\) terms, \(\mathbf a_A\) is replaced by its expression
accurate through 1PN order, whereas in the \(c^{-4}\) terms only the
Newtonian acceleration is required.  The same rule applies to
accelerations generated by the total time derivatives.

The required lower-order expansion is
\begin{align}
    a_A^i
    &=
    (a_A^{\mathrm{N}})^i
    +\frac{1}{c^2}(a_A^{\mathrm{1PN}})^i
    +O(c^{-4}),
    \label{eq:pn-acceleration-expansion}\\[1mm]
    (a_A^{\mathrm{N}})^i
    &=
    -
    \sum_{B\neq A}
    \frac{Gm_B}{r_{AB}^2}\,
    n_{AB}^i,
    \label{eq:newtonian-acceleration-main}\\[1mm]
    (a_A^{\mathrm{1PN}})^i
    &=
    -
    \sum_{B\neq A}
    \frac{Gm_B}{r_{AB}^2}\,
    n_{AB}^i
    \Bigg[
        v_A^2
        +2v_B^2
        -4\,\mathbf v_A\!\cdot\!\mathbf v_B
        -\frac{3}{2}
        \bigl(\mathbf n_{AB}\!\cdot\!\mathbf v_B\bigr)^2
    \nonumber\\
    &\qquad\qquad
        -4\sum_{C\neq A}\frac{Gm_C}{r_{AC}}
        -
        \sum_{C\neq B}
        \frac{Gm_C}{r_{BC}}
        \left(
            1
            +\frac{1}{2}
            \frac{r_{AB}}{r_{CB}}\,
            \mathbf n_{AB}\!\cdot\!\mathbf n_{CB}
        \right)
    \Bigg]
    \nonumber\\
    &\qquad
    -
    \frac{7}{2}
    \sum_{B\neq A}
    \sum_{C\neq B}
    \frac{G^2m_Bm_C}{r_{AB}r_{BC}^2}\,
    n_{BC}^i
    \nonumber\\
    &\qquad
    +
    \sum_{B\neq A}
    \frac{Gm_B}{r_{AB}^2}\,
    (v_A^i-v_B^i)
    \left(
        4\,\mathbf n_{AB}\!\cdot\!\mathbf v_A
        -3\,\mathbf n_{AB}\!\cdot\!\mathbf v_B
    \right).
    \label{eq:one-pn-acceleration-main}
\end{align}

Using Eqs.~\eqref{eq:pn-acceleration-expansion}--%
\eqref{eq:one-pn-acceleration-main} and the appendix replacements tagged in
Eq.~\eqref{eq:eom-2pn}, and then discarding terms beyond relative 2PN order,
we obtain an explicit acceleration in the particle variables.  After this
substitution, all contributions reduce to closed finite sums except for a
single nonclosed integral contained in \((\partial_i\hat X)_A\), which is
isolated below.

\subsection{Nonclosed Integral Contribution}
\label{sec:practical-nonclosed-contribution}

All contributions to Eq.~\eqref{eq:eom-2pn} reduce to closed finite sums
in the particle variables except for one part of
\((\partial_i\hat X)_A\). Denoting its free spatial index by \(k\), we
define
\begin{align}
    (I_k)_A
    \equiv
    \frac{G^3}{4\pi}
    \sum_{B\ne C}
    \sum_{D\ne B,C}
    m_Bm_Cm_D
    \operatorname{Pf}\int_{\mathbb R^3}
    &
    (\partial_B)_i(\partial_C)_j
    \ln(r'_B+r'_C+r_{BC})
    \nonumber\\
    &\times
    (\partial_A)_k
    \frac{1}{r'_A}
    (\partial_D)_i(\partial_D)_j
    \frac{1}{r'_D}\,
    d^3x' .
    \label{eq:nonclosed-integral-IA}
\end{align}

Here \(\mathbf x'\) is the integration variable,
\(r'_P=|\mathbf x'-\mathbf z_P|\), and
\((\partial_P)_i\equiv\partial/\partial z_P^i\). In particular,
\((\partial_A)_k\) acts on the dependence of \(1/r'_A\) on
\(\mathbf z_A\). The particle label \(A\) is held fixed as the particle
at which the EOM is evaluated, whereas \(B,C,D\) are summation labels.
The summations are restricted by \(B\ne C\) and \(D\ne B,C\).
After the \(D=A\) and \(D\ne A\) sectors are separated below,
\(\sum_{D\ne A,B,C}\) denotes the latter restriction.
The notation \(\operatorname{Pf}\int\) denotes the Hadamard finite part
of the spatial integral.

The singular second derivative of \(1/r'_D\) in
Eq.~\eqref{eq:nonclosed-integral-IA} is treated within the Hadamard
partie-finie pseudofunction framework of Blanchet and
Faye~\cite{Blanchet_2000}. In this framework,
\begin{align}
    (\partial_D)_i(\partial_D)_j\frac{1}{r'_D}
    =
    \operatorname{Pf}\left[
        \frac{
            3n_D'{}^i n_D'{}^j-\delta_{ij}
        }{(r'_D)^3}
    \right]
    -
    \frac{4\pi}{3}\,
    \delta_{ij}\delta_D ,
    \label{eq:distributional-second-derivative}
\end{align}
where
\begin{align}
    n_D'{}^i
    &=
    \frac{x'{}^i-z_D^i}{r'_D},
    &
    \delta_D
    &\equiv
    \delta^{(3)}(\mathbf x'-\mathbf z_D).
\end{align}
Here \(\delta_D\) is understood as the corresponding delta-supported
pseudofunction, whose action reduces to that of the ordinary Dirac delta
on smooth test functions.

To distinguish the pseudofunction extension of the singular kernel from
the finite part of a spatial integral, we define
\begin{align}
    \widehat{\mathcal T}_{ij}^{(D)}(\mathbf x')
    \equiv
    \frac{
        3n_D'{}^i n_D'{}^j-\delta_{ij}
    }{(r'_D)^3}.
\end{align}
Its action on a smooth test function \(\varphi\) is
\begin{align}
    \left\langle
        \operatorname{Pf}[\widehat{\mathcal T}_{ij}^{(D)}],\varphi
    \right\rangle
    \equiv
    \operatorname{Pf}\int
    \widehat{\mathcal T}_{ij}^{(D)}(\mathbf x')\,
    \varphi(\mathbf x')\,d^3x' .
\end{align}
Thus, \(\operatorname{Pf}[\widehat{\mathcal T}_{ij}^{(D)}]\) denotes the
Hadamard pseudofunction associated with the singular kernel, whereas
\(\operatorname{Pf}\int\) denotes the finite part of the corresponding
spatial integral. The delta-supported term in
Eq.~\eqref{eq:distributional-second-derivative} is part of the
pseudofunction derivative and cannot be omitted.

The derivation of Eq.~\eqref{eq:nonclosed-integral-IA} from the
non-compact part of \(\hat X\), together with the relation between its
field-point expression and particle finite value, is given in
Appendix~\ref{app:nonclosed-origin}. Its structure is closely related to
the unresolved integral appearing in the metric construction of
Ref.~\cite{10.1143/PTP.50.492} and to the nonlinear many-body sectors
studied in the ADM Hamiltonian formulation of
Ref.~\cite{heinze2026nbody2pnhamiltoniannumerical}.

The remaining non-explicit part of the 2PN EOM is therefore contained
entirely in \((I_k)_A\). In the following section, we separate the
\(D\ne A\) and \(D=A\) sectors, analyze their distinct singular
structures, and derive the corresponding integral forms used for
numerical evaluation.

\section{Singular Structure and Integral Reformulation
of the Nonclosed Contribution}
\label{sec:singular-integral-reformulation}

We now analyze the singular structure of the nonclosed integral
\((I_k)_A\) defined in Eq.~\eqref{eq:nonclosed-integral-IA}. For brevity,
we introduce the elementary potential and its smooth two-center coefficient
\begin{align}
    g'_{BC}
    \equiv
    \ln(r'_B+r'_C+r_{BC}),
    \qquad
    \mathcal K_{ij}^{(BC)}(\mathbf x')
    :=
    (\partial_B)_i(\partial_C)_j g'_{BC}.
\end{align}

For a fixed evaluation particle \(A\), the treatment of each term in the
sum over \(D\) depends on whether the singular center \(\mathbf z_D\)
coincides with the evaluation point \(\mathbf z_A\). We therefore
decompose
\begin{align}
    (I_k)_A
    =
    (I_k^{D=A})_A
    +
    (I_k^{D\ne A})_A .
\end{align}
The \(D\ne A\) and \(D=A\) sectors have different singular structures and
are regularized separately in the following subsections.

\subsection{\texorpdfstring{Distributional Split for
the \(D\neq A\) Sector}{Distributional Split for the D neq A Sector}}

We begin with the \(D\neq A\) contribution,
\begin{align}
    (I_k^{D\neq A})_A
    &=
    \frac{G^3}{4\pi}
    \sum_{B\ne C}
    \sum_{D\ne A,B,C}
    m_Bm_Cm_D
    \int_{\mathbb R^3}
    (\partial_B)_i(\partial_C)_j g'_{BC}\,
    (\partial_A)_k\frac{1}{r'_A}\,
    (\partial_D)_i(\partial_D)_j\frac{1}{r'_D}\,
    d^3x' .
\end{align}
Introduce
\begin{align}
    \mathcal K_{ij}^{(BC)}(\mathbf x')
    &:=
    (\partial_B)_i(\partial_C)_j g'_{BC},
    \\
    (\mathcal Y_{ijk}^{(BC)}(\mathbf x'))_A
    &:=
    \mathcal K_{ij}^{(BC)}(\mathbf x')\,
    (\partial_A)_k\frac{1}{r'_A},
    \\
    \mathcal T_{ij}^{(D)}
    &:=
    (\partial_D)_i(\partial_D)_j\frac{1}{r'_D}
    =
    \partial_i'\partial_j'\frac{1}{r'_D},
    \\
    \widehat{\mathcal T}_{ij}^{(D)}(\mathbf x')
    &:=
    \frac{
        3n_D'{}^i n_D'{}^j-\delta_{ij}
    }{
        (r'_D)^3
    }.
\end{align}

The factor \((\mathcal Y_{ijk}^{(BC)})_A\) is an ordinary locally
integrable function.  In particular,
\begin{align}
    (\partial_A)_k\frac{1}{r'_A}
    =
    \frac{n_A'{}^k}{(r'_A)^2}
\end{align}
is locally integrable in three dimensions, while
\(\mathcal K_{ij}^{(BC)}\) is locally bounded at the particle
positions.  Hence \((\mathcal Y_{ijk}^{(BC)})_A\) defines a regular
distribution and carries no delta-supported contact contribution of its
own.

The distributional singularity relevant to the present sector is
contained in \(\mathcal T_{ij}^{(D)}\).  Since \(D\neq A,B,C\),
\((\mathcal Y_{ijk}^{(BC)})_A\) is smooth in a neighborhood of
\(\mathbf z_D\), which is the singular support of
\(\mathcal T_{ij}^{(D)}\).  Away from \(\mathbf z_D\),
\(\mathcal T_{ij}^{(D)}\) coincides with the ordinary kernel
\(\widehat{\mathcal T}_{ij}^{(D)}\).

Using Eq.~\eqref{eq:distributional-second-derivative},
\begin{align}
    \mathcal T_{ij}^{(D)}
    &=
    \operatorname{Pf}
    \bigl[
        \widehat{\mathcal T}_{ij}^{(D)}
    \bigr]
    -
    \frac{4\pi}{3}\,
    \delta_{ij}\delta_D,
\end{align}
we obtain
\begin{align}
    \int_{\mathbb R^3}
    (\mathcal Y_{ijk}^{(BC)})_A\,
    \mathcal T_{ij}^{(D)}\,d^3x'
    &=
    \operatorname{Pf}\!\int_{\mathbb R^3}
    (\mathcal Y_{ijk}^{(BC)})_A\,
    \widehat{\mathcal T}_{ij}^{(D)}\,d^3x'
    \nonumber\\
    &\quad
    -
    \frac{4\pi}{3}\,
    \delta_{ij}
    (\mathcal Y_{ijk}^{(BC)}(\mathbf z_D))_A .
    \label{eq:DneqA-distributional-split}
\end{align}
The finite part is defined by spherical excision centered at
\(\mathbf z_D\).  The second term is the contact contribution associated
with the delta-supported part of the distributional second derivative.

Since all remaining factors are regular at \(\mathbf z_D\), the contact
term can be evaluated analytically.  Using
\(\mathbf n'_A(\mathbf z_D)=-\mathbf n_{AD}\) together with the trace of
\(\mathcal K_{ij}^{(BC)}\), one finds
\begin{align}
    -\frac{4\pi}{3}\,
    \delta_{ij}
    (\mathcal Y_{ijk}^{(BC)}(\mathbf z_D))_A
    &=
    \frac{2\pi}{3r_{AD}^2}
    \left(
        \frac{1}{r_{BD}r_{CD}}
        -
        \frac{1}{r_{BD}r_{BC}}
        -
        \frac{1}{r_{CD}r_{BC}}
    \right)
    n_{AD}^k .
    \label{eq:DneqA-contact-explicit}
\end{align}

We next rewrite the finite-part contribution in a directly integrable
centered-shell form.  Choose
\begin{align}
    0<\rho_D
    <
    \min\{r_{AD},r_{BD},r_{CD}\},
\end{align}
and define
\begin{align}
    B_{\rho_D}(\mathbf z_D)
    &:=
    \left\{
        \mathbf x'\in\mathbb R^3
        \,\middle|\,
        r_D'<\rho_D
    \right\}.
\end{align}
Since \((\mathcal Y_{ijk}^{(BC)})_A\) is smooth at \(\mathbf z_D\),
its local expansion may be written as
\begin{align}
    (\mathcal Y_{ijk}^{(BC)}
    (\mathbf z_D+r_D'\mathbf n_D'))_A
    &=
    (\mathcal Y_{ijk}^{(BC)}(\mathbf z_D))_A
    +
    r_D'n_D'{}^a
    \left.
    \partial_a'
    (\mathcal Y_{ijk}^{(BC)})_A
    \right|_{\mathbf z_D}
    +
    O\!\left((r_D')^2\right).
    \label{eq:DneqA-local-expansion}
\end{align}
The two lowest-order contributions vanish after integration over a
complete sphere,
\begin{align}
    \int_{S^2}
    \left(
        3n_D'{}^i n_D'{}^j-\delta_{ij}
    \right)d\Omega
    &=0,\label{eq:DneqA-angular-even}
    \\
    \int_{S^2}
    n_D'{}^a
    \left(
        3n_D'{}^i n_D'{}^j-\delta_{ij}
    \right)d\Omega
    &=0.
    \label{eq:DneqA-angular-average}
\end{align}
The first identity follows from the trace-free angular structure, while
the second follows from angular parity.  Consequently,
\begin{align}
    r_D'^2
    \int_{S^2}
    (\mathcal Y_{ijk}^{(BC)}
    (\mathbf z_D+r_D'\mathbf n_D'))_A\,
    \widehat{\mathcal T}_{ij}^{(D)}
    (\mathbf z_D+r_D'\mathbf n_D')\,
    d\Omega
    &=
    O(r_D')
    \qquad
    (r_D'\to0).
    \label{eq:DneqA-shell-behavior}
\end{align}
The remaining radial integral is therefore convergent at the origin.

The local finite part can thus be evaluated directly in the
centered-shell sense,
\begin{align}
    \operatorname{Pf}\!\int_{B_{\rho_D}(\mathbf z_D)}
    (\mathcal Y_{ijk}^{(BC)})_A\,
    \widehat{\mathcal T}_{ij}^{(D)}\,d^3x'
    &=
    \lim_{\varepsilon\to0^+}
    \int_{\varepsilon}^{\rho_D}
    r_D'^2\,dr_D'
    \int_{S^2}
    (\mathcal Y_{ijk}^{(BC)}
    (\mathbf z_D+r_D'\mathbf n_D'))_A
    \nonumber\\
    &\qquad\qquad\times
    \widehat{\mathcal T}_{ij}^{(D)}
    (\mathbf z_D+r_D'\mathbf n_D')\,
    d\Omega
    \nonumber\\
    &=
    \int_0^{\rho_D}
    r_D'^2\,dr_D'
    \int_{S^2}
    (\mathcal Y_{ijk}^{(BC)}
    (\mathbf z_D+r_D'\mathbf n_D'))_A
    \nonumber\\
    &\qquad\qquad\times
    \widehat{\mathcal T}_{ij}^{(D)}
    (\mathbf z_D+r_D'\mathbf n_D')\,
    d\Omega .
    \label{eq:DneqA-centered-shell-integral}
\end{align}
Here the complete angular integral is performed at each fixed radius
before the radial integration.  The angular cancellations in
Eqs.~\eqref{eq:DneqA-angular-even} and~\eqref{eq:DneqA-angular-average} render the resulting shell function
integrable as \(r_D'\to0\).

The explicit factorized expression in
Eq.~\eqref{eq:FABC-factorized} shows that the remaining
particle-centered singularities in the complementary region are locally
integrable and that the integral converges at spatial infinity.
The complementary contribution is therefore an ordinary spatial integral.

Combining the analytic contact term, the centered-shell local
contribution, and the complementary integral gives
\begin{align}
    (I_k^{D\neq A})_A
    &=
    \frac{G^3}{4\pi}
    \sum_{B\ne C}
    \sum_{D\ne A,B,C}
    m_Bm_Cm_D
    \Bigg\{
        -\frac{4\pi}{3}\,
        \delta_{ij}
        (\mathcal Y_{ijk}^{(BC)}(\mathbf z_D))_A
    \nonumber\\
    &\qquad\qquad
        +
        \int_0^{\rho_D}
        r_D'^2\,dr_D'
        \int_{S^2}
        (\mathcal Y_{ijk}^{(BC)}
        (\mathbf z_D+r_D'\mathbf n_D'))_A\,
        \widehat{\mathcal T}_{ij}^{(D)}
        (\mathbf z_D+r_D'\mathbf n_D')\,
        d\Omega
    \nonumber\\
    &\qquad\qquad
        +
        \int_{\mathbb R^3\setminus
        B_{\rho_D}(\mathbf z_D)}
        (\mathcal Y_{ijk}^{(BC)}(\mathbf x'))_A\,
        \widehat{\mathcal T}_{ij}^{(D)}(\mathbf x')\,
        d^3x'
    \Bigg\}.
    \label{eq:DneqA-centered-shell-form}
\end{align}
The local term is understood with the complete-sphere angular integration
performed at each radius, whereas the complementary term is an ordinary
improper integral.  Their sum is independent of the auxiliary radius
\(\rho_D\), since changing \(\rho_D\) only transfers complete spherical
shells between the two regions.

\subsection{\texorpdfstring{The Self-Singular Sector \(D=A\)}
{The Self-Singular Sector D Equals A}}

For \(D=A\), the nonclosed contribution becomes
\begin{align}
    (I_k^{D=A})_A
    &=
    \frac{G^3}{4\pi}\,m_A
    \sum_{\substack{B\ne C\\B\ne A,\ C\ne A}}
    m_Bm_C\,
    \operatorname{Pf}\!\int
    (\partial_B)_i(\partial_C)_j g'_{BC}\,
    (\partial_A)_k\frac{1}{r'_A}\,
    (\partial_A)_i(\partial_A)_j\frac{1}{r'_A}\,
    d^3x' .
    \label{eq:DA-original-contribution}
\end{align}
The same coefficient \(\mathcal K_{ij}^{(BC)}\) introduced above is used
here.  Since \(B,C\ne A\), it is smooth in a neighborhood of
\(\mathbf z_A\).

For \(r'_A\ne0\), the singular derivatives take their ordinary
punctured-space forms,
\begin{align}
    (\partial_A)_k\frac{1}{r'_A}
    &=
    \frac{n_A'{}^k}{(r'_A)^2},
    &
    (\partial_A)_i(\partial_A)_j\frac{1}{r'_A}
    &=
    \frac{
        3n_A'{}^i n_A'{}^j-\delta_{ij}
    }{(r'_A)^3}.
\end{align}
Their product is therefore
\begin{align}
    (\partial_A)_k\frac{1}{r'_A}\,
    (\partial_A)_i(\partial_A)_j\frac{1}{r'_A}
    &=
    \frac{
        n_A'{}^k
        \left(
            3n_A'{}^i n_A'{}^j-\delta_{ij}
        \right)
    }{(r'_A)^5}.
    \label{eq:DA-punctured-kernel}
\end{align}

The full distributional second derivative also contains the contact term
\begin{align}
    -\frac{4\pi}{3}\,
    \delta_{ij}\delta_A,
    \qquad
    \delta_A
    \equiv
    \delta^{(3)}(\mathbf x'-\mathbf z_A).
\end{align}
Its contribution is proportional to
\begin{align}
    \left(
        \mathcal K_{ii}^{(BC)}
        (\partial_A)_k\frac{1}{r'_A}
    \right)_A ,
\end{align}
which vanishes by angular averaging of the local expansion about
\(\mathbf z_A\), as shown in
Appendix~\ref{app:nonclosed-origin}.  The \(D=A\) finite part can
therefore be evaluated from the punctured-space kernel in
Eq.~\eqref{eq:DA-punctured-kernel}.

Introduce local coordinates centered at \(\mathbf z_A\),
\begin{align}
    \mathbf y
    &:=
    \mathbf x'-\mathbf z_A,
    &
    r
    &:=
    |\mathbf y|
    =
    r'_A,
    &
    \mathbf n
    &:=
    \frac{\mathbf y}{r},
\end{align}
and define
\begin{align}
    \mathcal Q_{ij}^{k}(\mathbf n)
    &:=
    n^k\left(3n^in^j-\delta_{ij}\right),
    \\
    \mathcal K_{ij,a_1\cdots a_p}^{(BC,A)}
    &:=
    \left.
        \partial'_{a_1}\cdots\partial'_{a_p}
        \mathcal K_{ij}^{(BC)}
    \right|_{\mathbf x'=\mathbf z_A}.
\end{align}
The smooth coefficient admits the local expansion
\begin{align}
    \mathcal K_{ij}^{(BC)}(\mathbf z_A+r\mathbf n)
    &=
    \mathcal K_{ij}^{(BC,A)}
    +
    r n^a\mathcal K_{ij,a}^{(BC,A)}
    +
    \frac{r^2}{2}
    n^an^b\mathcal K_{ij,ab}^{(BC,A)}
    +
    O(r^3).
    \label{eq:DA-local-Taylor-expansion}
\end{align}
Including the volume element \(r^2\,dr\,d\Omega\), the corresponding
local radial structure is
\begin{align}
    &\mathcal K_{ij}^{(BC)}(\mathbf z_A+r\mathbf n)\,
    \frac{\mathcal Q_{ij}^{k}(\mathbf n)}{r^5}\,
    r^2\,dr\,d\Omega
    \nonumber\\
    &\qquad=
    \Bigg[
        \frac{
            \mathcal K_{ij}^{(BC,A)}
            \mathcal Q_{ij}^{k}
        }{r^3}
        +
        \frac{
            \mathcal K_{ij,a}^{(BC,A)}
            n^a\mathcal Q_{ij}^{k}
        }{r^2}
        +
        \frac{
            \mathcal K_{ij,ab}^{(BC,A)}
            n^an^b\mathcal Q_{ij}^{k}
        }{2r}
        +
        O(1)
    \Bigg]dr\,d\Omega .
    \label{eq:DA-local-radial-structure}
\end{align}

The zeroth- and second-order Taylor contributions vanish on every
complete centered sphere by angular parity.  The linear Taylor term,
however, has a nonvanishing angular average and produces the unique
divergent radial contribution.  Its angular factor is
\begin{align}
    \int_{S^2}
    n^a\mathcal Q_{ij}^{k}(\mathbf n)\,d\Omega
    &=
    \frac{4\pi}{15}
    \left(
        -2\delta^{ak}\delta^{ij}
        +
        3\delta^{ai}\delta^{kj}
        +
        3\delta^{aj}\delta^{ki}
    \right),
\end{align}
while
\begin{align}
    \int_{\varepsilon}^{\rho_A}\frac{dr}{r^2}
    &=
    \frac{1}{\varepsilon}
    -
    \frac{1}{\rho_A}.
\end{align}
The coefficient of the resulting \(1/\varepsilon\) divergence is
therefore
\begin{align}
    \bigl(\mathcal C_k^{(BC)}\bigr)_A
    &:=
    \frac{4\pi}{15}
    \left[
        -2\,\partial_k'\mathcal K_{ii}^{(BC)}
        +
        3\,\partial_i'\mathcal K_{ik}^{(BC)}
        +
        3\,\partial_j'\mathcal K_{kj}^{(BC)}
    \right]_{\mathbf x'=\mathbf z_A}.
    \label{eq:DA-divergence-coefficient}
\end{align}

Choose \(\rho_A>0\) such that
\(B_{\rho_A}(\mathbf z_A)\) contains no other particle position.  Define
the complete self-singular integrand by
\begin{align}
    \bigl(\mathcal Z_k^{(BC)}(\mathbf x')\bigr)_A
    &:=
    \mathcal K_{ij}^{(BC)}(\mathbf x')\,
    \frac{
        \mathcal Q_{ij}^{k}(\mathbf n_A')
    }{
        (r'_A)^5
    }.
    \label{eq:DA-theory-Z-def}
\end{align}
The linear Taylor contribution is isolated through
\begin{align}
    \bigl(\mathcal P_{k,1}^{(BC)}(\mathbf y)\bigr)_A
    &:=
    y^a\mathcal K_{ij,a}^{(BC,A)}
    \frac{
        \mathcal Q_{ij}^{k}(\mathbf n)
    }{
        r^5
    }.
    \label{eq:DA-local-principal-part}
\end{align}
Its integral over a punctured ball is
\begin{align}
    \int_{\varepsilon<r_A'<\rho_A}
    \bigl(\mathcal P_{k,1}^{(BC)}(\mathbf y)\bigr)_A\,d^3x'
    &=
    \bigl(\mathcal C_k^{(BC)}\bigr)_A
    \left(
        \frac{1}{\varepsilon}
        -
        \frac{1}{\rho_A}
    \right).
    \label{eq:DA-principal-part-integral}
\end{align}

After removal of the linear contribution, the zeroth- and
second-order Taylor terms vanish on every complete centered sphere by
angular parity.  The first nonvanishing shell-integrated remainder is
locally integrable.  The local Hadamard finite part can therefore be
written directly in centered-shell form as
\begin{align}
    \operatorname{Pf}
    \int_{B_{\rho_A}(\mathbf z_A)}
    \bigl(\mathcal Z_k^{(BC)}(\mathbf x')\bigr)_A\,d^3x'
    &=
    \int_0^{\rho_A}
    r_A'^2\,dr_A'
    \int_{S^2}
    \Big[
        \bigl(
        \mathcal Z_k^{(BC)}
        (\mathbf z_A+r_A'\mathbf n)
        \bigr)_A
        \nonumber\\
    &\qquad\qquad
        -
        \bigl(
        \mathcal P_{k,1}^{(BC)}
        (r_A'\mathbf n)
        \bigr)_A
    \Big]\,d\Omega
    -
    \frac{
        \bigl(\mathcal C_k^{(BC)}\bigr)_A
    }{
        \rho_A
    }.
    \label{eq:DA-local-Hadamard-finite-part}
\end{align}
The complete angular integral is performed at each fixed radius before
the radial integration.  The centered-shell angular cancellations remove
the remaining lowest-order singular contributions, while the analytic
term
\(-(\mathcal C_k^{(BC)})_A/\rho_A\)
supplies the finite remainder associated with the isolated linear
divergence.

The complete \(D=A\) contribution is then
\begin{align}
    (I_k^{D=A})_A
    &=
    \frac{G^3}{4\pi}\,
    m_A
    \sum_{\substack{B\ne C\\B\ne A,\ C\ne A}}
    m_Bm_C
    \Bigg[
        \int_0^{\rho_A}
        r_A'^2\,dr_A'
        \int_{S^2}
        \Bigg(
            \regtag{
            \bigl(
            \mathcal Z_k^{(BC)}
            (\mathbf z_A+r_A'\mathbf n)
            \bigr)_A
            }{eq:app-Z-explicit}
            \nonumber\\
    &\qquad\qquad\qquad\qquad
            -
            \regtag{
            \bigl(
            \mathcal P_{k,1}^{(BC)}
            (r_A'\mathbf n)
            \bigr)_A
            }{eq:app-P-A1-explicit}
        \Bigg)d\Omega
        -
        \frac{
            \regtag{
            \bigl(\mathcal C_k^{(BC)}\bigr)_A
            }{eq:app-C-explicit}
        }{
            \rho_A
        }
    \nonumber\\
    &\qquad\qquad
        +
        \int_{\mathbb R^3\setminus B_{\rho_A}(\mathbf z_A)}
        \bigl(\mathcal Z_k^{(BC)}(\mathbf x')\bigr)_A\,d^3x'
    \Bigg].
    \label{eq:DA-Hadamard-finite-part}
\end{align}
The radius \(\rho_A\) is auxiliary: its dependence cancels between the
local finite-part contribution and the complementary integral.
Equation~\eqref{eq:DA-Hadamard-finite-part} therefore gives a direct
centered-shell representation of the regularized \(D=A\) sector.

The complete regularized nonclosed contribution is
\begin{align}
    (I_k)_A
    &=
    \regtag{(I_k^{D\neq A})_A}{eq:DneqA-centered-shell-form}
    +
    \regtag{(I_k^{D=A})_A}{eq:DA-Hadamard-finite-part}.
    \label{eq:nonclosed-regularized-complete}
\end{align}
The small equation numbers in the two final sectoral formulas and in
Eq.~\eqref{eq:nonclosed-regularized-complete} are cross-reference tags.  
To keep the derivation readable, such tags are used
only in the final results.  The explicit formulas for the compact kernels
and coefficients \((\mathcal Y_{ijk}^{(BC)})_A\),
\((\mathcal Z_k^{(BC)})_A\),
\((\mathcal P_{k,1}^{(BC)})_A\), and
\((\mathcal C_k^{(BC)})_A\) are collected in
Appendix~\ref{app:explicit-integrand-formulas}.

Equations~\eqref{eq:DneqA-centered-shell-form} and
\eqref{eq:DA-Hadamard-finite-part} are the two regularized forms used
below.  In the \(D\neq A\) sector, the lowest-order local contributions
vanish under complete centered-shell angular integration, leaving a
locally convergent radial integral together with the analytic
distributional contact term.  In the \(D=A\) sector, the unique linearly
divergent Taylor contribution is isolated analytically; after its removal,
the remaining centered-shell integral is locally convergent.  The
complementary regions in both sectors contain only ordinary locally
integrable particle-centered structures.

\section{Numerical Evaluation of the Regularized Nonclosed Integral}
\label{sec:numerical-evaluation-nonclosed}

This section specifies the numerical implementation of the regularized
representations derived in Sec.~\ref{sec:singular-integral-reformulation}.  For
\(D\neq A\), the delta-supported contact term in
Eq.~\eqref{eq:DneqA-centered-shell-form} is evaluated analytically, and
only the centered-shell inner integral and the complementary outer
integral require numerical quadrature.  For \(D=A\), the local finite part
is evaluated directly from Eq.~\eqref{eq:DA-local-Hadamard-finite-part},
including the analytic term
\(-(\mathcal C_k^{(BC)})_A/\rho_A\), while the complementary outer
integral is ordinary.

Throughout this section, the field point is fixed at
\begin{align}
    \mathbf x=\mathbf z_A.
\end{align}
The local balls are centered at the singular points and are chosen not to
contain any other particle position.  Define
\begin{align}
    d_Q^{\min}
    &:=
    \min_{E\ne Q}
    |\mathbf z_Q-\mathbf z_E|,
\end{align}
and choose
\begin{align}
    \rho_Q
    &:=
    \alpha_Q d_Q^{\min},
    \qquad
    0<\alpha_Q<1,
    \label{eq:local-radius-choice}
\end{align}
for \(Q=D\) in the \(D\neq A\) sector and \(Q=A\) in the \(D=A\)
sector.  The radii \(\rho_D\) and \(\rho_A\) are auxiliary splitting
parameters.  Their numerical values are checked by varying them while
holding the physical particle configuration fixed.

\subsection{Direct Evaluation of the Local Contributions}
\label{sec:direct-local-numerical}

The regularized local contributions derived in
Eqs.~\eqref{eq:DneqA-centered-shell-integral} and
\eqref{eq:DA-local-Hadamard-finite-part} are evaluated directly by
numerical quadrature.  In both sectors, spherical coordinates are centered
at the corresponding singular point, and the complete angular integral is
performed at each fixed radius before the remaining radial integration.

For \(D\neq A\), the local contribution is evaluated from
\begin{align}
    \operatorname{Pf}\!\int_{B_{\rho_D}(\mathbf z_D)}
    (\mathcal Y_{ijk}^{(BC)})_A\,
    \widehat{\mathcal T}_{ij}^{(D)}\,d^3x'
    &=
    \int_0^{\rho_D}
    r_D'^2\,dr_D'
    \int_{S^2}
    (\mathcal Y_{ijk}^{(BC)}
    (\mathbf z_D+r_D'\mathbf n))_A
    \nonumber\\
    &\qquad\qquad\times
    \widehat{\mathcal T}_{ij}^{(D)}
    (\mathbf z_D+r_D'\mathbf n)\,
    d\Omega .
    \label{eq:DneqA-direct-local-integral}
\end{align}
The corresponding distributional contact contribution,
\begin{align}
    -\frac{4\pi}{3}\,
    \delta_{ij}
    (\mathcal Y_{ijk}^{(BC)}(\mathbf z_D))_A,
\end{align}
is evaluated analytically and added to the centered-shell local
contribution and the complementary spatial integral.

For \(D=A\), the local finite part is evaluated from
\begin{align}
    \operatorname{Pf}
    \int_{B_{\rho_A}(\mathbf z_A)}
    (\mathcal Z_k^{(BC)}(\mathbf x'))_A\,d^3x'
    &=
    \int_0^{\rho_A}
    r_A'^2\,dr_A'
    \int_{S^2}
    \Big[
        (\mathcal Z_k^{(BC)}
        (\mathbf z_A+r_A'\mathbf n))_A
    \nonumber\\
    &\qquad\qquad
        -
        (\mathcal P_{k,1}^{(BC)}
        (r_A'\mathbf n))_A
    \Big]\,d\Omega
    -
    \frac{
        (\mathcal C_k^{(BC)})_A
    }{
        \rho_A
    }.
    \label{eq:DA-direct-local-integral}
\end{align}
The numerical quadrature is therefore applied directly to the
centered-shell remainder, while the finite contribution associated with
the isolated linear divergence is included analytically through the term
\(-(\mathcal C_k^{(BC)})_A/\rho_A\).

\subsection{Particle-Centered Treatment of the Outer Contributions}
\label{sec:outer-patch-numerical}

After the primary singularities have been treated by the preceding
centered-shell and finite-part prescriptions, the remaining outer
contributions are locally integrable spatial integrals. They may still
contain localized peaks near other particle positions, which are handled
by a particle-centered patch decomposition.

For any primary center \(Q\), define
\begin{align}
    d_Q^{\max}
    &:=
    \max_{E\ne Q}
    |\mathbf z_Q-\mathbf z_E|,
    &
    R_{\rm far}^{(Q)}
    &:=
    \lambda_{\rm far}d_Q^{\max},
    \label{eq:outer-far-radius-general}
\end{align}
where \(\lambda_{\rm far}>1\) sets the finite outer boundary. The
corresponding truncated outer domains are
\begin{align}
    \Omega_D^{\rm out}(\lambda_{\rm far})
    &:=
    \left\{
        \mathbf x'
        \,\middle|\,
        \rho_D<r_D'<R_{\rm far}^{(D)}
    \right\},
    \qquad
    D\ne A,
    \label{eq:DneqA-outer-truncated-domain}
    \\
    \Omega_A^{\rm out}(\lambda_{\rm far})
    &:=
    \left\{
        \mathbf x'
        \,\middle|\,
        \rho_A<r_A'<R_{\rm far}^{(A)}
    \right\},
    \qquad
    D=A.
    \label{eq:DA-outer-truncated-domain}
\end{align}

At fixed \(\lambda_{\rm far}\), the truncated outer contributions are
\begin{align}
    \bigl(I_k^{D\ne A,\mathrm{out}}
    (\lambda_{\rm far})\bigr)_A
    &:=
    \frac{G^3}{4\pi}
    \sum_{B\ne C}
    \sum_{D\ne A,B,C}
    m_Bm_Cm_D
    \int_{\Omega_D^{\rm out}(\lambda_{\rm far})}
    (\mathcal Y_{ijk}^{(BC)})_A
    \widehat{\mathcal T}_{ij}^{(D)}
    \,d^3x',
    \label{eq:DneqA-outer-contribution}
    \\
    \bigl(I_k^{D=A,\mathrm{out}}
    (\lambda_{\rm far})\bigr)_A
    &:=
    \frac{G^3}{4\pi}\,m_A
    \sum_{\substack{B\ne C\\B\ne A,\ C\ne A}}
    m_Bm_C
    \int_{\Omega_A^{\rm out}(\lambda_{\rm far})}
    \bigl(\mathcal Z_k^{(BC)}\bigr)_A
    \,d^3x'.
    \label{eq:DA-outer-contribution}
\end{align}
The complete outer contributions are recovered in the limit
\(\lambda_{\rm far}\to\infty\). In the numerical calculation,
\(\lambda_{\rm far}\) is chosen sufficiently large, and convergence with
respect to the outer boundary is checked separately.

The patch centers are chosen at particle positions where localized peaks
may remain. For the \(D\ne A\) sector,
\begin{align}
    \mathscr S_{D\ne A}^{\rm patch}
    &:=
    \{A,B,C\},
    \label{eq:DneqA-patch-center-set}
\end{align}
while for the \(D=A\) sector,
\begin{align}
    \mathscr S_{D=A}^{\rm patch}
    &:=
    \{B,C\}.
    \label{eq:DA-patch-center-set}
\end{align}
In the latter case, the self-singular position \(\mathbf z_A\) has already
been removed by the inner ball.

For each \(P\in\mathscr S^{\rm patch}\), choose a patch radius
\begin{align}
    R_P^{\rm patch}
    &=
    \eta_{\rm patch}L_P,
    \qquad
    0<\eta_{\rm patch}<1,
    \label{eq:particle-patch-radius-general}
\end{align}
where \(L_P\) is a local geometric scale chosen so that the patch lies
inside the relevant outer domain and different patch supports are
disjoint. Let \(\phi_P\) be a compactly supported smooth weight satisfying
\begin{align}
    0
    \le
    \phi_P(\mathbf x')
    \le
    1,
    \qquad
    \phi_P
    =
    1
    \ \text{near }\mathbf z_P,
    \qquad
    \phi_P
    =
    0
    \ \text{outside the \(P\)-centered patch}.
\end{align}
For a given patch-center set, define
\begin{align}
    \Phi_{\mathscr S}(\mathbf x')
    &:=
    \sum_{P\in\mathscr S^{\rm patch}}
    \phi_P(\mathbf x').
    \label{eq:patch-weight-general}
\end{align}
Since the patch supports are disjoint,
\begin{align}
    0
    \le
    \Phi_{\mathscr S}(\mathbf x')
    \le
    1.
\end{align}

For any outer integrand \(F_k\), outer domain \(\Omega^{\rm out}\), and
associated patch-center set \(\mathscr S^{\rm patch}\), the exact
decomposition is
\begin{align}
    \int_{\Omega^{\rm out}}
    F_k(\mathbf x')\,d^3x'
    &=
    \int_{\Omega^{\rm out}}
    \bigl[1-\Phi_{\mathscr S}(\mathbf x')\bigr]
    F_k(\mathbf x')\,d^3x'
    \nonumber\\
    &\quad
    +
    \sum_{P\in\mathscr S^{\rm patch}}
    \int_{\Omega^{\rm out}}
    \phi_P(\mathbf x')
    F_k(\mathbf x')\,d^3x'.
    \label{eq:outer-patch-decomposition-general}
\end{align}
The first term is evaluated in spherical coordinates centered at the
primary point, \(\mathbf z_D\) for \(D\ne A\) and \(\mathbf z_A\) for
\(D=A\), whereas each patch term is evaluated in coordinates centered at
the corresponding particle position \(\mathbf z_P\). The restriction to
the original outer domain is retained throughout. Thus, the decomposition
changes only the local integration coordinates and does not alter either
the integration domain or the value of the integral.

At finite \(\lambda_{\rm far}\), the two sectors are assembled as
\begin{align}
    \bigl(I_k^{D\ne A}(\lambda_{\rm far})\bigr)_A
    &=
    \bigl(I_k^{D\ne A,\mathrm{in}}\bigr)_A
    +
    \bigl(I_k^{D\ne A,\mathrm{out}}
    (\lambda_{\rm far})\bigr)_A,
    \\
    \bigl(I_k^{D=A}(\lambda_{\rm far})\bigr)_A
    &=
    \bigl(I_k^{D=A,\mathrm{in}}\bigr)_A
    +
    \bigl(I_k^{D=A,\mathrm{out}}
    (\lambda_{\rm far})\bigr)_A.
\end{align}
Here the inner contribution denotes the complete regularized contribution
associated with the primary ball \(r_Q'<\rho_Q\), while the outer
contribution is the ordinary complementary integral over
\(\rho_Q<r_Q'<R_{\rm far}^{(Q)}\).

The complete nonclosed contribution is therefore
\begin{align}
    (I_k)_A
    =
    \lim_{\lambda_{\rm far}\to\infty}
    \left[
        \bigl(I_k^{D\ne A}(\lambda_{\rm far})\bigr)_A
        +
        \bigl(I_k^{D=A}(\lambda_{\rm far})\bigr)_A
    \right].
    \label{eq:nonclosed-final-assembly}
\end{align}

Explicit formulas for the kernels and derivatives entering the numerical
integrands are collected in
Appendix~\ref{app:explicit-integrand-formulas}.

\FloatBarrier

\section{Numerical Results for Representative Three-Body Systems}
\label{sec:numerical-results-benchmarks}

The numerical analysis below uses the Sun--Jupiter--Saturn (SJS) and
Sun--Mercury--Venus (SMV) systems as controlled benchmarks for isolating and
quantifying the nonclosed 2PN contribution. Its purpose is to determine both
the instantaneous magnitude of the nonclosed acceleration and the leading
finite-time response that this small acceleration produces in the pairwise
relative distances. The two systems probe different mass hierarchies and
orbital time scales. They are benchmark three-body models rather than complete
planetary ephemerides.

The nonclosed contribution \((I_i)_A\) derived above defines the 2PN
acceleration coefficient
\begin{align}
    (a_A^{\mathrm{2PN,nc}})^i
    &=
    4(I_i)_A ,
    \label{eq:nc-acceleration-definition}
\end{align}
where the label \(\mathrm{2PN,nc}\) identifies the nonclosed 2PN
coefficient. The corresponding physical acceleration entering the EOM is
\(c^{-4}\mathbf a_A^{\mathrm{2PN,nc}}\).

To display the PN ordering explicitly, we write the trajectory and the EOM
through 2PN order as
\begin{align*}
    \mathbf z_A
    &=
    \mathbf z_A^{\mathrm{N}}
    +\frac{1}{c^2}\mathbf z_A^{\mathrm{1PN}}
    +\frac{1}{c^4}\mathbf z_A^{\mathrm{2PN}}
    +\cdots,
    \\
    \ddot{\mathbf z}_A
    &=
    \mathbf a_A^{\mathrm{N}}
    +\frac{1}{c^2}\mathbf a_A^{\mathrm{1PN}}
    +\frac{1}{c^4}\mathbf a_A^{\mathrm{2PN,cl}}
    +\frac{1}{c^4}\mathbf a_A^{\mathrm{2PN,nc}}
    +\cdots,
\end{align*}
Here \(\mathbf z_A^{\mathrm{1PN}}\) and
\(\mathbf z_A^{\mathrm{2PN}}\) are the corresponding displacement
coefficients, the velocities expand analogously, and ``cl'' denotes the
closed analytic 2PN contribution. The
superscript ``N'' denotes evaluation on the Newtonian motion, with
\(\{\mathbf z_C^{\mathrm{N}}(t)\}\) denoting the positions of all bodies.
All displacement and acceleration quantities carrying the PN labels
\(\mathrm{1PN}\) or \(\mathrm{2PN}\) denote the corresponding coefficients
and do not include the displayed powers of \(c^{-1}\). Thus all PN factors
are shown explicitly in the trajectory and the EOM.

Using
\(\mathbf z_B-\mathbf z_B^{\mathrm{N}}
=c^{-2}\mathbf z_B^{\mathrm{1PN}}
+c^{-4}\mathbf z_B^{\mathrm{2PN}}+\cdots\),
the position-dependent quantity \((I_i)_A\) expands as
\begin{align}
    (I_i)_A
    \bigl[\{\mathbf z_B(t)\}\bigr]
    &=
    (I_i)_A
    \bigl[\{\mathbf z_B^{\mathrm{N}}(t)\}\bigr]
    \nonumber\\
    &\quad
    +\frac{1}{c^2}
    \sum_B
    \left.
    \frac{\partial (I_i)_A}
         {\partial z_B^j}
    \right|_{\{\mathbf z_C^{\mathrm{N}}(t)\}}
    (z_B^{\mathrm{1PN}})^j
    +O(c^{-4}) .
    \label{eq:nc-expansion-about-newtonian}
\end{align}
Because the nonclosed coefficient enters the acceleration with the explicit
factor \(c^{-4}\), its first correction in
Eq.~\eqref{eq:nc-expansion-about-newtonian} is of 3PN order. Consequently,
the nonclosed coefficient entering the 2PN EOM is evaluated consistently
on the Newtonian trajectory:
\begin{align}
    (a_A^{\mathrm{2PN,nc}})^i
    \bigl[\{\mathbf z_B^{\mathrm{N}}(t)\}\bigr]
    &=
    4
    (I_i)_A
    \bigl[\{\mathbf z_B^{\mathrm{N}}(t)\}\bigr].
    \label{eq:nc-acceleration-along-newtonian}
\end{align}

The equation governing the 2PN perturbation follows by applying the same
expansion to both sides of the EOM. Before isolating any source, its 2PN part
has the order-complete structure
\begingroup
\small
\begin{align*}
    (\ddot z_A^{\mathrm{2PN}})^i
    ={}&
    \sum_B
    \left.
    \frac{\partial(a_A^{\mathrm{N}})^i}{\partial z_B^j}
    \right|_{\mathrm{N}}
    (z_B^{\mathrm{2PN}})^j
    +\frac{1}{2}\sum_{B,C}
    \left.
    \frac{\partial^2(a_A^{\mathrm{N}})^i}
         {\partial z_B^j\partial z_C^k}
    \right|_{\mathrm{N}}
    (z_B^{\mathrm{1PN}})^j
    (z_C^{\mathrm{1PN}})^k
    \\
    &
    +\sum_B
    \left.
    \frac{\partial(a_A^{\mathrm{1PN}})^i}{\partial z_B^j}
    \right|_{\mathrm{N}}
    (z_B^{\mathrm{1PN}})^j
    +\sum_B
    \left.
    \frac{\partial(a_A^{\mathrm{1PN}})^i}{\partial v_B^j}
    \right|_{\mathrm{N}}
    (v_B^{\mathrm{1PN}})^j
    \\
    &
    +(a_A^{\mathrm{2PN,cl}})^i
    [\mathbf z^{\mathrm{N}},\mathbf v^{\mathrm{N}}]
    +(a_A^{\mathrm{2PN,nc}})^i[\mathbf z^{\mathrm{N}}],
\end{align*}
\endgroup
All vertical bars and bracketed sources denote evaluation on the Newtonian
trajectory. Thus the 1PN and closed 2PN terms remain in the complete 2PN
equation. Because this equation is linear in the unknown 2PN displacement
coefficient, its solution can be decomposed according to the source terms.
The coefficient
\(\mathbf z_A^{\mathrm{2PN,nc}}\) driven by the nonclosed source obeys
\begin{align}
    (\ddot z_A^{\mathrm{2PN,nc}})^i
    &=
    \sum_B
    \left.
    \frac{\partial (a_A^{\mathrm{N}})^i}
         {\partial z_B^j}
    \right|_{\{\mathbf z_C^{\mathrm{N}}(t)\}}
    (z_B^{\mathrm{2PN,nc}})^j
    +
    (a_A^{\mathrm{2PN,nc}})^i
    \bigl[\{\mathbf z_C^{\mathrm{N}}(t)\}\bigr],
    \label{eq:nc-leading-perturbation}
\end{align}
with
\begin{align}
    \mathbf z_A^{\mathrm{2PN,nc}}(t_0)
    &=
    \dot{\mathbf z}_A^{\mathrm{2PN,nc}}(t_0)
    =
    \mathbf 0 .
    \label{eq:nc-perturbation-initial}
\end{align}

The Jacobian in Eq.~\eqref{eq:nc-leading-perturbation} is the Newtonian
linear operator inherited from the full 2PN equation. With zero initial
data, the solution isolates the orbital correction generated by the
nonclosed source; it does not replace the 1PN or remaining 2PN terms.

\subsection{Benchmark Systems and Numerical Setup}
\label{subsec:benchmark-systems-setup}

For both benchmarks, the Newtonian trajectory and the perturbation equations
are integrated with the same sixth-order Runge--Kutta method. The physical
nonclosed acceleration \(c^{-4}\mathbf a_A^{\mathrm{2PN,nc}}\) is evaluated
on a uniform time grid along the Newtonian trajectory.
During the integration of the perturbation equations, its Cartesian components
are linearly interpolated in time as required. The integration interval and
temporal resolution are chosen separately for the two systems according to
their characteristic orbital timescales.

\subsubsection{Sun--Jupiter--Saturn}
\label{subsubsec:sjs-setup}

We consider an isolated Newtonian SJS system in barycentric coordinates.
The initial conditions and gravitational parameters are obtained from
JPL Horizons at
\begin{align}
    t_0
    &=
    \text{A.D. 2026-May-24 00:00:00 TDB}
    \nonumber\\
    &=
    \mathrm{JDTDB}\ 2461184.5 .
    \label{eq:benchmark-initial-epoch}
\end{align}
The initial states are geometric barycentric vectors in the ICRF, with the
Solar System barycenter (SSB) as the coordinate origin. The corresponding
initial data, converted to SI units, are listed in
Table~\ref{tab:sjs-initial-data}.

\begin{table}[H]
    \centering
    \caption{Initial gravitational parameters and barycentric Cartesian
    states of the SJS benchmark system at
    A.D. 2026-May-24 00:00:00 TDB.}
    \label{tab:sjs-initial-data}
    \small
    \begingroup
    \renewcommand{\arraystretch}{1.12}
    \setlength{\tabcolsep}{7pt}
    \begin{tabular}{@{}lr@{}}
        \toprule
        Body
        & \(Gm_A\,[\mathrm{m^3\,s^{-2}}]\) \\
        \midrule
        Sun
        & \(1.3271244004193938\times10^{20}\) \\
        Jupiter
        & \(1.2668653190000000\times10^{17}\) \\
        Saturn
        & \(3.7931206234000000\times10^{16}\) \\
        \bottomrule
    \end{tabular}

    \medskip
    \textit{Barycentric positions}\par\smallskip
    \begin{tabular}{@{}lrrr@{}}
        \toprule
        Body
        & \(x\,[\mathrm m]\)
        & \(y\,[\mathrm m]\)
        & \(z\,[\mathrm m]\) \\
        \midrule
        Sun
        & \(-3.095837313147680\times10^{8}\)
        & \(-7.489398930617912\times10^{8}\)
        & \(-3.064826640026351\times10^{8}\) \\
        Jupiter
        & \(-4.019774857930104\times10^{11}\)
        & \(6.180632299322687\times10^{11}\)
        & \(2.747105997245013\times10^{11}\) \\
        Saturn
        & \(1.407065173046752\times10^{12}\)
        & \(1.669538698752182\times10^{11}\)
        & \(8.355072453573833\times10^{9}\) \\
        \bottomrule
    \end{tabular}

    \medskip
    \textit{Barycentric velocities}\par\smallskip
    \begin{tabular}{@{}lrrr@{}}
        \toprule
        Body
        & \(v_x\,[\mathrm{m\,s^{-1}}]\)
        & \(v_y\,[\mathrm{m\,s^{-1}}]\)
        & \(v_z\,[\mathrm{m\,s^{-1}}]\) \\
        \midrule
        Sun
        & \(1.158372726365104\times10^{1}\)
        & \(2.611481833291995\)
        & \(8.718465139860870\times10^{-1}\) \\
        Jupiter
        & \(-1.138918272874147\times10^{4}\)
        & \(-5.670966742687678\times10^{3}\)
        & \(-2.153569555792905\times10^{3}\) \\
        Saturn
        & \(-1.600057284666138\times10^{3}\)
        & \(8.828644385719750\times10^{3}\)
        & \(3.716154362714075\times10^{3}\) \\
        \bottomrule
    \end{tabular}
    \endgroup
\end{table}

The SJS calculation covers \(657360\,\mathrm d\), or approximately
\(1799.753\) Julian years. This interval contains about 152 Jupiter orbits,
61 Saturn orbits, and 91 Jupiter--Saturn synodic cycles, providing repeated
sampling of the outer-planet configuration. A one-day integration step is
used, and the nonclosed acceleration is evaluated on a uniform 180-day grid.

\subsubsection{Sun--Mercury--Venus}
\label{subsubsec:smv-setup}

We next apply the same calculation to an isolated SMV system. The initial
epoch and reference frame are the same as for the SJS benchmark in
Eq.~\eqref{eq:benchmark-initial-epoch}, and the corresponding SI initial
data are listed in Table~\ref{tab:smv-initial-data}.

\begin{table}[H]
    \centering
    \caption{Initial gravitational parameters and barycentric Cartesian
    states of the SMV benchmark system at
    A.D. 2026-May-24 00:00:00 TDB.}
    \label{tab:smv-initial-data}
    \small
    \begingroup
    \renewcommand{\arraystretch}{1.12}
    \setlength{\tabcolsep}{7pt}
    \begin{tabular}{@{}lr@{}}
        \toprule
        Body
        & \(Gm_A\,[\mathrm{m^3\,s^{-2}}]\) \\
        \midrule
        Sun
        & \(1.3271244004193938\times10^{20}\) \\
        Mercury
        & \(2.2031868550000000\times10^{13}\) \\
        Venus
        & \(3.2485859200000000\times10^{14}\) \\
        \bottomrule
    \end{tabular}

    \medskip
    \textit{Barycentric positions}\par\smallskip
    \begin{tabular}{@{}lrrr@{}}
        \toprule
        Body
        & \(x\,[\mathrm m]\)
        & \(y\,[\mathrm m]\)
        & \(z\,[\mathrm m]\) \\
        \midrule
        Sun
        & \(-3.095837313147680\times10^{8}\)
        & \(-7.489398930617912\times10^{8}\)
        & \(-3.064826640026351\times10^{8}\) \\
        Mercury
        & \(-1.800434706524910\times10^{10}\)
        & \( 3.728756609302444\times10^{10}\)
        & \( 2.184662669802891\times10^{10}\) \\
        Venus
        & \(-8.916143086263946\times10^{10}\)
        & \( 5.212615231908512\times10^{10}\)
        & \( 2.910709324822512\times10^{10}\) \\
        \bottomrule
    \end{tabular}

    \medskip
    \textit{Barycentric velocities}\par\smallskip
    \begin{tabular}{@{}lrrr@{}}
        \toprule
        Body
        & \(v_x\,[\mathrm{m\,s^{-1}}]\)
        & \(v_y\,[\mathrm{m\,s^{-1}}]\)
        & \(v_z\,[\mathrm{m\,s^{-1}}]\) \\
        \midrule
        Sun
        & \( 1.158372726365104\times10^{1}\)
        & \( 2.611481833291995\)
        & \( 8.718465139860870\times10^{-1}\) \\
        Mercury
        & \(-5.492866555058516\times10^{4}\)
        & \(-1.657144435026688\times10^{4}\)
        & \(-3.159131977531049\times10^{3}\) \\
        Venus
        & \(-1.977506887204981\times10^{4}\)
        & \(-2.705105949419921\times10^{4}\)
        & \(-1.092071145968270\times10^{4}\) \\
        \bottomrule
    \end{tabular}
    \endgroup
\end{table}

The SMV calculation covers \(18260\,\mathrm d\), or approximately
\(49.993\) Julian years. This interval contains about 208 Mercury orbits,
81 Venus orbits, and 126 Mercury--Venus synodic cycles, providing repeated
sampling of the inner-planet configuration. A \(0.125\)-day integration
step is used, and the nonclosed acceleration is evaluated on a uniform
10-day grid.

\FloatBarrier

\subsection{Instantaneous Nonclosed 2PN Acceleration}
\label{subsec:instantaneous-nonclosed}

We characterize the nonclosed contribution by its absolute acceleration
scale and its magnitude relative to the complete 2PN acceleration. For each
body \(A\), we define
\begin{align}
    q_A(t)
    &\equiv
    \left|
        \frac{1}{c^4}\mathbf a_A^{\mathrm{2PN,nc}}
        \bigl[\{\mathbf z_B^{\mathrm{N}}(t)\}\bigr]
    \right|
    \nonumber\\
    &=
    \frac{1}{c^4}\left[
        \delta_{ij}
        (a_A^{\mathrm{2PN,nc}})^i(t)
        (a_A^{\mathrm{2PN,nc}})^j(t)
    \right]^{1/2}.
    \label{eq:nc-acceleration-norm}
\end{align}

For its relative magnitude within the 2PN sector, we write
\begin{align}
    \mathbf a_A^{\mathrm{2PN}}
    &=
    \mathbf a_A^{\mathrm{2PN,cl}}
    +
    \mathbf a_A^{\mathrm{2PN,nc}},
    \label{eq:complete-2pn-split}
\end{align}
where all accelerations on both sides are 2PN coefficients. The common
factor \(c^{-4}\) cancels from the ratio of the corresponding physical
accelerations, so we define
\begin{align}
    \chi_A(t)
    &\equiv
    \frac{
        \left|\mathbf a_A^{\mathrm{2PN,nc}}(t)\right|
    }{
        \left|\mathbf a_A^{\mathrm{2PN}}(t)\right|
    }
    \nonumber\\
    &=
    \frac{
        \left|\mathbf a_A^{\mathrm{2PN,nc}}(t)\right|
    }{
        \left|
            \mathbf a_A^{\mathrm{2PN,cl}}(t)
            +
            \mathbf a_A^{\mathrm{2PN,nc}}(t)
        \right|
    }.
    \label{eq:nc-fraction-complete-2pn}
\end{align}
The denominator is the complete 2PN acceleration coefficient and does not
include the Newtonian or 1PN contributions.

For sampled epochs \(t_n\), we summarize \(q_A\) and \(\chi_A\) by their
root-mean-square (rms) and largest sampled values,
\begin{align}
    q_{A,\mathrm{rms}}
    &\equiv
    \left[
        \frac{1}{N_{\rm samp}}
        \sum_{n=1}^{N_{\rm samp}}
        q_A^2(t_n)
    \right]^{1/2},
    &
    q_{A,\max}
    &\equiv
    \max_n q_A(t_n),
    \label{eq:nc-norm-summary-definitions}
    \\
    \chi_{A,\mathrm{rms}}
    &\equiv
    \left[
        \frac{1}{N_{\rm samp}}
        \sum_{n=1}^{N_{\rm samp}}
        \chi_A^2(t_n)
    \right]^{1/2},
    &
    \chi_{A,\max}
    &\equiv
    \max_n \chi_A(t_n).
    \label{eq:nc-fraction-summary-definitions}
\end{align}
Here \(N_{\rm samp}\) is the number of samples and
\(\Delta t_{\mathrm{nc}}\) their uniform spacing. The rms values characterize
the typical magnitude over the sampled interval, while the maxima refer
only to the sampled grid and are not continuous-time extrema. For SJS,
\((N_{\rm samp},\Delta t_{\mathrm{nc}})=(3653,180\,\mathrm d)\); for SMV,
\((N_{\rm samp},\Delta t_{\mathrm{nc}})=(1827,10\,\mathrm d)\).
The corresponding sampled statistics for both benchmark systems are summarized in Table~\ref{tab:nc-instantaneous-summary}.

\begin{figure}[H]
    \centering
    \begin{subfigure}{0.98\linewidth}
        \centering
        \includegraphics[width=\linewidth,height=0.44\textheight,keepaspectratio]
        {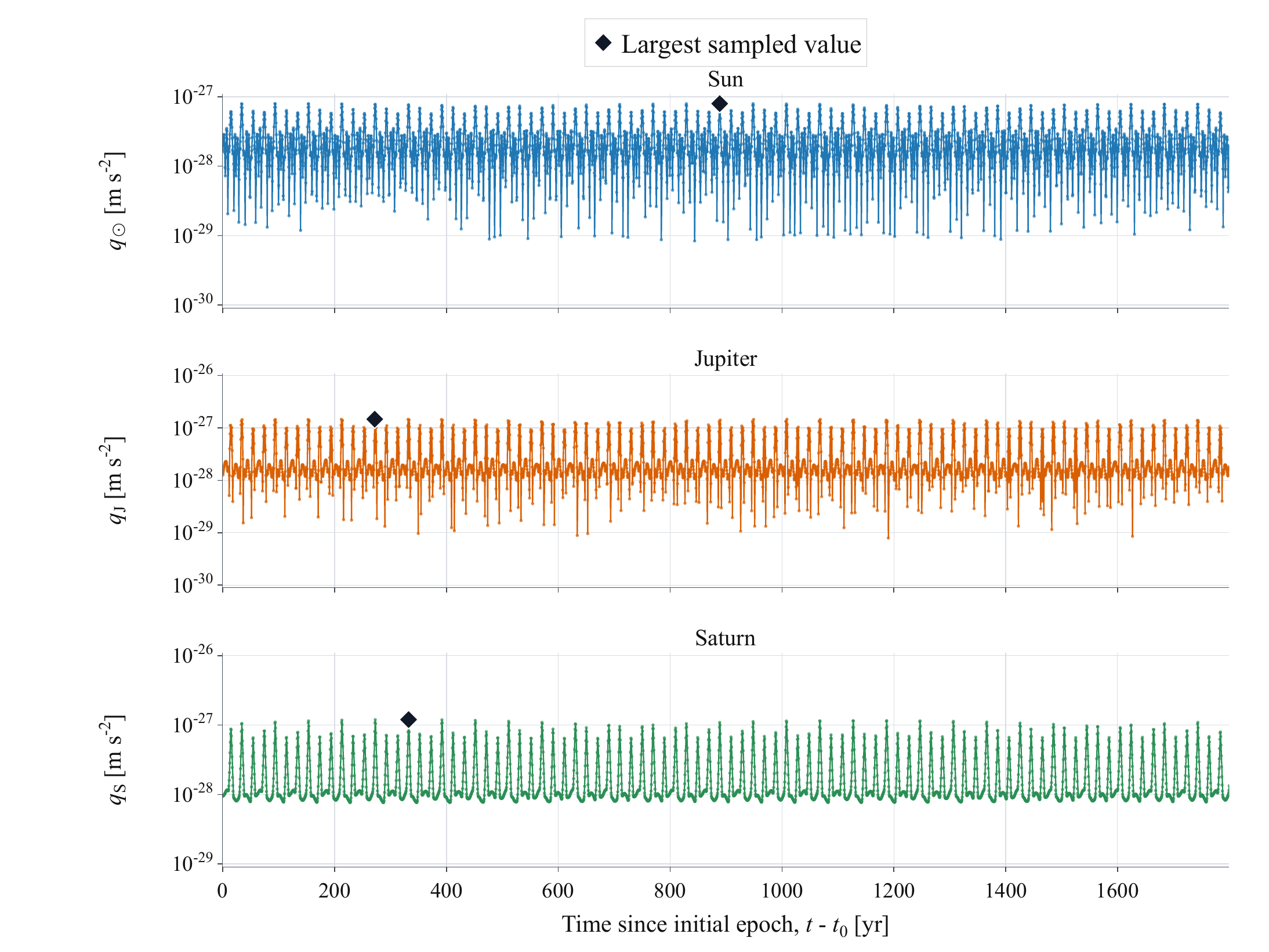}
        \caption{SJS benchmark.}
        \label{fig:sjs-nc-acceleration-norm-long}
    \end{subfigure}
    \caption{Instantaneous nonclosed acceleration norm
    \(q_A(t)=|c^{-4}\mathbf a_A^{\mathrm{2PN,nc}}(t)|\) for the two benchmark systems.
    Diamonds mark the largest sampled values. Each panel has its own
    logarithmic vertical scale. Panel (a) shows the SJS benchmark, and panel (b) shows the SMV benchmark.}
    \label{fig:nc-acceleration-norm-benchmarks}
\end{figure}

\begin{figure}[H]
    \ContinuedFloat
    \centering
    \setcounter{subfigure}{1}
    \begin{subfigure}{0.98\linewidth}
        \centering
        \includegraphics[width=\linewidth,height=0.44\textheight,keepaspectratio]
        {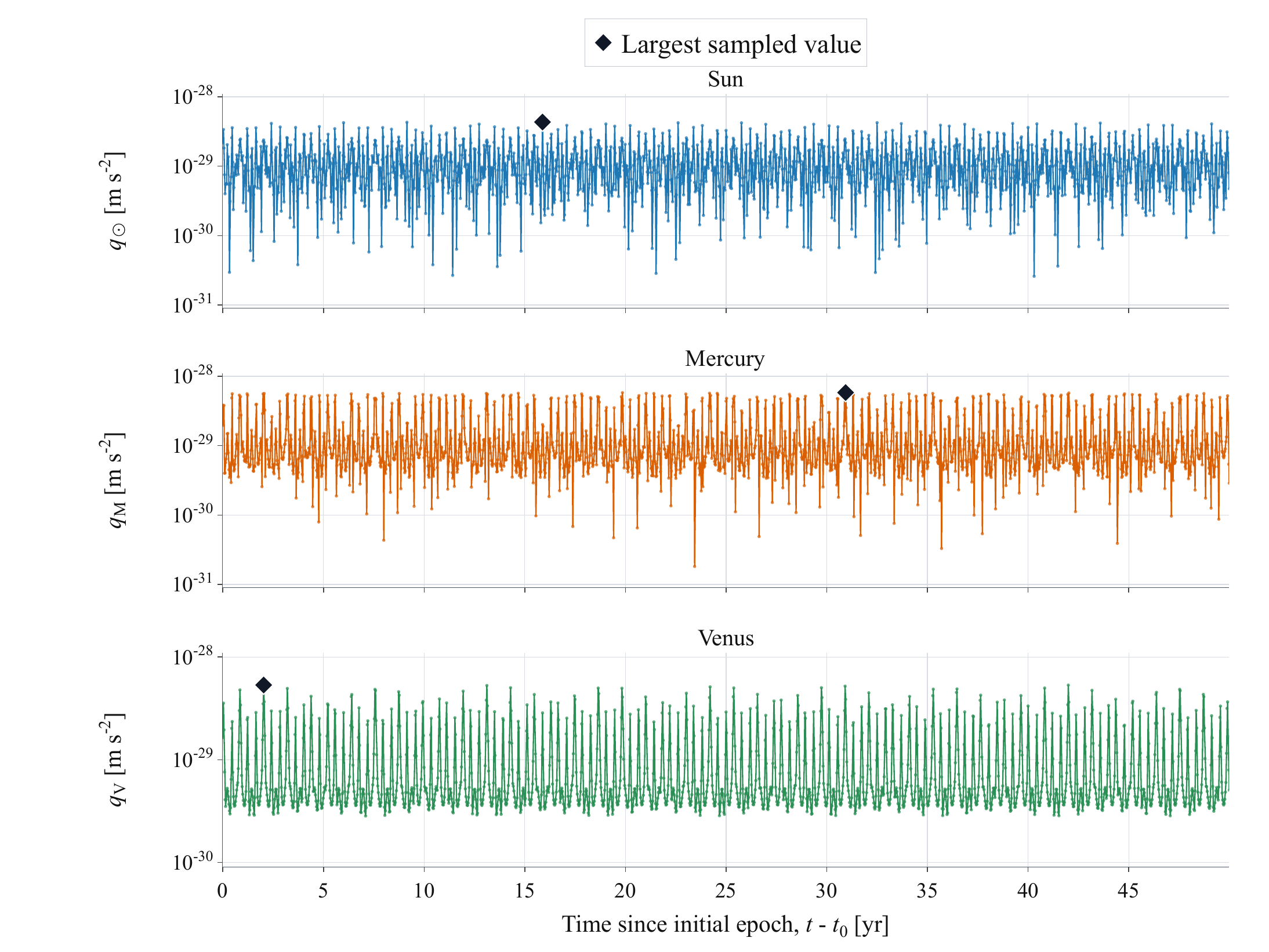}
        \caption{SMV benchmark.}
        \label{fig:smv-nc-acceleration-norm}
    \end{subfigure}
    \caption[]{Instantaneous nonclosed acceleration norm for the two
    benchmark systems (continued).}
\end{figure}

Figure~\ref{fig:nc-acceleration-norm-benchmarks} compares the absolute
magnitude of the nonclosed acceleration in the two benchmarks. In SJS,
\(q_A(t)\) exhibits recurrent variations while remaining at a broadly
similar scale over the \(1799.753\)-yr interval, with rms values of a few
\(10^{-28}\,\mathrm{m\,s^{-2}}\) and sampled maxima reaching the
\(10^{-27}\,\mathrm{m\,s^{-2}}\) level. The SMV time series shows more
rapid recurrent variations than the SJS time series, while remaining at a
smaller overall scale over the \(49.993\)-yr interval, with rms values of a
few \(10^{-29}\,\mathrm{m\,s^{-2}}\) and sampled maxima below
\(6\times10^{-29}\,\mathrm{m\,s^{-2}}\).

\begin{figure}[H]
    \centering
    \begin{subfigure}{0.98\linewidth}
        \centering
        \includegraphics[width=\linewidth,height=0.42\textheight,keepaspectratio]
        {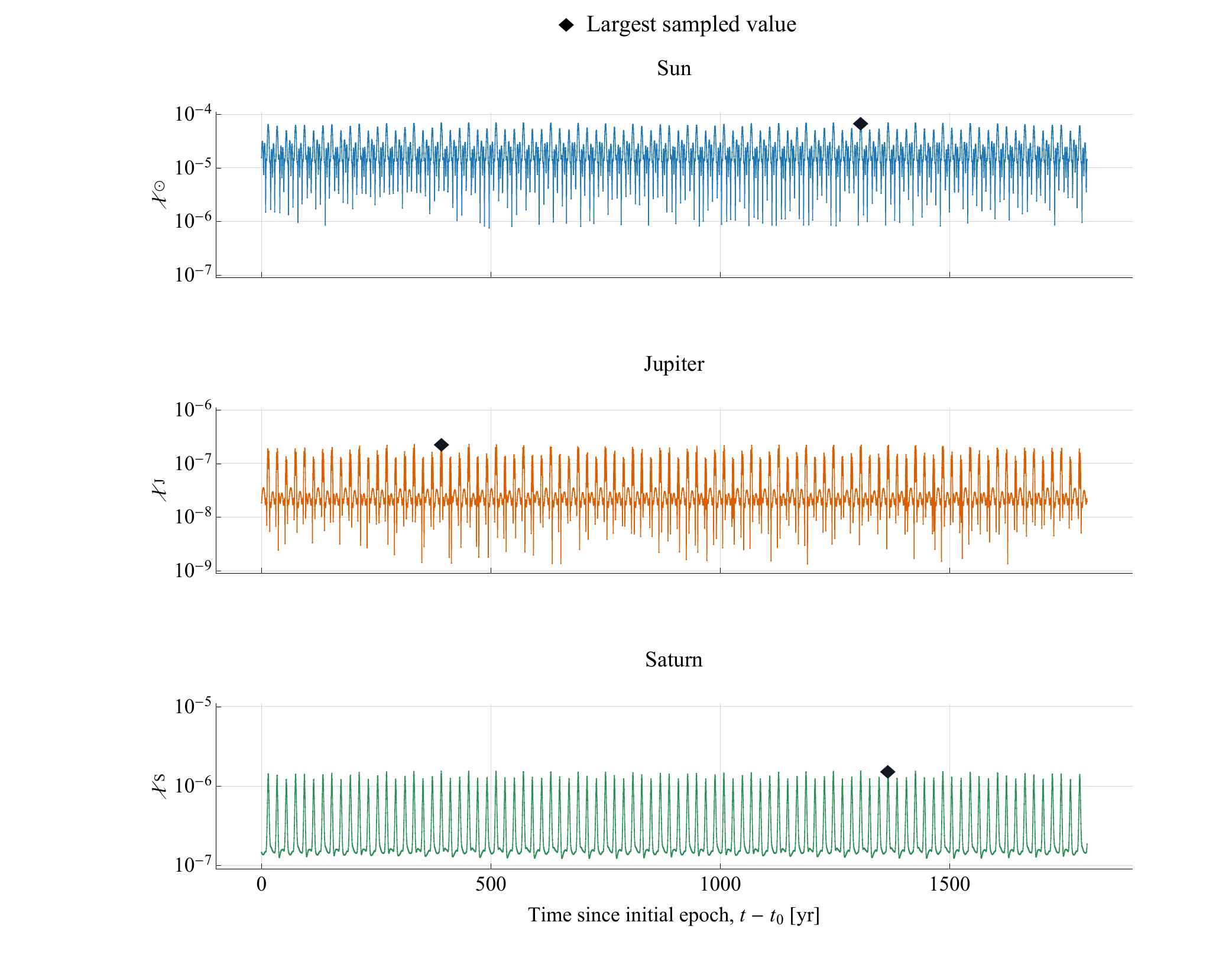}
        \caption{SJS benchmark.}
        \label{fig:sjs-nc-fraction-complete-2pn}
    \end{subfigure}

    \vspace{0.3ex}

    \begin{subfigure}{0.98\linewidth}
        \centering
        \includegraphics[width=\linewidth,height=0.42\textheight,keepaspectratio]
        {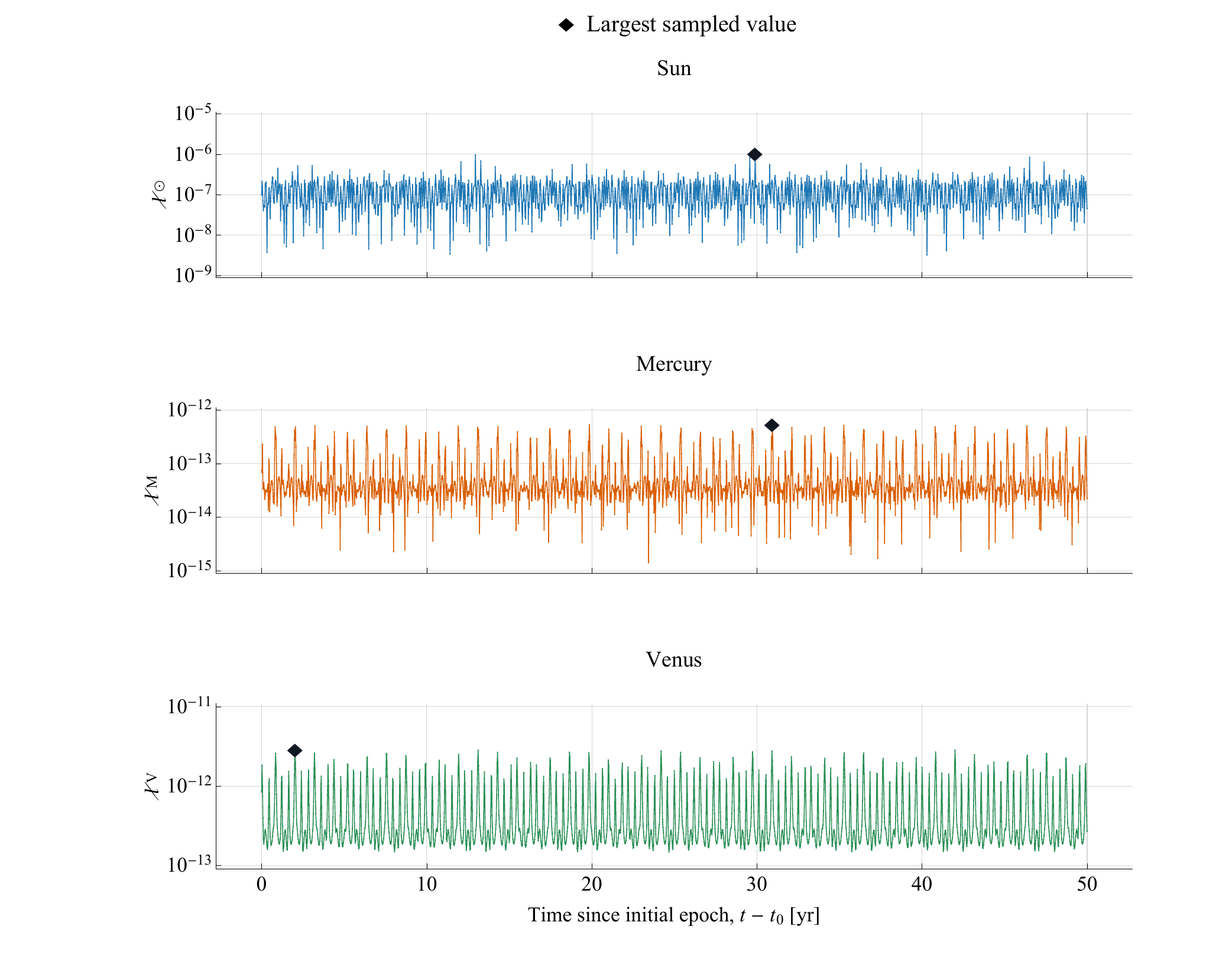}
        \caption{SMV benchmark.}
        \label{fig:smv-nc-fraction-complete-2pn}
    \end{subfigure}
    \caption{Norm ratio
    \(\chi_A(t)=|\mathbf a_A^{\mathrm{2PN,nc}}(t)|/
    |\mathbf a_A^{\mathrm{2PN}}(t)|\) for the two benchmark systems.
    Diamonds mark the largest sampled values. Each panel has its own
    logarithmic vertical scale. Panel (a) shows the SJS benchmark, and panel (b) shows the SMV benchmark.}
    \label{fig:nc-fraction-complete-2pn-benchmarks}
\end{figure}

Figure~\ref{fig:nc-fraction-complete-2pn-benchmarks} compares the relative
magnitude of the nonclosed contribution within the complete 2PN
acceleration. In both benchmarks, \(\chi_A(t)\) remains small throughout the
sampled interval. The largest relative contribution occurs for the Sun,
reaching \(6.911\times10^{-5}\), or \(6.911\times10^{-3}\%\), in SJS and
\(1.036\times10^{-6}\), or \(1.036\times10^{-4}\%\), in SMV. The
corresponding planetary ratios are smaller, especially in SMV.

\begin{table}[H]
    \centering
    \small
    \begingroup
    \renewcommand{\arraystretch}{1.12}
    \setlength{\tabcolsep}{4pt}
    \begin{tabular}{@{}llcccc@{}}
        \toprule
        System
        & Body
        & \(q_{A,\mathrm{rms}}\)
        & \(q_{A,\max}\)
        & \(\chi_{A,\mathrm{rms}}\)
        & \(\chi_{A,\max}\) \\
        \midrule
        SJS & Sun
        & \(2.954\times10^{-28}\)
        & \(7.970\times10^{-28}\)
        & \(2.594\times10^{-5}\)
        & \(6.911\times10^{-5}\) \\
        & Jupiter
        & \(4.202\times10^{-28}\)
        & \(1.468\times10^{-27}\)
        & \(5.992\times10^{-8}\)
        & \(2.286\times10^{-7}\) \\
        & Saturn
        & \(3.019\times10^{-28}\)
        & \(1.198\times10^{-27}\)
        & \(4.663\times10^{-7}\)
        & \(1.542\times10^{-6}\) \\
        \midrule
        SMV & Sun
        & \(1.434\times10^{-29}\)
        & \(4.322\times10^{-29}\)
        & \(1.571\times10^{-7}\)
        & \(1.036\times10^{-6}\) \\
        & Mercury
        & \(2.008\times10^{-29}\)
        & \(5.814\times10^{-29}\)
        & \(1.106\times10^{-13}\)
        & \(5.288\times10^{-13}\) \\
        & Venus
        & \(1.356\times10^{-29}\)
        & \(5.339\times10^{-29}\)
        & \(7.170\times10^{-13}\)
        & \(2.848\times10^{-12}\) \\
        \bottomrule
    \end{tabular}
    \endgroup
    \caption{Summary of the instantaneous nonclosed acceleration and its
    norm ratio to the complete 2PN acceleration. The SJS and SMV values are
    obtained from the 180-day and 10-day sampling grids, respectively. The
    two \(q_A\) columns are in \(\mathrm{m\,s^{-2}}\).}
    \label{tab:nc-instantaneous-summary}
\end{table}

In neither benchmark does \(q_A(t)\) or \(\chi_A(t)\) show an evident
systematic increase or decrease over the sampled interval. Both the absolute
magnitude of the nonclosed acceleration and its relative contribution to
the complete 2PN acceleration are smaller in SMV than in SJS.

\FloatBarrier
\subsection{Leading Finite-Time Relative-Distance Perturbation}
\label{subsec:finite-time-response}

Using Eqs.~\eqref{eq:nc-leading-perturbation} and
\eqref{eq:nc-perturbation-initial}, we characterize the orbital perturbation
generated by the nonclosed 2PN acceleration through the leading change in
the pairwise separations. For any pair \(A\ne B\), we define
\begin{align}
    \delta r_{AB}^{\mathrm{nc}}(t)
    &\equiv
    \frac{1}{c^4}
    \mathbf n_{AB}^{\mathrm{N}}(t)
    \!\cdot\!
    \left[
        \mathbf z_A^{\mathrm{2PN,nc}}(t)
        -
        \mathbf z_B^{\mathrm{2PN,nc}}(t)
    \right],
    \label{eq:relative-distance-response}
\end{align}
where
\begin{align}
    \mathbf n_{AB}^{\mathrm{N}}(t)
    &\equiv
    \frac{
        \mathbf z_A^{\mathrm{N}}(t)-\mathbf z_B^{\mathrm{N}}(t)
    }{
        r_{AB}^{\mathrm{N}}(t)
    },
    &
    r_{AB}^{\mathrm{N}}(t)
    &\equiv
    \left|
        \mathbf z_A^{\mathrm{N}}(t)-\mathbf z_B^{\mathrm{N}}(t)
    \right|.
    \label{eq:newtonian-pair-direction}
\end{align}
Thus, \(\delta r_{AB}^{\mathrm{nc}}\) is the physical first-order change in
the pairwise distance generated by the nonclosed contribution; the explicit
factor \(c^{-4}\) converts the 2PN displacement coefficient into a distance.

For the one-day perturbation-output grid \(t_m\), we define the sampled
maximum absolute perturbation by
\begin{align}
    \Delta r_{AB,\max}
    &\equiv
    \max_m
    \left|
        \delta r_{AB}^{\mathrm{nc}}(t_m)
    \right|.
    \label{eq:response-sampled-maximum}
\end{align}
These values are sampled maxima rather than continuous-time extrema.
The sampled maximum absolute relative-distance perturbations and their occurrence times are summarized in Table~\ref{tab:nc-response-summary}.

The physical nonclosed acceleration
\(c^{-4}\mathbf a_A^{\mathrm{2PN,nc}}\) is supplied on the 180-day SJS and
10-day SMV grids used above and is linearly interpolated in time. Numerically,
we integrate the equivalent \(c^{-4}\)-rescaled form of the perturbation
equations, so the propagated Cartesian perturbations are obtained directly
in SI units. The sixth-order Runge--Kutta integrations use
one-day and \(0.125\)-day steps for SJS and SMV, respectively, over the
\(1799.753\)-yr and \(49.993\)-yr benchmark intervals. In both cases,
\(\delta r_{AB}^{\mathrm{nc}}\) is recorded at one-day intervals.

\begin{figure}[H]
    \centering
    \captionsetup[subfigure]{skip=0pt}
    \begin{subfigure}{0.98\linewidth}
        \centering
        \includegraphics[width=\linewidth,height=0.44\textheight,keepaspectratio]
        {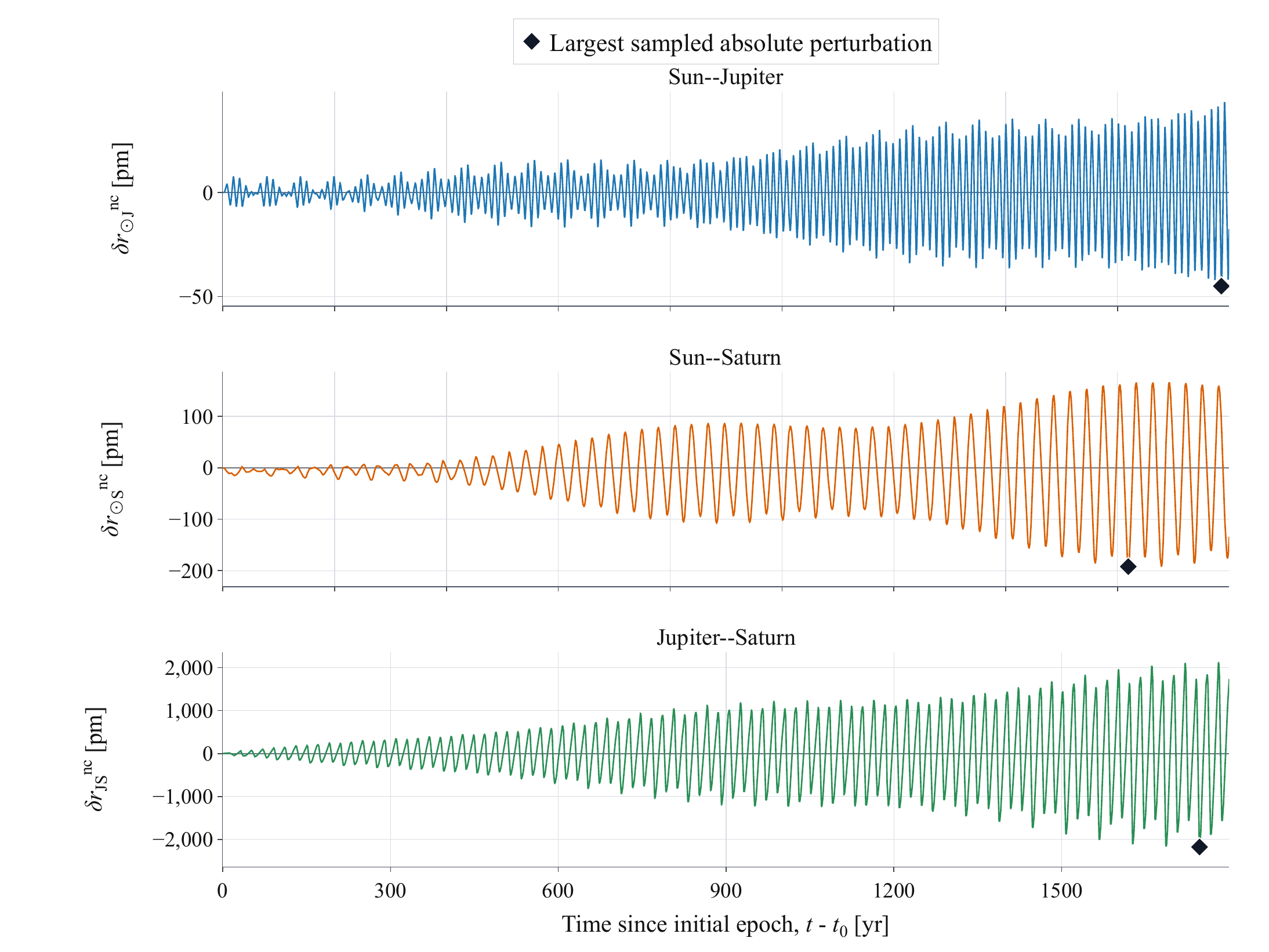}
        \caption{SJS benchmark.}
        \label{fig:sjs-nc-relative-distance-response}
    \end{subfigure}

    \vspace{0.3ex}

    \begin{subfigure}{0.98\linewidth}
        \centering
        \includegraphics[width=\linewidth,height=0.44\textheight,keepaspectratio]
        {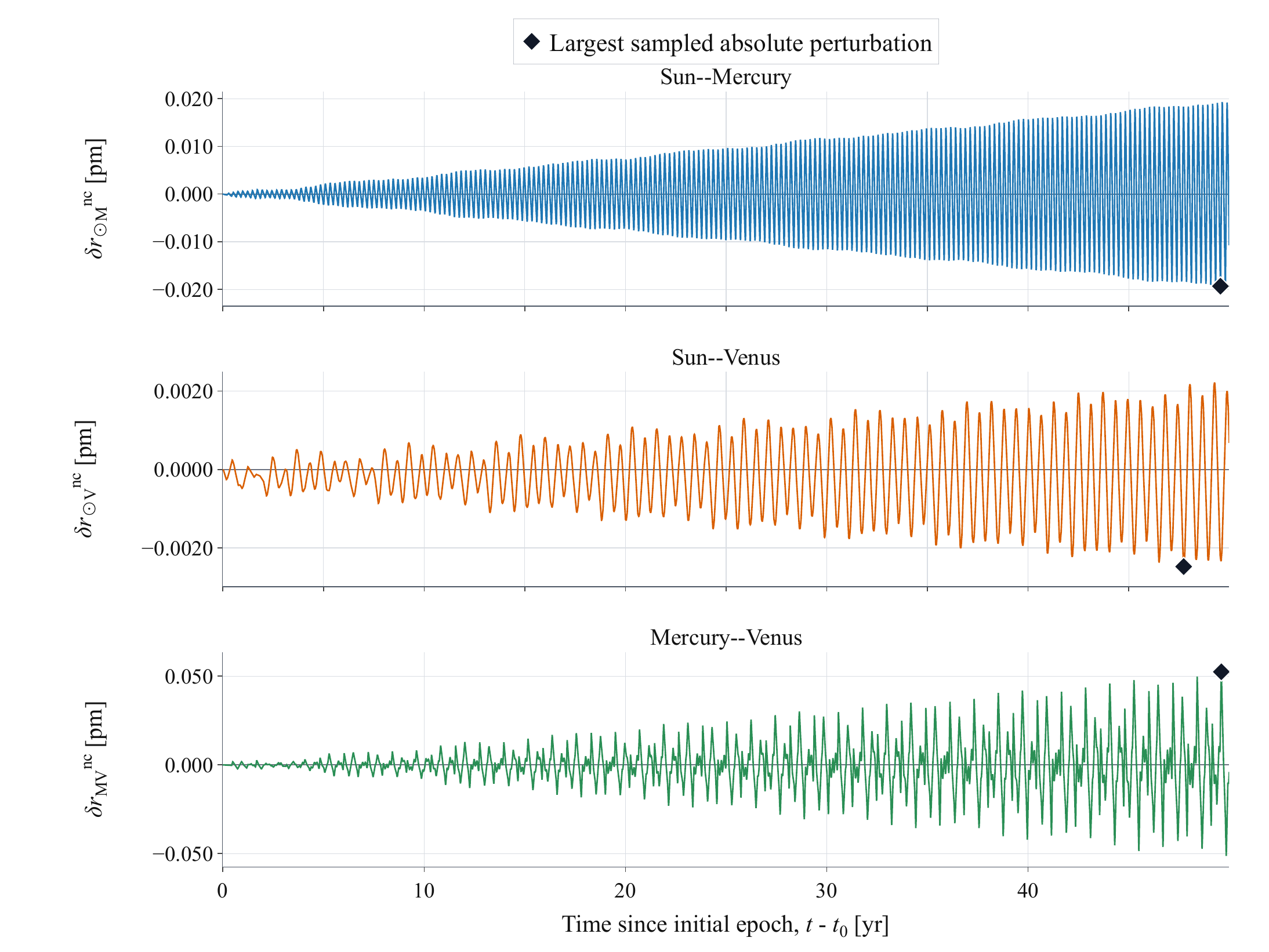}
        \caption{SMV benchmark.}
        \label{fig:smv-nc-relative-distance-response}
    \end{subfigure}
    \caption{Leading finite-time relative-distance perturbations generated by
    the nonclosed 2PN acceleration in the SJS and SMV benchmarks.
    Diamonds mark the largest sampled absolute perturbations. Panel (a) shows the SJS benchmark, and panel (b) shows the SMV benchmark.}
    \label{fig:nc-relative-distance-response-benchmarks}
\end{figure}

Figure~\ref{fig:nc-relative-distance-response-benchmarks} shows oscillatory
perturbations with repeated sign changes in both benchmarks, rather than
monotonic accumulation. Over the respective benchmark intervals, the SJS
perturbations span the picometer-to-nanometer range, with the
Jupiter--Saturn perturbation reaching \(2.176407\,\mathrm{nm}\), whereas all
three SMV perturbations remain below \(0.1\,\mathrm{pm}\). Since the two
benchmarks correspond to different systems and substantially different
integration intervals, these values characterize only their respective
finite-time perturbations and do not define a duration-independent ranking.

\begin{table}[H]
    \centering
    \small
    \begingroup
    \renewcommand{\arraystretch}{1.14}
    \setlength{\tabcolsep}{9pt}
    \begin{tabular}{@{}llrr@{}}
        \toprule
        System
        & Pair
        & \(\Delta r_{AB,\max}\)
        & \(t_{\max}-t_0\) \\
        &
        & \([\mathrm{pm}]\)
        & \([\mathrm{yr}]\) \\
        \midrule
        SJS & Sun--Jupiter
        & 45.054
        & 1785.714 \\
        & Sun--Saturn
        & 192.084
        & 1619.244 \\
        & Jupiter--Saturn
        & 2176.407
        & 1746.609 \\
        \midrule
        SMV & Sun--Mercury
        & 0.01932
        & 49.555 \\
        & Sun--Venus
        & 0.002475
        & 47.729 \\
        & Mercury--Venus
        & 0.05245
        & 49.599 \\
        \bottomrule
    \end{tabular}
    \endgroup
    \caption{Sampled maximum absolute relative-distance perturbations and
    their epochs for the SJS and SMV benchmarks. The perturbations are
    sampled on the one-day output grid.}
    \label{tab:nc-response-summary}
\end{table}

The larger perturbation amplitudes reached at later times are not accompanied
by a systematic increase in the instantaneous forcing norms shown in
Fig.~\ref{fig:nc-acceleration-norm-benchmarks}. They therefore reflect the
finite-time propagation of the nonclosed perturbation along the Newtonian
trajectories rather than growth of the forcing itself. Within the adopted benchmark
intervals, the perturbation remains oscillatory and shows no evidence of monotonic
or secular growth; no conclusion about unbounded long-term growth can be drawn 
from these finite-time calculations.

\FloatBarrier
\section{Conclusion}
\label{sec:conclusion}

In this work, we have derived the harmonic-gauge 2PN EOM for a general
system of \(N\) point masses within Hadamard regularization. The result is
expressed as the sum of a closed analytic contribution and a nonclosed
spatial-integral contribution \((I_k)_A\). By analyzing the distributional
and singular structures of the latter, we have separated the \(D\neq A\)
and \(D=A\) sectors and obtained a regularized semianalytic and
seminumerical representation that can be evaluated for a prescribed
\(N\)-body configuration. This completes the treatment of the residual
nonclosed contribution required by the harmonic-gauge 2PN EOM.

The numerical analysis was carried out for two representative three-body
benchmarks, SJS and SMV.
For SJS, the instantaneous nonclosed acceleration has an rms scale of a
few \(10^{-28}\,\mathrm{m\,s^{-2}}\), with sampled maxima reaching the
\(10^{-27}\,\mathrm{m\,s^{-2}}\) level, while its norm ratio to the
complete 2PN acceleration remains below \(7\times10^{-5}\) over the
\(1799.753\)-yr benchmark interval. The corresponding finite-time
relative-distance perturbations range from tens of picometers to a couple of nanometers,
with the largest sampled Jupiter--Saturn perturbation reaching
\(2.176407\,\mathrm{nm}\). For SMV, over \(49.993\) yr, the instantaneous
nonclosed acceleration has an rms scale of a few
\(10^{-29}\,\mathrm{m\,s^{-2}}\), the largest sampled norm ratio to the
complete 2PN acceleration is \(1.036\times10^{-6}\), and all three sampled
relative-distance perturbations remain below \(0.1\,\mathrm{pm}\).

In both benchmarks, the instantaneous nonclosed acceleration shows no
evident systematic increase or decrease over the sampled interval. The
finite-time perturbations remain oscillatory, although larger sampled amplitudes
can occur at later times within the adopted intervals. This behavior reflects
the finite-time dynamical propagation of the nonclosed perturbation. Within the
computed intervals, no evidence of monotonic or secular growth is found.
Because the two calculations involve different configurations and very
different integration intervals, the numerical perturbation scales characterize
the respective benchmarks and should not be interpreted as
duration-independent properties of the nonclosed contribution.

These results are intended as controlled evaluations of the leading
instantaneous and finite-time effects of the nonclosed 2PN contribution,
rather than as complete ephemeris residuals or observability forecasts.
They show that the nonclosed term is small in both representative
configurations considered here, while also demonstrating that its
finite-time dynamical response cannot be inferred from its instantaneous
norm alone. Any assessment of whether the nonclosed contribution can be
neglected in a particular application must therefore depend on the required
accuracy, the integration time, and the configuration being modeled.

Several extensions of the present work are natural. First, the numerical
evaluation of the nonclosed contribution can be further accelerated and
incorporated into a complete Solar System dynamical model. Combined with
fitting the initial conditions, mass parameters, and other relevant
parameters to observational data, this would make it possible to assess the
effect of the complete 2PN dynamics at the ephemeris level. Second, the
finite-time response
generated by the nonclosed contribution can be studied more systematically
over wider integration intervals and a broader range of \(N\)-body
configurations, with particular attention to the dynamical origin and
evolution of the response. Third, a systematic comparison with the ADM formulation is a natural direction for future work. Such a comparison would involve the complete 2PN dynamics and the appropriate transformation between ADM and harmonic-coordinate variables. Fourth, the harmonic-coordinate many-body
formulation can be extended beyond general relativity by parametrizing the
second-order post-Newtonian interactions. Such a development would provide a
route toward parametrized 2PN \(N\)-body EOM and, ultimately, toward
higher-order Solar System tests of relativistic gravity.

\noindent\textit{Note added.} After submission of this manuscript, Heinze, Sch\"afer, and Br\"ugmann presented a fully analytic expression for the general \(N\)-body 2PN ADM Hamiltonian~\cite{heinze2026fullyanalyticexpression2pn}. This analytic form may facilitate future comparisons with the present harmonic-coordinate formulation.

\section*{Acknowledgments}

This work was supported by the National Key Research and Development
Program of China (No.~2021YFC2203003), the National Natural Science
Foundation of China (Grant No.~12247101), the Fundamental
Research Funds for the Central Universities (Grant
No.~lzujbky-2025-jdzx07), the Natural Science Foundation of Gansu Province
(Nos.~22JR5RA389 and 25JRRA799), and the ``111 Center'' under Grant No.~B20063.

We acknowledge the use of OpenAI's GPT-5.6 Sol model as an auxiliary tool for cross-checking derivations and calculations independently obtained by the authors and for assisting with language editing. All analytical and numerical results and all scientific conclusions were independently derived, reviewed, and verified by the authors, who take full responsibility for the content of this work.

\addtocontents{toc}{\protect\setcounter{tocdepth}{1}}
\appendix
\setcounter{tocdepth}{1}
\setupappendixstyle

\section{2PN Metric Potentials}
\label{app:2pn-metric-potentials}
In this appendix, we collect the explicit expressions for the potentials entering the 2PN metric and EOM. 
These formulas are obtained by substituting the point-particle source terms into the potential definitions 
and applying the Hadamard regularization rules described in the main text. The inverse wave operator is the 
instantaneous PN inverse \(\Box^{-1}_{\rm inst}\), namely the even-power near-zone expansion of the retarded 
inverse used for the 2PN conservative dynamics. The notation used below follows the conventions 
introduced in Sec.~\ref{sec:notation-conventions}.

In the appendixes, several auxiliary symbols are introduced only to shorten
long intermediate formulas.  Because of the large number of quantities
appearing in the full 2PN expressions, some of these local auxiliary symbols
may coincide in notation with symbols used elsewhere in the main text.  Such
symbols should be understood as appendix-local abbreviations within the
formula in which they are defined; they do not denote, and should not be
identified with, the quantities carrying the same symbols in the main text.
For field-point two-body logarithms in this appendix, we use
\begin{align}
    S_{AB}
    &\equiv r_A+r_B+r_{AB},
    &
    g_{AB}
    &\equiv \ln S_{AB}.
\end{align}

\subsection{Potential \(V\)}
\begin{align}
    V
    &= G\sum_A \frac{m_A}{r_A}
    + \frac{G}{2c^2}\sum_A m_A\Bigg[
    \frac{4 v_A^{2} - (\mathbf n_A\!\cdot\!\mathbf v_A)^{2}}{r_A}
    - 2G\sum_{B\ne A}\frac{m_B}{r_A r_{AB}}
    - \mathbf n_A\!\cdot\!\mathbf a_A^{\mathrm{N}}
    \Bigg] \nonumber\\
    &\quad
    + \frac{G}{c^{4}}\sum_A m_A \Bigg\{
    \frac{7}{8}\frac{v_A^{4}}{r_A}
    + \frac{1}{2}\frac{G v_A^{2}}{r_A}\sum_{B\ne A}\frac{m_B}{r_{AB}}
    - 4G\sum_{B\ne A}\frac{m_B}{r_{AB}}\frac{\mathbf v_B\!\cdot\!\mathbf v_A}{r_A}
    + 4G\sum_{B\ne A}\frac{m_B v_B^{2}}{r_{AB} r_A} \nonumber\\[2pt]
    &\qquad
    + G^{2}\frac{1}{r_A}\Bigg[
    \frac{3}{2}\sum_{B\ne A}\frac{m_B^{2}}{r_{AB}^{2}}
    + 3\sum_{\substack{B<C\\ B,C\ne A}}\frac{m_B m_C}{r_{AB} r_{AC}}
    - 2\sum_{B\ne A}\frac{m_B}{r_{AB}}\sum_{C\ne B}\frac{m_C}{r_{BC}}
    \Bigg] \nonumber\\[2pt]
    &\qquad
    - G\frac{1}{r_A}\sum_{B\ne A}\Bigg\{
    \frac{m_B}{r_{AB}}\!\left(\frac{3}{2}v_B^{2}
    - G\sum_{C\ne B}\frac{m_C}{r_{BC}}\right)
    \nonumber\\[2pt]
    &\qquad\qquad
    + \frac{m_B}{2}\Bigg[
    \sum_{C\ne B}\frac{Gm_C}{r_{BC}^{2}}\,
    \frac{-r_{AB}^{2}+r_{AC}^{2}-r_{BC}^{2}}{2\,r_{AB}\,r_{BC}}
    + \frac{v_B^{2}-(\mathbf n_{AB}\!\cdot\!\mathbf v_B)^{2}}{r_{AB}}
    \Bigg]\!\Bigg\} \nonumber\\[4pt]
    &\qquad
    + \frac{1}{2}r_A\Bigg[
    3\,(a_A^{\mathrm{N}})^2 + 3\,\mathbf v_A\!\cdot\! \dot {\mathbf a}^{\mathrm{N}}_A
    \nonumber\\[2pt]
    &\qquad\qquad
    + G\sum_{B\ne A} m_B\Bigg(
    \frac{\mathbf n_{AB} \cdot (\mathbf a_A^{\mathrm{N}}-\mathbf a_B^{\mathrm{N}})}{r_{AB}^{2}}
    + \frac{(\mathbf v_A-\mathbf v_B)^{2}-3\big(\mathbf n_{AB} \cdot (\mathbf v_A-\mathbf v_B)\big)^{2}}{r_{AB}^{3}}
    \Bigg)
    \Bigg] \nonumber\\[2pt]
    &\qquad
    - (\mathbf n_A\!\cdot\!\mathbf v_A)\Bigg[
    3\,\mathbf v_A\!\cdot\!\mathbf a^{\mathrm{N}}_A
    + G\sum_{B\ne A} m_B\,\frac{\mathbf n_{AB}\!\cdot\!(\mathbf v_A-\mathbf v_B)}{r_{AB}^{2}}
    \Bigg] \nonumber\\[2pt]
    &\qquad
    + \frac{1}{2}\left(
    \frac{v_A^{2}-(\mathbf n_A\!\cdot\!\mathbf v_A)^{2}}{r_A}
    - \mathbf n_A\!\cdot\!\mathbf a^{\mathrm{N}}_A
    \right)\left(
    \frac{3}{2}v_A^2 - G\sum_{B\ne A}\frac{m_B}{r_{AB}}
    \right) \nonumber\\[4pt]
    &\qquad
    + \frac{1}{8}\Bigg[
    \frac{3}{r_A}\Big(v_A^{2}-(\mathbf n_A\!\cdot\!\mathbf v_A)^{2}\Big)^{2}
    - 6\big(\mathbf n_A\!\cdot\!\mathbf a^{\mathrm{N}}_A\big)\Big(v_A^{2}-(\mathbf n_A\!\cdot\!\mathbf v_A)^{2}\Big)
    - 12\,(\mathbf n_A\!\cdot\!\mathbf v_A)\,(\mathbf v_A\!\cdot\!\mathbf a^{\mathrm{N}}_A) \nonumber\\[2pt]
    &\qquad\qquad
    + 4 r_A\,(\mathbf n_A\!\cdot\!\mathbf v_A)\,(\mathbf n_A\!\cdot\!\dot{\mathbf a}^{\mathrm{N}}_A)
    + 3 r_A\big(\mathbf n_A\!\cdot\!\mathbf a^{\mathrm{N}}_A\big)^{2}
    + 3 r_A\,(a_A^{\mathrm{N}})^{2}
    + 4 r_A\,(\mathbf v_A\!\cdot\!\dot{\mathbf a}^{\mathrm{N}}_A)
    - r_A^{2}(\mathbf n_A\!\cdot\!\ddot{\mathbf a}^{\mathrm{N}}_A)
    \Bigg] \nonumber\\[2pt]
    &\qquad
    + \frac{1}{2}\sum_{B\neq A} \frac{G m_B}{r_{AB}^2}\,(\mathbf n_A\!\cdot\!\mathbf n_{AB})
    \Bigg[
    v_A^2 + 2 v_B^2 - 4\,\mathbf v_A\!\cdot\!\mathbf v_B
    - \frac{3}{2}\big(\mathbf n_{AB}\!\cdot\!\mathbf v_B\big)^2
    - 4 \sum_{C\neq A} \frac{G m_C}{r_{AC}} \nonumber\\[4pt]
    &\qquad- \sum_{C\neq B} \frac{G m_C}{r_{BC}}
    \left( 1 + \frac{1}{2}\frac{r_{AB}}{r_{CB}}\,\mathbf n_{AB}\!\cdot\!\mathbf n_{CB} \right)
    \Bigg] 
    + \frac{7}{4}\sum_{B\neq A}\sum_{C\neq B}
    \frac{G^2 m_B m_C}{r_{AB} r_{BC}^{2}}\, (\mathbf n_A\!\cdot\!\mathbf n_{BC}) \nonumber\\[4pt]
    &\qquad- \frac{1}{2}\sum_{B\neq A}\frac{G m_B}{r_{AB}^2}\,
    \big(\mathbf n_A\!\cdot\!(\mathbf v_A - \mathbf v_B)\big)\,
    \Big( 4\,\mathbf n_{AB}\!\cdot\!\mathbf v_A - 3\,\mathbf n_{AB}\!\cdot\!\mathbf v_B \Big)
    \Bigg\}
    + O(c^{-6}).
\end{align}

where the Newtonian acceleration and its time derivatives are given by
\begin{align}
    (a_{A}^{\mathrm{N}})^i
    &= - \sum_{B\neq A} \frac{G m_B}{r_{AB}^2}\, n_{AB}^i,
    \label{eq:appendix-newtonian-acceleration}\\
    (\dot a_A^{\mathrm{N}})^i
    &= - \sum_{B\neq A} \frac{G m_B}{r_{AB}^3}
    \Big[
    (v_A^i - v_B^i)
    - 3\big(n_{AB}\!\cdot\!(\mathbf v_A-\mathbf v_B)\big)\,n_{AB}^i
    \Big],
    \label{eq:appendix-newtonian-jerk}\\
    (\ddot a_{A}^{\mathrm{N}})^i
    &= - \sum_{B\neq A} \frac{G m_B}{r_{AB}^3}
    \Bigg[
    \bigg(
    - \sum_{C\neq A} \frac{G m_C}{r_{AC}^2}\, n_{AC}^i
    + \sum_{C\neq B} \frac{G m_C}{r_{BC}^2}\, n_{BC}^i
    \bigg)
    - \frac{6}{r_{AB}}\big(n_{AB}\!\cdot\!(\mathbf v_A-\mathbf v_B)\big)\,(v_A^i - v_B^i) \nonumber\\[-0.5ex]
    &\qquad
    - \frac{3}{r_{AB}}
    \Big(
    |\mathbf v_A-\mathbf v_B|^2
    + r_{AB}\,n_{AB}\!\cdot\!\big((\mathbf a_A^{\mathrm{N}})-(\mathbf a_B^{\mathrm{N}})\big)
    - 5\big(n_{AB}\!\cdot\!(\mathbf v_A-\mathbf v_B)\big)^2
    \Big)\,n_{AB}^i
    \Bigg].
    \label{eq:appendix-newtonian-snap}
\end{align}

\subsection{Potential \(V_i\)}
\begin{align}
    V_i
    &= G\sum_A \frac{m_A v_A^i}{r_A}
    + \frac{G}{c^{2}} \sum_A m_A\Bigg\{
    v_A^i\Bigg[
    \frac{ v_A^{2}}{r_A}
    - \frac{1}{2}\frac{(\mathbf n_A\!\cdot\!\mathbf v_A)^{2}}{r_A}
    - G\sum_{B\neq A}\frac{m_B}{r_A r_{AB}}
    - \frac{1}{2} \mathbf n_A\!\cdot\!\mathbf a_A^{\mathrm{N}}
    \Bigg] \nonumber\\
    &\qquad\qquad
    -(\mathbf n_A\!\cdot\!\mathbf v_A)(a_A^{\mathrm{N}})^i
    +\frac{1}{2}r_A(\dot a_A^{\mathrm{N}})^i
    \Bigg\}
    + O(c^{-4}).
\end{align}

\subsection{Potential \(\hat{W}_{ij}\)}
\begin{align}
    \hat{W}_{ij}
    &= G\sum_A \frac{m_A}{r_A}\Bigl(v_A^i v_A^j-\delta^{ij}v_A^2\Bigr)
    -\frac{G^2}{4}\sum_A m_A^2\,\frac{\delta^{ij}-n_A^i n_A^j}{r_A^2} \nonumber\\
    &\quad
    - G^2\sum_{A<B} m_A m_B \Biggl[
    2\,\frac{n_{AB}^i n_{AB}^j-\delta^{ij}}{r_{AB}S_{AB}}
    +\frac{
    2n_{AB}^i n_{AB}^j
    +2n_{AB}^{(i}n_B^{j)}
    -2n_{AB}^{(i}n_A^{j)}
    -2n_A^{(i}n_B^{j)}
    }{S_{AB}^2}
    \Biggr]
    +O(c^{-2}).
\end{align}

\subsection{Potential \(\hat{R}_{i}\)}
\begin{align}
    \hat{R}_{i}
    &= G^2 \sum_{A<B} m_A m_B
    \frac{1}{r_{AB}}
    \left(\frac{1}{r_A}-\frac{1}{r_B}\right)(v_A^i-v_B^i)
    -\frac{G^2}{8}\sum_A m_A^2
    \frac{v_A^i-n_A^i n_A^j v_A^j}{r_A^2} \nonumber\\
    &\quad
    - G^2 \sum_{A\ne B} m_A m_B
    \Biggl(2 v_B^j-\frac{3}{2}v_A^j\Biggr)
    \left[
    \frac{n_{AB}^i n_{AB}^j-\delta^{ij}}{r_{AB}S_{AB}}
    +\frac{(n_{AB}^j-n_A^j)(n_{AB}^i+n_B^i)}{S_{AB}^2}
    \right]
    +O(c^{-2}).
\end{align}

\subsection{Potential \(\hat{X}\)}
\begin{align}
    \hat{X}
    &= G^2 \sum_A \sum_{B \ne A} \frac{m_A m_B}{r_A r_{AB}}\, v_A^2 \nonumber\\
    &\quad
    +G^2 \sum_A m_A^2 \Bigg[
    -\frac{1}{2}\,\frac{(a_A^{\mathrm{N}})^i n_A^i}{r_A}
    -\frac{1}{8}\,\frac{v_A^2}{r_A^2}
    +\frac{1}{8}\,\frac{(\mathbf v_A\!\cdot\!\mathbf n_A)^2}{r_A^2}
    \Bigg] \nonumber\\
    &\quad
    + G^2 \sum_{A<B} m_A m_B \Bigg[
    \big((a_B^{\mathrm{N}})^i(\partial_B)_i + (a_A^{\mathrm{N}})^i(\partial_A)_i\big)\,g_{AB} \nonumber\\
    &\qquad
    + \Big(v_B^i v_B^j + v_A^i v_A^j - 2v_B^i v_A^j - \delta_{ij}v_B^2\Big)\,
    (\partial_A)_i(\partial_A)_j\,g_{AB} \nonumber\\
    &\qquad
    + \Big(v_B^i v_B^j + v_A^i v_A^j - 2v_A^i v_B^j - \delta_{ij}v_A^2\Big)\,
    (\partial_B)_i(\partial_B)_j\,g_{AB} \nonumber\\
    &\qquad
    + \Big(\tfrac{3}{2}v_A^i v_B^j - 2v_A^j v_B^i\Big)\,
    (\partial_A)_i(\partial_B)_j\,g_{AB} \nonumber\\
    &\qquad
    + \Big(\tfrac{3}{2}v_B^i v_A^j - 2v_B^j v_A^i\Big)\,
    (\partial_B)_i(\partial_A)_j\,g_{AB}
    \Bigg] \nonumber\\
    &\quad
    + \frac{G^3}{12} \sum_A \frac{m_A^3}{r_A^3}
    - G^3 \sum_A \sum_{B \ne A} \frac18 m_A^2 m_B
    \left(\frac12 K_{AB}+\frac{1}{r_B r_{AB}^2}\right) \nonumber\\
    &\quad
    - G^3 \sum_{A \ne B} m_A m_B
    \left[
    \frac12 m_A H_{AB}
    +\frac12 m_B H_{BA}
    +\sum_{C \ne A,B} m_C\,\mathcal I_{ABC}
    \right]
    + O(c^{-2}).
\end{align}
where
\begin{align*}
    \mathcal I_{ABC}
    &\equiv
    \Delta^{-1}\!\left[\partial_i \partial_j \frac{1}{r_C}\,(\partial_A)_i(\partial_B)_j g_{AB}\right],\\
    K_{AB}
    &= -\frac{1}{r_B^3}
    +\frac{1}{r_B r_{AB}^2}
    -\frac{1}{r_A^2 r_B}
    +\frac{r_B}{2r_A^2 r_{AB}^2}
    +\frac{r_{AB}^2}{2r_A^2 r_B^3}
    +\frac{r_A^2}{2r_B^3 r_{AB}^2}, \\[1mm]
    H_{AB}
    &= -\frac{1}{2 r_A^3}
    -\frac{1}{4 r_{AB}^3}
    -\frac{1}{4 r_A^2 r_{AB}}
    -\frac{r_B}{2 r_A^2 r_{AB}^2}
    +\frac{r_B}{2 r_A^3 r_{AB}}
    +\frac{3r_B^2}{4 r_A^2 r_{AB}^3}
    +\frac{r_B^2}{2 r_A^3 r_{AB}^2}
    -\frac{r_B^3}{2 r_A^3 r_{AB}^3}.
\end{align*}

\section{Derivation of the Nonclosed Integral Contribution}
\label{app:nonclosed-origin}

This appendix justifies the particle expression used for the nonclosed
integral contribution in the main text.  The point is the following.  The
quantity first arises as a field-point integral \(I_k(\mathbf x)\), defined
for a field point \(\mathbf x\) away from the particles.  The value entering
the EOM for particle \(A\) should then be the Hadamard finite
part of this field-point function as \(\mathbf x\to\mathbf z_A\).  In the
main text, however, the result is written in the direct particle form
obtained by replacing the field-point kernel with
\((\partial_A)_k(1/r'_A)\).  We show that these two procedures give the same
finite part.

Throughout this appendix, \(A,B,C,D\) denote particle labels, while
\(i,j,k,a,b\) denote Cartesian components; repeated Cartesian indices are
summed.  For any particle label \(P\), we write
\[
    r'_P:=|\mathbf x'-\mathbf z_P|,
    \qquad
    \mathbf n'_P:=\frac{\mathbf x'-\mathbf z_P}{r'_P},
    \qquad
    r_{PQ}:=|\mathbf z_P-\mathbf z_Q|.
\]
The derivative \((\partial_P)_i\) acts on the particle position
\(\mathbf z_P\), while \(\partial_k\) and \(\partial'_k\) act on
\(\mathbf x\) and \(\mathbf x'\), respectively.  The three-dimensional
Dirac distribution supported at \(\mathbf z_D\) is denoted by
\[
    \delta_D:=\delta^{(3)}(\mathbf x'-\mathbf z_D),
\]
and \(\delta_{ij}\) denotes the Kronecker delta.  The symbol
\(\operatorname{Pf}\) denotes the Hadamard finite-part operation applied to
the following singular kernel or singular integral.

\subsection{Field-Point Integral and Target Identity}

Starting from the nonclosed part of \(\hat X\) obtained in
Appendix~\ref{app:2pn-metric-potentials}, differentiation with respect to the field
point \(x^k\) gives
\begin{align}
    I_k(\mathbf x)
    &=
    \frac{G^3}{4\pi}
    \sum_{B\neq C}\sum_{D\neq B,C}
    m_Bm_Cm_D\,
    \operatorname{Pf}
    \int
    (\partial_B)_i(\partial_C)_j
    \ln(r'_B+r'_C+r_{BC})
    \notag\\
    &\quad\times
    \partial_k\frac{1}{|\mathbf x-\mathbf x'|}
    \,
    (\partial_D)_i(\partial_D)_j\frac{1}{r'_D}\,
    d^3x' .
    \label{eq:app-Ik-field}
\end{align}
The finite-part operation is needed because
\((\partial_D)_i(\partial_D)_j(1/r'_D)\) is a singular distribution
centered at \(\mathbf z_D\).  For a field point away from the particles, the
additional local singularity at \(\mathbf x'=\mathbf x\) is locally
integrable in three spatial dimensions.

The particle value associated with particle \(A\) is defined by the
Hadamard finite part of the field-point function at \(\mathbf z_A\):
\begin{align}
    (I_k)_A
    &:=
    \frac{1}{4\pi}
    \int_{S^2}d\Omega_A\,
    \bigl[
    I_k(\mathbf z_A+\varepsilon\mathbf n_A)
    \bigr]_{\varepsilon^0}.
    \label{eq:app-IA-Hadamard}
\end{align}
Here \(d\Omega_A\) is the solid angle element of the unit vector
\(\mathbf n_A\), and \([\cdots]_{\varepsilon^0}\) denotes the coefficient of
\(\varepsilon^0\) in the expansion as \(\varepsilon\to0\).

The direct particle expression used in the main text is
\begin{align}
    I_k(\mathbf z_A)
    &:=
    \frac{G^3}{4\pi}
    \sum_{B\neq C}\sum_{D\neq B,C}
    m_Bm_Cm_D\,
    \operatorname{Pf}
    \int
    (\partial_B)_i(\partial_C)_j
    \ln(r'_B+r'_C+r_{BC})
    \notag\\
    &\quad\times
    (\partial_A)_k\frac{1}{r'_A}
    \,
    (\partial_D)_i(\partial_D)_j\frac{1}{r'_D}\,
    d^3x' .
    \label{eq:app-Ik-direct}
\end{align}
Since
\begin{align}
    (\partial_A)_k\frac{1}{r'_A}
    =
    \frac{n_A'{}^k}{(r'_A)^2},
    \label{eq:app-dA-1overrA}
\end{align}
Eq.~\eqref{eq:app-Ik-direct} is a finite-part integral in the single
integration variable \(\mathbf x'\).  The goal of this appendix is to prove
\begin{align}
    (I_k)_A=I_k(\mathbf z_A).
    \label{eq:app-target-equivalence}
\end{align}

\subsection{Distributional Split of the \(D\)-Centered Kernel}

The singular \(D\)-centered factor is the distributional second derivative
of \(1/r'_D\):
\begin{align}
    (\partial_D)_i(\partial_D)_j\frac{1}{r'_D}
    =
    \operatorname{Pf}
    \frac{3n_D'{}^i n_D'{}^j-\delta_{ij}}{(r'_D)^3}
    -
    \frac{4\pi}{3}\delta_{ij}\,\delta_D .
    \label{eq:app-Tij-distribution}
\end{align}
Introduce the abbreviations
\begin{align}
    \mathcal K_k(\mathbf x,\mathbf x')
    &:=
    \partial_k\frac{1}{|\mathbf x-\mathbf x'|},
    \notag\\
    \mathcal K_{ij}^{(BC)}(\mathbf x')
    &:=
    (\partial_B)_i(\partial_C)_j
    \ln(r'_B+r'_C+r_{BC}),
    \notag\\
    Q_D^{ij}(\mathbf n'_D)
    &:=
    3n_D'{}^i n_D'{}^j-\delta_{ij}.
    \label{eq:app-dA-kernel}
\end{align}
Then the field-point integral naturally splits into a finite-part source
piece and a contact piece:
\begin{align}
    I_k(\mathbf x)
    =
    I_k^{\rm Pf}(\mathbf x)
    +
    I_k^\delta(\mathbf x),
    \label{eq:app-Ik-split}
\end{align}
with
\begin{align}
    I_k^{\rm Pf}(\mathbf x)
    &=
    \frac{G^3}{4\pi}
    \sum_{B\neq C}\sum_{D\neq B,C}
    m_Bm_Cm_D\,
    \operatorname{Pf}
    \int
    \mathcal K_{ij}^{(BC)}(\mathbf x')\,
    \mathcal K_k(\mathbf x,\mathbf x')\,
    \frac{Q_D^{ij}(\mathbf n'_D)}{(r'_D)^3}\,
    d^3x',
    \label{eq:app-Ik-Pf-part}
    \\
    I_k^\delta(\mathbf x)
    &=
    -\frac{G^3}{3}
    \sum_{B\neq C}\sum_{D\neq B,C}
    m_Bm_Cm_D\,
    \mathcal K_k(\mathbf x,\mathbf z_D)\,
    \mathcal K_{ii}^{(BC)}(\mathbf z_D).
    \label{eq:app-Ik-delta-part}
\end{align}
The trace in the contact term is an ordinary finite value because
\(D\neq B,C\).  Explicitly,
\begin{align}
    \mathcal K_{ii}^{(BC)}(\mathbf z_D)
    =
    \frac{1}{2}
    \left(
    \frac{1}{r_{BD}r_{CD}}
    -
    \frac{1}{r_{BD}r_{BC}}
    -
    \frac{1}{r_{CD}r_{BC}}
    \right),
    \qquad
    D\neq B,C .
    \label{eq:app-Aii-D}
\end{align}

We shall compare the two pieces in Eq.~\eqref{eq:app-Ik-split} separately.
The source part \(I_k^{\rm Pf}\) carries the punctured finite-part kernel,
whereas \(I_k^\delta\) is a finite sum of ordinary field kernels evaluated
at particle positions.

It remains only to recall why the finite-part source kernel can be
represented by a punctured limit.  Fix \(\mathbf x\) and choose
\(\rho_D>0\) so that the ball
\[
    \mathcal B_D(\rho_D)
    :=
    \{\mathbf x'\mid r'_D<\rho_D\}
\]
contains neither \(\mathbf z_B\), nor \(\mathbf z_C\), nor the singular
point \(\mathbf x'=\mathbf x\) of \(\mathcal K_k(\mathbf x,\mathbf x')\).
Inside this ball define
\[
    \mathcal F_{kij}^{(BCD)}(\mathbf x,\mathbf x')
    :=
    \mathcal K_{ij}^{(BC)}(\mathbf x')\,
    \mathcal K_k(\mathbf x,\mathbf x') .
\]
Then \(\mathcal F_{kij}^{(BCD)}\) is smooth in \(\mathbf x'\) near
\(\mathbf z_D\).  Writing
\(\mathbf x'=\mathbf z_D+r\mathbf n'_D\), Taylor's formula with integral
remainder gives
\begin{align}
    \mathcal F_{kij}^{(BCD)}
    (\mathbf x,\mathbf z_D+r\mathbf n'_D)
    &=
    \mathcal F_{kij}^{(BCD)}(\mathbf x,\mathbf z_D)
    +
    r n_D'{}^a
    \partial'_a\mathcal F_{kij}^{(BCD)}(\mathbf x,\mathbf z_D)
    \notag\\
    &\quad
    +
    r^2 n_D'{}^a n_D'{}^b
    \int_0^1(1-s)\,
    \partial'_a\partial'_b
    \mathcal F_{kij}^{(BCD)}
    (\mathbf x,\mathbf z_D+s r\mathbf n'_D)\,ds .
    \label{eq:app-F-local-expansion}
\end{align}
For a small shell \(\lambda<r'_D<\lambda'\), the only possible logarithmic
contribution comes from the zeroth-order term:
\begin{align}
    \int_{\lambda}^{\lambda'}
    \frac{dr}{r}
    \int d\Omega'_D\,
    Q_D^{ij}(\mathbf n'_D)
    \mathcal F_{kij}^{(BCD)}(\mathbf x,\mathbf z_D).
\end{align}
It vanishes because
\begin{align}
    \int d\Omega'_D\,
    Q_D^{ij}(\mathbf n'_D)
    =
    \int d\Omega'_D\,
    \left(3n_D'{}^i n_D'{}^j-\delta_{ij}\right)
    =
    0 .
    \label{eq:app-Q-angular-zero}
\end{align}
The remaining terms are integrable in the limit \(\lambda\to0\).  Hence
\begin{align}
    \operatorname{Pf}
    \int
    \mathcal F_{kij}^{(BCD)}(\mathbf x,\mathbf x')\,
    \frac{Q_D^{ij}(\mathbf n'_D)}{(r'_D)^3}\,
    d^3x'
    =
    \lim_{\lambda\to0}
    \int_{r'_D>\lambda}
    \mathcal F_{kij}^{(BCD)}(\mathbf x,\mathbf x')\,
    \frac{Q_D^{ij}(\mathbf n'_D)}{(r'_D)^3}\,
    d^3x' .
    \label{eq:app-Pf-to-punctured}
\end{align}

\subsection{Angular Average of the Field-Point Kernel}

The bridge between the field-point construction and the direct particle
formula is the angular average of the field-point kernel around
\(\mathbf z_A\).  Set
\[
    \mathbf x=\mathbf z_A+\varepsilon\mathbf n_A,
    \qquad
    \varepsilon>0,
\]
and average over \(\mathbf n_A\).  For fixed \(\mathbf x'\), write
\[
    \mathbf x'=\mathbf z_A+r'_A\mathbf n'_A .
\]
The Newtonian shell theorem gives
\begin{align}
    \frac{1}{4\pi}
    \int_{S^2}
    \frac{d\Omega_A}
    {|\mathbf z_A+\varepsilon\mathbf n_A-\mathbf x'|}
    =
    \frac{1}{\varepsilon}\Theta(\varepsilon-r'_A)
    +
    \frac{1}{r'_A}\Theta(r'_A-\varepsilon),
    \label{eq:app-poisson-average}
\end{align}
where \(\Theta\) is the Heaviside step function.  Since
\(\partial_k=-\partial'_k\) when acting on
\(|\mathbf x-\mathbf x'|^{-1}\), differentiating
Eq.~\eqref{eq:app-poisson-average} yields
\begin{align}
    \frac{1}{4\pi}
    \int_{S^2}
    d\Omega_A\,
    \mathcal K_k(\mathbf z_A+\varepsilon\mathbf n_A,\mathbf x')
    =
    \frac{n_A'{}^k}{(r'_A)^2}
    \Theta(r'_A-\varepsilon).
    \label{eq:app-angular-identity}
\end{align}
The terms proportional to \(\delta(r'_A-\varepsilon)\) cancel because the
two coefficients in Eq.~\eqref{eq:app-poisson-average} agree on the surface
\(r'_A=\varepsilon\).

Equation~\eqref{eq:app-angular-identity} is the key identity.  It says that
the angular average of the field-point kernel is precisely the direct
particle kernel \((\partial_A)_k(1/r'_A)\), with the small ball
\(r'_A<\varepsilon\) excluded.

We also need to pass the source finite-part limit through the angular
average.  Fix \(\varepsilon>0\) small enough that the sphere
\(\mathbf z_A+\varepsilon\mathbf n_A\) contains no particle position.  For
each \(D\), choose \(\eta>0\) so that
\[
    \mathcal F_{kij}^{(BCD)}
    (\mathbf z_A+\varepsilon\mathbf n_A,\mathbf x')
\]
is smooth for \(r'_D<\eta\), uniformly in \(\mathbf n_A\).  This is
possible because the field sphere and the relevant \(D\)-centered source
ball can be separated at fixed \(\varepsilon\).  The Taylor formula
\eqref{eq:app-F-local-expansion} then gives, for
\(0<\lambda<\lambda'<\eta\),
\begin{align}
    &\sup_{\mathbf n_A\in S^2}
    \Biggl|
    \int_{\lambda<r'_D<\lambda'}
    \mathcal F_{kij}^{(BCD)}
    (\mathbf z_A+\varepsilon\mathbf n_A,\mathbf x')\,
    \frac{Q_D^{ij}(\mathbf n'_D)}{(r'_D)^3}\,
    d^3x'
    \Biggr|
    \notag\\
    &\qquad\le
    C_1(\varepsilon,\eta)(\lambda'-\lambda)
    +
    C_2(\varepsilon,\eta)\bigl[(\lambda')^2-\lambda^2\bigr],
    \label{eq:app-uniform-estimate}
\end{align}
where \(C_1\) and \(C_2\) are finite for fixed \(\varepsilon,\eta\) and do
not depend on \(\lambda,\lambda'\) or \(\mathbf n_A\).  The vanishing of the
zeroth angular moment in Eq.~\eqref{eq:app-Q-angular-zero} removes the
logarithmic shell term.  Therefore the punctured source integrals form a
uniformly Cauchy family in \(\mathbf n_A\), and the source finite-part
limit may be interchanged with the angular average.

Combining this interchange with Eq.~\eqref{eq:app-angular-identity} gives
\begin{align}
    \frac{1}{4\pi}
    \int d\Omega_A\,
    I_k^{\rm Pf}(\mathbf z_A+\varepsilon\mathbf n_A)
    &=
    \frac{G^3}{4\pi}
    \sum_{B\neq C}\sum_{D\neq B,C}
    m_Bm_Cm_D\,
    \operatorname{Pf}
    \int_{r'_A>\varepsilon}
    \mathcal K_{ij}^{(BC)}(\mathbf x')\,
    \notag\\
    &\quad\times
    \frac{n_A'{}^k}{(r'_A)^2}
    \frac{Q_D^{ij}(\mathbf n'_D)}{(r'_D)^3}\,
    d^3x' .
    \label{eq:app-Ibar-Pf}
\end{align}
Thus the averaged field-point source term has exactly the direct particle
integrand, with the \(A\)-centered ball \(r'_A<\varepsilon\) removed.

\subsection{Comparison with the Direct Particle Expression}

We now compare Eq.~\eqref{eq:app-Ibar-Pf} with the finite-part source part
of the direct expression:
\begin{align}
    I_k^{\rm Pf}(\mathbf z_A)
    =
    \frac{G^3}{4\pi}
    \sum_{B\neq C}\sum_{D\neq B,C}
    m_Bm_Cm_D\,
    \operatorname{Pf}
    \int
    \mathcal K_{ij}^{(BC)}(\mathbf x')\,
    \frac{n_A'{}^k}{(r'_A)^2}
    \frac{Q_D^{ij}(\mathbf n'_D)}{(r'_D)^3}\,
    d^3x' .
    \label{eq:app-Ik-Pf-direct}
\end{align}
The only difference between Eqs.~\eqref{eq:app-Ibar-Pf} and
\eqref{eq:app-Ik-Pf-direct} is the exclusion of the ball
\(r'_A<\varepsilon\) before the coefficient of \(\varepsilon^0\) is taken.
This exclusion is precisely the Hadamard finite-part operation at
\(\mathbf z_A\).

More explicitly, if \(D\neq A\), then the \(D\)-centered factor
\(Q_D^{ij}/(r'_D)^3\) is smooth near \(\mathbf z_A\).  The removed ball
\(r'_A<\varepsilon\) is therefore only the usual \(A\)-centered puncture
used to define the finite part of the direct expression.  Taking the
coefficient of \(\varepsilon^0\) gives the same Hadamard finite part.

If \(D=A\), the source puncture and the field-point puncture are centered at
the same particle.  In the source finite part, one removes
\(r'_A<\lambda\); in Eq.~\eqref{eq:app-Ibar-Pf}, the angular average also
removes \(r'_A<\varepsilon\).  Together they amount to the single puncture
\[
    r'_A>\max(\lambda,\varepsilon).
\]
Taking first \(\lambda\to0\) at fixed \(\varepsilon\), and then extracting
the coefficient of \(\varepsilon^0\), gives exactly the direct Hadamard
finite part at \(\mathbf z_A\).  Therefore
\begin{align}
    \left[
    \frac{1}{4\pi}
    \int d\Omega_A\,
    I_k^{\rm Pf}(\mathbf z_A+\varepsilon\mathbf n_A)
    \right]_{\varepsilon^0}
    =
    I_k^{\rm Pf}(\mathbf z_A).
    \label{eq:app-Pf-equivalence}
\end{align}

It remains to compare the contact terms.  From
Eq.~\eqref{eq:app-Ik-delta-part},
\begin{align}
    I_k^\delta(\mathbf x)
    =
    -\frac{G^3}{3}
    \sum_{B\neq C}\sum_{D\neq B,C}
    m_Bm_Cm_D\,
    \mathcal K_k(\mathbf x,\mathbf z_D)\,
    \mathcal K_{ii}^{(BC)}(\mathbf z_D).
    \label{eq:app-contact-field-again}
\end{align}
For \(D\neq A\), Eq.~\eqref{eq:app-angular-identity} at
\(\mathbf x'=\mathbf z_D\) gives
\begin{align}
    \frac{1}{4\pi}
    \int d\Omega_A\,
    \mathcal K_k(\mathbf z_A+\varepsilon\mathbf n_A,\mathbf z_D)
    =
    -\frac{n_{AD}^k}{r_{AD}^2}
    \Theta(r_{AD}-\varepsilon),
    \label{eq:app-contact-DneqA}
\end{align}
where
\[
    \mathbf n_{AD}
    :=
    \frac{\mathbf z_A-\mathbf z_D}{r_{AD}} .
\]
For \(D=A\), the angular average of the odd radial vector vanishes:
\begin{align}
    \frac{1}{4\pi}
    \int d\Omega_A\,
    \mathcal K_k(\mathbf z_A+\varepsilon\mathbf n_A,\mathbf z_A)
    =
    0 .
    \label{eq:app-contact-DeqA}
\end{align}
Hence
\begin{align}
    \left[
    \frac{1}{4\pi}
    \int d\Omega_A\,
    I_k^\delta(\mathbf z_A+\varepsilon\mathbf n_A)
    \right]_{\varepsilon^0}
    =
    \frac{G^3}{3}
    \sum_{B\neq C}
    \sum_{\substack{D\neq B,C\\D\neq A}}
    m_Bm_Cm_D\,
    \frac{n_{AD}^k}{r_{AD}^2}\,
    \mathcal K_{ii}^{(BC)}(\mathbf z_D).
    \label{eq:app-contact-FP}
\end{align}

On the other hand, the direct contact contribution is
\begin{align}
    I_k^\delta(\mathbf z_A)
    =
    -\frac{G^3}{3}
    \sum_{B\neq C}\sum_{D\neq B,C}
    m_Bm_Cm_D\,
    \bigl(\mathcal K_k(\mathbf x,\mathbf z_D)\bigr)_A\,
    \mathcal K_{ii}^{(BC)}(\mathbf z_D),
    \label{eq:app-contact-direct}
\end{align}
where \(\bigl(\mathcal K_k(\mathbf x,\mathbf z_D)\bigr)_A\) denotes the
Hadamard finite value of the field kernel at particle \(A\).  For
\(D\neq A\),
\begin{align}
    \bigl(\mathcal K_k(\mathbf x,\mathbf z_D)\bigr)_A
    =
    \partial_k\frac{1}{|\mathbf x-\mathbf z_D|}
    \Bigg|_{\mathbf x=\mathbf z_A}
    =
    -\frac{n_{AD}^k}{r_{AD}^2},
    \label{eq:app-d1rD-at-A}
\end{align}
whereas for \(D=A\) the Hadamard finite value of the odd radial vector is
zero.  Thus
\begin{align}
    I_k^\delta(\mathbf z_A)
    =
    \frac{G^3}{3}
    \sum_{B\neq C}
    \sum_{\substack{D\neq B,C\\D\neq A}}
    m_Bm_Cm_D\,
    \frac{n_{AD}^k}{r_{AD}^2}\,
    \mathcal K_{ii}^{(BC)}(\mathbf z_D).
    \label{eq:app-contact-direct-final}
\end{align}
Comparing Eqs.~\eqref{eq:app-contact-FP} and
\eqref{eq:app-contact-direct-final}, we obtain
\begin{align}
    \left[
    \frac{1}{4\pi}
    \int d\Omega_A\,
    I_k^\delta(\mathbf z_A+\varepsilon\mathbf n_A)
    \right]_{\varepsilon^0}
    =
    I_k^\delta(\mathbf z_A).
    \label{eq:app-contact-equivalence}
\end{align}

Equations~\eqref{eq:app-Pf-equivalence} and
\eqref{eq:app-contact-equivalence} prove
\begin{align}
    (I_k)_A
    =
    I_k(\mathbf z_A).
    \label{eq:app-IkA-equals-direct}
\end{align}
Therefore the nonclosed integral contribution may be written as
\begin{align}
    (I_k)_A
    &=
    \frac{G^3}{4\pi}
    \sum_{B\neq C}\sum_{D\neq B,C}
    m_Bm_Cm_D\,
    \operatorname{Pf}
    \int
    (\partial_B)_i(\partial_C)_j
    \ln(r'_B+r'_C+r_{BC})
    \notag\\
    &\quad\times
    (\partial_A)_k\frac{1}{r'_A}
    \,
    (\partial_D)_i(\partial_D)_j\frac{1}{r'_D}\,
    d^3x' .
    \label{eq:app-final-nonclosed-integral}
\end{align}
This is the expression used in the main text.  The equality shows that the
direct particle notation does not introduce an additional prescription; it
is exactly the Hadamard finite part of the field-point integral at
\(\mathbf z_A\).

\section{Regularized Quantities Entering the 2PN EOM}
\label{app:regularized-quantities-2pn-eom}

\begingroup
\allowdisplaybreaks[3]
\setlength{\jot}{1.0ex}

The formulas below are written for a fixed evaluation particle \(A\), with
dummy labels summed as displayed.  Time derivatives are taken after assigning
Hadamard finite values.  Explicit accelerations are Newtonian unless stated
otherwise; \(a^{\mathrm{N}}\), \(\dot a^{\mathrm{N}}\), and \(\ddot a^{\mathrm{N}}\) are given by
Eqs.~\eqref{eq:newtonian-acceleration-main},
\eqref{eq:appendix-newtonian-jerk}, and
\eqref{eq:appendix-newtonian-snap}.  If order reduction generates a 1PN
acceleration, Eq.~\eqref{eq:one-pn-acceleration-main} is used.  Parentheses
on spatial indices denote unit-weight symmetrization, e.g.
\(X^{(i}Y^{j)}=(X^iY^j+X^jY^i)/2\).  The nonclosed term \((I_i)_A\) is
defined by Eq.~\eqref{eq:nonclosed-integral-IA}.

For any distinct particle labels appearing below,
\begin{align}
    P_{XY}^{ij}
    &\equiv \delta^{ij}-n_{XY}^i n_{XY}^j .
    \label{eq:reg-basic-PS-def}
\end{align}
At the fixed evaluation particle \(A\), for \(X\ne A\),
\begin{align}
    \alpha_X
    &\equiv \mathbf n_{AX}\!\cdot\!\mathbf v_X,
    &
    \beta_X
    &\equiv \mathbf n_{AX}\!\cdot\!\mathbf a_X^{\mathrm{N}},
    &
    \gamma_X
    &\equiv \mathbf n_{AX}\!\cdot\!\dot{\mathbf a}_X^{\mathrm{N}},
    \nonumber\\
    \chi_X
    &\equiv \mathbf n_{AX}\!\cdot\!\ddot{\mathbf a}_X^{\mathrm{N}},
    &
    q_X
    &\equiv v_X^2-\alpha_X^2 .
    \label{eq:reg-time-alpha-beta}
\end{align}
Here \(\chi_X\) is used only for the projected second time derivative of
the Newtonian acceleration; \(\delta_D\) continues to denote a Dirac delta
supported at particle \(D\).
For three-body kernels we use
\begin{align}
    S_{ABC}&\equiv r_{AB}+r_{AC}+r_{BC},
    &
    g_{BC}&\equiv\ln S_{ABC},
    \label{eq:reg-gBC-def}\\
    N_{ABC}^i&\equiv n_{AB}^i+n_{AC}^i,
    &
    L_{ABC}^i&\equiv n_{BC}^i-n_{AB}^i,
    &
    M_{ABC}^i&\equiv n_{BC}^i+n_{AC}^i .
    \nonumber
\end{align}
We also define
\begin{align}
    \mathcal T_{AX}^{ijk}
    &\equiv
    \frac{
    n_{AX}^iP_{AX}^{jk}
    +P_{AX}^{ij}n_{AX}^k
    +P_{AX}^{ik}n_{AX}^j
    }{r_{AX}^2},
    \qquad X\ne A .
\end{align}
For total time derivatives,
\begin{align}
    u_{XY}^i
    &\equiv v_X^i-v_Y^i,
    &
    \dot r_{XY}
    &\equiv
    \mathbf n_{XY}\!\cdot\!\mathbf u_{XY},
    &
    \dot n_{XY}^i
    &\equiv
    \frac{u_{XY}^i-\dot r_{XY}n_{XY}^i}{r_{XY}} .
    \label{eq:reg-time-basic-kinematics}
\end{align}
In particular,
\begin{align}
    \dot S_{ABC}
    &\equiv
    \dot r_{AB}+\dot r_{AC}+\dot r_{BC},
    \nonumber\\
    \dot L_{ABC}^i
    &\equiv
    \dot n_{BC}^i-\dot n_{AB}^i,
    &
    \dot M_{ABC}^i
    &\equiv
    \dot n_{BC}^i+\dot n_{AC}^i,
    \nonumber\\
    \dot P_{XY}^{ij}
    &\equiv
    -\dot n_{XY}^i n_{XY}^j-n_{XY}^i\dot n_{XY}^j .
\end{align}
Equivalently,
\begin{align}
    \dot\alpha_X
    &\equiv
    \frac{(\mathbf v_A-\mathbf v_X)\!\cdot\!\mathbf v_X
    -\dot r_{AX}\alpha_X}{r_{AX}}
    +\beta_X,
    \nonumber\\
    \dot\beta_X
    &\equiv
    \frac{(\mathbf v_A-\mathbf v_X)\!\cdot\!\mathbf a_X^{\mathrm{N}}
    -\dot r_{AX}\beta_X}{r_{AX}}
    +\gamma_X,
    \qquad X\ne A .
    \label{eq:reg-time-alpha-beta-expanded}
\end{align}
The kernels used repeatedly below are
\begin{align}
    \mathcal A_{AX}^i
    &\equiv
    \frac{\big[-4v_X^2+\alpha_X^2\big]n_{AX}^i
    -2\alpha_XP_{AX}^{ij}v_X^j}{r_{AX}^2}
    +2G\sum_{Y\ne X}\frac{m_Yn_{AX}^i}{r_{AX}^2r_{XY}}
    -\frac{P_{AX}^{ij}(a_X^{\mathrm{N}})^j}{r_{AX}},
    \qquad X\ne A,
    \nonumber\\
    \mathcal Q_{AB}^{ij}
    &\equiv \frac{P_{AB}^{ij}}{r_{AB}^2},
    \nonumber\\
    \mathcal Z_{AB}^{ij}
    &\equiv \frac{2n_{AB}^i n_{AB}^j-\delta^{ij}}{r_{AB}^2}.
    \label{eq:reg-time-QZ-def}
\end{align}
For \(B,C\ne A\),
\begin{align}
    \mathcal L_{ABC}^{ij}
    &\equiv
    -\frac{P_{BC}^{ij}}{r_{BC}S_{ABC}}
    +\frac{L_{ABC}^jM_{ABC}^i}{S_{ABC}^2},
    \nonumber\\
    \mathcal N_{ABC}^{ij}
    &\equiv
    2n_{BC}^i n_{BC}^j
    +2n_{BC}^{(i}n_{AC}^{j)}
    -2n_{BC}^{(i}n_{AB}^{j)}
    -2n_{AB}^{(i}n_{AC}^{j)},
    \nonumber\\
    \mathcal Y_{ABC}^{ij}
    &\equiv
    -2\,\frac{P_{BC}^{ij}}{r_{BC}S_{ABC}}
    +\frac{\mathcal N_{ABC}^{ij}}{S_{ABC}^2}.
    \label{eq:reg-time-LNY-def}
\end{align}
Their total time derivatives are
\begin{align}
    \dot{\mathcal Q}_{AB}^{ij}
    &\equiv
    -\frac{
    2\dot r_{AB}P_{AB}^{ij}
    +2P_{AB}^{k(i}n_{AB}^{j)}(v_A^k-v_B^k)
    }{r_{AB}^3},
    \nonumber\\
    \dot{\mathcal Z}_{AB}^{ij}
    &\equiv
    \frac{
    4P_{AB}^{k(i}n_{AB}^{j)}(v_A^k-v_B^k)
    -2\dot r_{AB}(2n_{AB}^in_{AB}^j-\delta^{ij})
    }{r_{AB}^3},
    \label{eq:reg-time-dotQ-dotZ}
    \\
    \dot{\mathcal L}_{ABC}^{ij}
    &\equiv
    -\frac{\dot P_{BC}^{ij}}{r_{BC}S_{ABC}}
    +\frac{P_{BC}^{ij}
    (\dot r_{BC}S_{ABC}+r_{BC}\dot S_{ABC})}
    {r_{BC}^2S_{ABC}^2}
    \nonumber\\
    &\quad
    +\frac{
    \dot L_{ABC}^jM_{ABC}^i
    +L_{ABC}^j\dot M_{ABC}^i
    }{S_{ABC}^2}
    -2\frac{\dot S_{ABC}}{S_{ABC}^3}
    L_{ABC}^jM_{ABC}^i,
    \nonumber\\
    \dot{\mathcal N}_{ABC}^{ij}
    &\equiv
    4\dot n_{BC}^{(i}n_{BC}^{j)}
    +2\dot n_{BC}^{(i}n_{AC}^{j)}
    +2n_{BC}^{(i}\dot n_{AC}^{j)}
    -2\dot n_{BC}^{(i}n_{AB}^{j)}
    -2n_{BC}^{(i}\dot n_{AB}^{j)}
    \nonumber\\
    &\quad
    -2\dot n_{AB}^{(i}n_{AC}^{j)}
    -2n_{AB}^{(i}\dot n_{AC}^{j)},
    \nonumber\\
    \dot{\mathcal Y}_{ABC}^{ij}
    &\equiv
    -2\frac{\dot P_{BC}^{ij}}{r_{BC}S_{ABC}}
    +2\frac{P_{BC}^{ij}
    (\dot r_{BC}S_{ABC}+r_{BC}\dot S_{ABC})}
    {r_{BC}^2S_{ABC}^2}
    +\frac{\dot{\mathcal N}_{ABC}^{ij}}{S_{ABC}^2}
    -2\frac{\dot S_{ABC}}{S_{ABC}^3}\mathcal N_{ABC}^{ij}.
    \label{eq:reg-time-dotL-dotY}
\end{align}

\subsection{\((V)_A\)}
\begin{align}
    (V)_A
    &=
    G\sum_{B\ne A}\frac{m_B}{r_{AB}}
    +\frac{G}{2c^2}\sum_{B\ne A}m_B
    \left[
    \frac{4v_B^2-\alpha_B^2}{r_{AB}}
    -2G\sum_{C\ne B}\frac{m_C}{r_{AB}r_{BC}}
    -\beta_B
    \right]
    +O(c^{-4}).
    \label{eq:reg-V-A}
\end{align}

\subsection{\(\frac{d}{dt}(V)_A\)}
\begin{align}
    \frac{d}{dt}(V)_A
    &=
    -G\sum_{B\ne A}m_B\frac{\dot r_{AB}}{r_{AB}^2}
    +\frac{G}{2c^2}\sum_{B\ne A}m_B
    \Bigg\{
    \frac{8\,\mathbf v_B\!\cdot\!\mathbf a_B^{\mathrm{N}}
    -2\alpha_B\dot\alpha_B}{r_{AB}}
    -\frac{(4v_B^2-\alpha_B^2)\dot r_{AB}}{r_{AB}^2}
    \nonumber\\
    &\qquad\qquad
    +2G\sum_{C\ne B}m_C
    \left[
    \frac{\dot r_{AB}}{r_{AB}^2r_{BC}}
    +\frac{\dot r_{BC}}{r_{AB}r_{BC}^2}
    \right]
    -\dot\beta_B
    \Bigg\}
    +O(c^{-4}).
    \label{eq:reg-dt-V-A}
\end{align}

\subsection{\((V_i)_A\)}
\begin{align}
    (V_i)_A
    &=
    G\sum_{B\ne A}\frac{m_Bv_B^i}{r_{AB}}
    +\frac{G}{c^{2}}\sum_{B\ne A}m_B\Bigg\{
    v_B^i
    \left[
    \frac{v_B^2}{r_{AB}}
    -\frac{\alpha_B^2}{2r_{AB}}
    -G\sum_{C\ne B}\frac{m_C}{r_{AB}r_{BC}}
    -\frac12\beta_B
    \right]
    \nonumber\\
    &\qquad\qquad
    -\alpha_B(a_B^{\mathrm{N}})^i
    +\frac12r_{AB}(\dot a_B^{\mathrm{N}})^i
    \Bigg\}
    +O(c^{-4}).
    \label{eq:reg-Vi-A}
\end{align}

\subsection{\(\frac{d}{dt}(V_i)_A\)}
\begin{align}
    \frac{d}{dt}(V_i)_A
    &=
    G\sum_{B\ne A}m_B
    \left[
    \frac{(a_B^{\mathrm{N}})^i}{r_{AB}}
    -\frac{\dot r_{AB}}{r_{AB}^2}v_B^i
    \right]
    \nonumber\\
    &\quad
    +\frac{G}{c^2}\sum_{B\ne A}m_B
    \Bigg[
    \frac{(a_B^{\mathrm{1PN}})^i}{r_{AB}}+(a_B^{\mathrm{N}})^i
    \left(
    \frac{v_B^2}{r_{AB}}
    -\frac{\alpha_B^2}{2r_{AB}}
    -G\sum_{C\ne B}\frac{m_C}{r_{AB}r_{BC}}
    -\frac12\beta_B
    \right)
    \nonumber\\
    &\qquad
    +v_B^i
    \Bigg(
    \frac{2\,\mathbf v_B\!\cdot\!\mathbf a_B^{\mathrm{N}}
    -\alpha_B\dot\alpha_B}{r_{AB}}
    -\frac{\left(v_B^2-\frac12\alpha_B^2\right)\dot r_{AB}}{r_{AB}^2}
    \nonumber\\
    &\qquad\qquad
    +G\sum_{C\ne B}m_C
    \left[
    \frac{\dot r_{AB}}{r_{AB}^2r_{BC}}
    +\frac{\dot r_{BC}}{r_{AB}r_{BC}^2}
    \right]
    -\frac12\dot\beta_B
    \Bigg)
    \nonumber\\
    &\qquad
    -\dot\alpha_B(a_B^{\mathrm{N}})^i
    -\alpha_B(\dot a_B^{\mathrm{N}})^i
    +\frac12\dot r_{AB}(\dot a_B^{\mathrm{N}})^i
    +\frac12 r_{AB}(\ddot a_B^{\mathrm{N}})^i
    \Bigg]
    +O(c^{-4}).
    \label{eq:reg-dt-Vi-A}
\end{align}

\subsection{\((\hat R_i)_A\)}
\begin{align}
    (\hat R_i)_A
    &=
    G^2\sum_{\substack{B<C\\ B,C\ne A}}m_Bm_C
    \frac{1}{r_{BC}}
    \left(\frac1{r_{AB}}-\frac1{r_{AC}}\right)(v_B^i-v_C^i)
    -G^2m_A\sum_{B\ne A}\frac{m_B}{r_{AB}^2}(v_A^i-v_B^i)
    \nonumber\\
    &\quad
    -\frac{G^2}{8}\sum_{B\ne A}m_B^2\,\mathcal Q_{AB}^{ij}v_B^j
    -G^2\sum_{\substack{B\ne C\\ B,C\ne A}}m_Bm_C
    \left(2v_C^j-\frac32v_B^j\right)\mathcal L_{ABC}^{ij}
    \nonumber\\
    &\quad
    -\frac{G^2}{4}m_A\sum_{B\ne A}m_B
    (v_A^j+v_B^j)\mathcal Z_{AB}^{ij}
    +O(c^{-2}).
    \label{eq:reg-Rhat-A}
\end{align}

\subsection{\(\frac{d}{dt}(\hat R_i)_A\)}
\begin{align}
    \frac{d}{dt}(\hat R_i)_A
    &=
    G^2\sum_{\substack{B<C\\B,C\ne A}}m_Bm_C
    \Bigg[
    -\frac{\dot r_{BC}}{r_{BC}^2}
    \left(\frac1{r_{AB}}-\frac1{r_{AC}}\right)(v_B^i-v_C^i)
    \nonumber\\
    &\qquad
    +\frac1{r_{BC}}
    \left(-\frac{\dot r_{AB}}{r_{AB}^2}
    +\frac{\dot r_{AC}}{r_{AC}^2}\right)(v_B^i-v_C^i)
    +\frac1{r_{BC}}
    \left(\frac1{r_{AB}}-\frac1{r_{AC}}\right)
    \bigl((a_B^{\mathrm{N}})^i-(a_C^{\mathrm{N}})^i\bigr)
    \Bigg]
    \nonumber\\
    &\quad
    -G^2m_A\sum_{B\ne A}m_B
    \left[
    \frac{(a_A^{\mathrm{N}})^i-(a_B^{\mathrm{N}})^i}{r_{AB}^2}
    -2\frac{\dot r_{AB}}{r_{AB}^3}(v_A^i-v_B^i)
    \right]
    \nonumber\\
    &\quad
    -\frac{G^2}{8}\sum_{B\ne A}m_B^2
    \left[
    \dot{\mathcal Q}_{AB}^{ij}v_B^j
    +\mathcal Q_{AB}^{ij}(a_B^{\mathrm{N}})^j
    \right]
    \nonumber\\
    &\quad
    -G^2\sum_{\substack{B\ne C\\B,C\ne A}}m_Bm_C
    \left[
    \left(2(a_C^{\mathrm{N}})^j-\frac32(a_B^{\mathrm{N}})^j\right)\mathcal L_{ABC}^{ij}
    +\left(2v_C^j-\frac32v_B^j\right)
    \dot{\mathcal L}_{ABC}^{ij}
    \right]
    \nonumber\\
    &\quad
    -\frac{G^2}{4}m_A\sum_{B\ne A}m_B
    \left[
    \bigl((a_A^{\mathrm{N}})^j+(a_B^{\mathrm{N}})^j\bigr)\mathcal Z_{AB}^{ij}
    +(v_A^j+v_B^j)\dot{\mathcal Z}_{AB}^{ij}
    \right]
    +O(c^{-2}).
    \label{eq:reg-dt-Rhat-A}
\end{align}

\subsection{\((\hat W_{ij})_A\)}
\begin{align}
    (\hat W_{ij})_A
    &=
    G\sum_{B\ne A}\frac{m_B}{r_{AB}}
    \Bigl(v_B^iv_B^j-\delta^{ij}v_B^2\Bigr)
    -\frac{G^2}{4}\sum_{B\ne A}m_B^2\,\mathcal Q_{AB}^{ij}
    \nonumber\\
    &\quad
    -G^2\sum_{\substack{B<C\\ B,C\ne A}}m_Bm_C\,\mathcal Y_{ABC}^{ij}
    -G^2m_A\sum_{B\ne A}m_B\,\mathcal Z_{AB}^{ij}
    +O(c^{-2}).
    \label{eq:reg-What-A}
\end{align}

\subsection{\(\frac{d}{dt}(\hat W_{ij})_A\)}
\begin{align}
    \frac{d}{dt}(\hat W_{ij})_A
    &=
    G\sum_{B\ne A}m_B
    \left[
    \frac{
    2v_B^{(i}(a_B^{\mathrm{N}})^{j)}
    -2\delta^{ij}\,\mathbf v_B\!\cdot\!\mathbf a_B^{\mathrm{N}}}
    {r_{AB}}
    -\frac{\dot r_{AB}}{r_{AB}^2}
    \bigl(v_B^iv_B^j-\delta^{ij}v_B^2\bigr)
    \right]
    \nonumber\\
    &\quad
    -\frac{G^2}{4}\sum_{B\ne A}m_B^2\,
    \dot{\mathcal Q}_{AB}^{ij}
    -G^2\sum_{\substack{B<C\\B,C\ne A}}m_Bm_C\,
    \dot{\mathcal Y}_{ABC}^{ij}
    -G^2m_A\sum_{B\ne A}m_B\,
    \dot{\mathcal Z}_{AB}^{ij}
    +O(c^{-2}).
    \label{eq:reg-dt-What-A}
\end{align}

\subsection{\((V^2)_A\)}
\begin{align}
    (V^2)_A
    &=
    G^2\left[
    \sum_{B\ne A}\frac{m_B^2}{r_{AB}^2}
    +2\sum_{\substack{B<C\\ B,C\ne A}}\frac{m_B m_C}{r_{AB}\,r_{AC}}
    \right]
    +O(c^{-2}).
    \label{eq:reg-V2-A}
\end{align}

\subsection{\(\frac{d}{dt}(V^2)_A\)}
\begin{align}
    \frac{d}{dt}(V^2)_A
    &=
    -2G^2\sum_{B\ne A}m_B^2
    \frac{\dot r_{AB}}{r_{AB}^3}
    -2G^2\sum_{\substack{B<C\\B,C\ne A}}m_Bm_C
    \left[
    \frac{\dot r_{AB}}{r_{AB}^2r_{AC}}
    +\frac{\dot r_{AC}}{r_{AB}r_{AC}^2}
    \right]
    +O(c^{-2}).
    \label{eq:reg-dt-V2-A}
\end{align}

\subsection{\((V\,V_i)_A\)}
\begin{align}
    (V\,V_i)_A
    &=
    G^2\left[
    \sum_{B\ne A}\frac{m_B^2\,v_B^i}{r_{AB}^2}
    +\sum_{\substack{B<C\\ B,C\ne A}}
    \frac{m_B m_C\,(v_B^i+v_C^i)}{r_{AB}\,r_{AC}}
    \right]
    +O(c^{-2}).
    \label{eq:reg-VVi-A}
\end{align}

\subsection{\(\frac{d}{dt}(V\,V_i)_A\)}
\begin{align}
    \frac{d}{dt}(V\,V_i)_A
    &=
    G^2\sum_{B\ne A}m_B^2
    \left[
    \frac{(a_B^{\mathrm{N}})^i}{r_{AB}^2}
    -2\frac{\dot r_{AB}}{r_{AB}^3}v_B^i
    \right]
    \nonumber\\
    &\quad
    +G^2\sum_{\substack{B<C\\B,C\ne A}}m_Bm_C
    \Bigg[
    \frac{(a_B^{\mathrm{N}})^i+(a_C^{\mathrm{N}})^i}{r_{AB}r_{AC}}
    \nonumber\\
    &\qquad\qquad
    -(v_B^i+v_C^i)
    \left(
    \frac{\dot r_{AB}}{r_{AB}^2r_{AC}}
    +\frac{\dot r_{AC}}{r_{AB}r_{AC}^2}
    \right)
    \Bigg]
    +O(c^{-2}).
    \label{eq:reg-dt-VVi-A}
\end{align}

\subsection{\((\partial_i V)_A\)}
\begin{align}
    (\partial_i V)_A
    &=
    -\,G\sum_{B\ne A}\frac{m_B\,n_{AB}^i}{r_{AB}^2}
    +\frac{G}{2c^2}\sum_{B\ne A}m_B\,\mathcal A_{AB}^i
    +\frac{G}{c^4}\sum_{B\ne A}m_B\,\mathcal B_{AB}^i
    +O(c^{-6}).
    \label{eq:reg-dV-A}
\end{align}

\begin{align}
    \Phi_B
    &\equiv \frac{3}{2}v_B^2
    - G\sum_{C\ne B}\frac{m_C}{r_{BC}}, \\
    \Psi_B
    &\equiv
    3\,\mathbf v_B\!\cdot\!\mathbf a_B^{\mathrm{N}}
    +G\sum_{C\ne B}m_C\,
    \frac{\mathbf n_{BC}\!\cdot\!(\mathbf v_B-\mathbf v_C)}{r_{BC}^2}, \\
    \Xi_B
    &\equiv
    3(a_B^{\mathrm{N}})^2
    +3\,\mathbf v_B\!\cdot\!\dot{\mathbf a}_B^{\mathrm{N}}
    +G\sum_{C\ne B}m_C\!\left[
    \frac{\mathbf n_{BC}\!\cdot\!(\mathbf a_B^{\mathrm{N}}-\mathbf a_C^{\mathrm{N}})}{r_{BC}^2}
    +\frac{(\mathbf v_B-\mathbf v_C)^2
    -3\big(\mathbf n_{BC}\!\cdot\!(\mathbf v_B-\mathbf v_C)\big)^2}{r_{BC}^3}
    \right], \\
    \Omega_B
    &\equiv
    \frac{3}{2}\sum_{C\ne B}\frac{m_C^2}{r_{BC}^2}
    +3\sum_{\substack{C<D\\ C,D\ne B}}\frac{m_Cm_D}{r_{BC}r_{BD}}
    -2\sum_{C\ne B}\frac{m_C}{r_{BC}}\sum_{D\ne C}\frac{m_D}{r_{CD}}, \\
    \Sigma_B
    &\equiv
    \sum_{C\ne B}\Bigg\{
    \frac{m_C}{r_{BC}}
    \left(
    \frac{3}{2}v_C^2
    -G\sum_{D\ne C}\frac{m_D}{r_{CD}}
    \right)
    +\frac{m_C}{2}\Bigg[
    \sum_{D\ne C}\frac{Gm_D}{r_{CD}^2}\,
    \frac{-r_{BC}^2+r_{BD}^2-r_{CD}^2}{2r_{BC}r_{CD}}
    \nonumber\\
    &\qquad\qquad
    +\frac{v_C^2-(\mathbf n_{BC}\!\cdot\!\mathbf v_C)^2}{r_{BC}}
    \Bigg]
    \Bigg\}, \\
    \Lambda_{BC}
    &\equiv
    v_B^2+2v_C^2-4\,\mathbf v_B\!\cdot\!\mathbf v_C
    -\frac{3}{2}\big(\mathbf n_{BC}\!\cdot\!\mathbf v_C\big)^2
    \nonumber\\
    &\quad
    -4\sum_{D\ne B}\frac{Gm_D}{r_{BD}}
    \nonumber\\
    &\quad
    -\sum_{D\ne C}\frac{Gm_D}{r_{CD}}
    \nonumber\\
    &\qquad
    \left(
    1+\frac{1}{2}\frac{r_{BC}}{r_{DC}}\,
    \mathbf n_{BC}\!\cdot\!\mathbf n_{DC}
    \right), \\
    \mathcal M_{BC}
    &\equiv
    4\,\mathbf n_{BC}\!\cdot\!\mathbf v_B
    -3\,\mathbf n_{BC}\!\cdot\!\mathbf v_C.
\end{align}

\begin{align}
    T_B
    &\equiv
    \frac78\,v_B^4
    +\frac12\,Gv_B^2\sum_{C\ne B}\frac{m_C}{r_{BC}}
    -4G\sum_{C\ne B}\frac{m_C}{r_{BC}}\,(\mathbf v_C\!\cdot\!\mathbf v_B)
    +4G\sum_{C\ne B}\frac{m_Cv_C^2}{r_{BC}}
    +G^2\Omega_B
    -G\Sigma_B .
\end{align}

\begin{align}
    \mathcal B_{AB}^i
    &=
    -\frac{n_{AB}^i}{r_{AB}^2}\,T_B
    +\frac12\,n_{AB}^i\,\Xi_B
    -\frac{P_{AB}^{ij}v_B^j}{r_{AB}}\,\Psi_B
    \nonumber\\[1mm]
    &\quad
    +\frac12\,\Phi_B
    \Bigg[
    -\frac{q_B\,n_{AB}^i+2\alpha_B\,P_{AB}^{ij}v_B^j}{r_{AB}^2}
    -\frac{P_{AB}^{ij}(a_B^{\mathrm{N}})^j}{r_{AB}}
    \Bigg]
    \nonumber\\[1mm]
    &\quad
    +\frac18\Bigg\{
    \frac{3}{r_{AB}^2}
    \Big[
    -4\alpha_B q_B\,P_{AB}^{ij}v_B^j
    -q_B^2\,n_{AB}^i
    \Big]
    \nonumber\\[1mm]
    &\qquad\qquad
    -\frac{6}{r_{AB}}
    \Big[
    q_B\,P_{AB}^{ij}(a_B^{\mathrm{N}})^j
    -2\alpha_B\beta_B\,P_{AB}^{ij}v_B^j
    \Big]
    \nonumber\\[1mm]
    &\qquad\qquad
    -\frac{12\,(\mathbf v_B\!\cdot\!\mathbf a_B^{\mathrm{N}})}{r_{AB}}\,
    P_{AB}^{ij}v_B^j
    +4\gamma_B\,P_{AB}^{ij}v_B^j
    +4\alpha_B\,P_{AB}^{ij}(\dot a_B^{\mathrm{N}})^j
    +6\beta_B\,P_{AB}^{ij}(a_B^{\mathrm{N}})^j
    \nonumber\\[1mm]
    &\qquad\qquad
    +\Big[
    4\alpha_B\gamma_B
    +3\beta_B^2
    +3(a_B^{\mathrm{N}})^2
    +4\,\mathbf v_B\!\cdot\!\dot{\mathbf a}_B^{\mathrm{N}}
    -r_{AB}\chi_B
    \Big]n_{AB}^i
    -r_{AB}(\ddot a_B^{\mathrm{N}})^i
    \Bigg\}
    \nonumber\\[1mm]
    &\quad
    +\frac12\sum_{C\ne B}\frac{Gm_C}{r_{BC}^2}\,
    \frac{P_{AB}^{ij}n_{BC}^j}{r_{AB}}\,
    \Lambda_{BC}
    \nonumber\\[1mm]
    &\quad
    +\frac74\sum_{C\ne B}\sum_{D\ne C}
    \frac{G^2m_Cm_D}{r_{BC}\,r_{CD}^{2}}\,
    \frac{P_{AB}^{ij}n_{CD}^j}{r_{AB}}
    \nonumber\\[1mm]
    &\quad
    -\frac12\sum_{C\ne B}\frac{Gm_C}{r_{BC}^2}\,
    \frac{P_{AB}^{ij}(v_B^j-v_C^j)}{r_{AB}}\,
    \mathcal M_{BC}.
\end{align}

\subsection{\((V\,\partial_i V)_A\)}
\begin{align}
    (V\,\partial_i V)_A
    &=
    -G^2
    \left(\sum_{B\ne A}\frac{m_B}{r_{AB}}\right)
    \left(\sum_{C\ne A}\frac{m_Cn_{AC}^i}{r_{AC}^2}\right)
    \nonumber\\
    &\quad
    +\frac{G^2}{2c^2}
    \left(\sum_{B\ne A}\frac{m_B}{r_{AB}}\right)
    \sum_{C\ne A}m_C\,\mathcal A_{AC}^i
    \nonumber\\
    &\quad
    -\frac{G^2}{2c^2}
    \left(\sum_{C\ne A}\frac{m_Cn_{AC}^i}{r_{AC}^2}\right)
    \sum_{B\ne A}m_B
    \left[
    \frac{4v_B^2-\alpha_B^2}{r_{AB}}
    -2G\sum_{D\ne B}\frac{m_D}{r_{AB}r_{BD}}
    -\beta_B
    \right]
    +O(c^{-4}) .
    \label{eq:reg-VdV-A}
\end{align}

\subsection{\((\partial_i V_j)_A\)}
\begin{align}
    (\partial_i V_j)_A
    &=
    -G\sum_{B\ne A}\frac{m_Bn_{AB}^iv_B^j}{r_{AB}^2}
    \nonumber\\
    &\quad
    +\frac{G}{c^2}\sum_{B\ne A}m_B\Bigg\{
    \frac{v_B^j}{r_{AB}^2}
    \left[
    \left(
    -v_B^2+\frac12\alpha_B^2
    +G\sum_{C\ne B}\frac{m_C}{r_{BC}}\right)n_{AB}^i
    -\alpha_BP_{AB}^{ik}v_B^k
    \right]
    \nonumber\\
    &\qquad
    -\frac{1}{2r_{AB}}v_B^jP_{AB}^{ik}(a_B^{\mathrm{N}})^k
    -\frac{1}{r_{AB}}P_{AB}^{ik}v_B^k(a_B^{\mathrm{N}})^j
    +\frac12n_{AB}^i(\dot a_B^{\mathrm{N}})^j
    \Bigg\}
    +O(c^{-4}) .
    \label{eq:reg-dVi-A}
\end{align}

\subsection{\((\partial_i \hat R_j)_A\)}
\begin{align}
    (\partial_i \hat R_j)_A
    &=
    G^2m_A\sum_{B\ne A}
    \frac{m_Bn_{AB}^i}{r_{AB}^3}(v_A^j-v_B^j)
    \nonumber\\
    &\quad
    +G^2\sum_{\substack{B<C\\ B,C\ne A}}
    \frac{m_Bm_C}{r_{BC}}
    \left(
    -\frac{n_{AB}^i}{r_{AB}^2}
    +\frac{n_{AC}^i}{r_{AC}^2}
    \right)(v_B^j-v_C^j)
    \nonumber\\
    &\quad
    +\frac{G^2}{8}\sum_{B\ne A}\frac{m_B^2}{r_{AB}^3}
    \left[
    2n_{AB}^iP_{AB}^{jk}v_B^k
    +\alpha_BP_{AB}^{ij}
    +n_{AB}^jP_{AB}^{ik}v_B^k
    \right]
    \nonumber\\
    &\quad
    -G^2\sum_{B\ne A}m_Am_B
    \left(2v_B^k-\frac32v_A^k\right)\mathcal H_{AB}^{ijk}
    \nonumber\\
    &\quad
    -G^2\sum_{B\ne A}m_Am_B
    \left(2v_A^k-\frac32v_B^k\right)\mathcal H_{AB}^{ikj}
    \nonumber\\
    &\quad
    -G^2\sum_{\substack{B\ne C\\ B,C\ne A}}
    m_Bm_C
    \left(2v_C^k-\frac32v_B^k\right)\mathcal K_{A;BC}^{ijk}
    +O(c^{-2}).
    \label{eq:reg-dRhat-A}
\end{align}
\begin{align}
    \mathcal H_{AB}^{ijk}
    &=
    \frac1{r_{AB}^3}
    \left(
    \frac14\delta^{jk}n_{AB}^i
    +\frac14\delta^{ij}n_{AB}^k
    +\frac12\delta^{ik}n_{AB}^j
    -n_{AB}^in_{AB}^jn_{AB}^k
    \right), \\
    \mathcal H_{AB}^{ikj}
    &=
    \frac1{r_{AB}^3}
    \left(
    \frac14\delta^{jk}n_{AB}^i
    +\frac12\delta^{ij}n_{AB}^k
    +\frac14\delta^{ik}n_{AB}^j
    -n_{AB}^in_{AB}^jn_{AB}^k
    \right).
\end{align}

\begin{align}
    \mathcal K_{A;BC}^{ijk}
    &=
    \frac{P_{BC}^{jk}N_{ABC}^i}{r_{BC}S_{ABC}^2}
    -\frac{P_{AB}^{ik}M_{ABC}^j}{r_{AB}S_{ABC}^2}
    +\frac{L_{ABC}^kP_{AC}^{ij}}{r_{AC}S_{ABC}^2}
    -2\frac{N_{ABC}^iM_{ABC}^jL_{ABC}^k}{S_{ABC}^3}.
\end{align}

\subsection{\((\partial_i \hat X)_A\)}
For each pair \(B\ne A\), the projections \(\alpha_A,\alpha_B\) use \(\mathbf n_{AB}\).
\begin{gather*}
    r_{XY}=|\mathbf z_X-\mathbf z_Y|,\qquad
    n_{XY}^i=\frac{z_X^i-z_Y^i}{r_{XY}},\qquad
    P_{XY}^{ij}=\delta^{ij}-n_{XY}^in_{XY}^j,\quad X\ne Y,\\
    \alpha_A=\mathbf n_{AB}\cdot\mathbf v_A,\qquad
    \alpha_B=\mathbf n_{AB}\cdot\mathbf v_B,\qquad
    \beta_B=\mathbf n_{AB}\cdot\mathbf a_B^{\mathrm N},\\
    (a_X^{\mathrm N})^i=-G\sum_{Y\ne X}\frac{m_Y n_{XY}^i}{r_{XY}^2},\qquad q_B=v_B^2-\alpha_B^2,\\
    S_{ABC}=r_{AB}+r_{AC}+r_{BC},\qquad N_{ABC}^i=n_{AB}^i+n_{AC}^i,\\
    L_{ABC}^i=n_{BC}^i-n_{AB}^i,\qquad M_{ABC}^i=n_{BC}^i+n_{AC}^i,\\
    \mathcal T_{XY}^{ijk}=\frac{n_{XY}^iP_{XY}^{jk}+P_{XY}^{ij}n_{XY}^k+P_{XY}^{ik}n_{XY}^j}{r_{XY}^2},\qquad X\ne Y.
\end{gather*}
\begin{align}
    \Pi_{BC}^{jk}
    &\equiv
    v_C^jv_C^k+v_B^jv_B^k-2v_C^jv_B^k-\delta^{jk}v_C^2,\\
    \widetilde\Pi_{BC}^{jk}
    &\equiv
    v_C^jv_C^k+v_B^jv_B^k-2v_B^jv_C^k-\delta^{jk}v_B^2,\\
    \Theta_{BC}^{jk}
    &\equiv
    \frac32\,v_B^jv_C^k-2v_B^kv_C^j,\\
    \Theta_{CB}^{jk}
    &\equiv
    \frac32\,v_C^jv_B^k-2v_C^kv_B^j.
\end{align}

\begin{equation}
\label{eq:reg-dXhat-A}
\begin{aligned}
    (\partial_i \hat X)_A
    &=
    -\,G^2\sum_{B\ne A}\frac{m_B\,n_{AB}^i}{r_{AB}^2}
    \sum_{C\ne B}\frac{m_C\,v_B^2}{r_{BC}}
    \\[1mm]
    &\quad
    +G^2\sum_{B\ne A}m_B^2
    \left[
    -\frac{P_{AB}^{ij}(a_B^{\mathrm{N}})^j-\beta_B n_{AB}^i}{2\,r_{AB}^2}
    +\frac{q_B\,n_{AB}^i+\alpha_B P_{AB}^{ij}v_B^j}{4\,r_{AB}^3}
    \right]
    \\[1mm]
    &\quad
    +G^2\sum_{\substack{B<C\\ B,C\ne A}}m_Bm_C\Big[
    (a_C^{\mathrm{N}})^j\,\widetilde{\mathcal G}_{A;BC}^{\,ij}
    +(a_B^{\mathrm{N}})^j\,\mathcal G_{A;BC}^{\,ij}
    +\Pi_{BC}^{jk}\,\mathcal G_{A;BC}^{\,ijk}
    +\widetilde\Pi_{BC}^{jk}\,\widetilde{\mathcal G}_{A;BC}^{\,ijk}
    \\
    &\qquad\qquad\qquad\qquad
    +\Theta_{BC}^{jk}\,\mathcal J_{A;BC}^{\,ijk}
    +\Theta_{CB}^{jk}\,\mathcal J_{A;CB}^{\,ijk}
    \Big]
    \\[1mm]
    &\quad
    +G^2m_A\sum_{B\ne A}m_B\,\mathscr{C}_{AB}^{i}
    \\[1mm]
    &\quad
    -\frac{G^3}{4}\sum_{B\ne A}\frac{m_B^3\,n_{AB}^i}{r_{AB}^4}
    +\frac{G^3}{8}\sum_{\substack{B\ne C\\ C\ne A}}
    \frac{m_B^2m_C\,n_{AC}^i}{r_{AC}^2\,r_{BC}^2}
    -\frac{G^3}{16}\sum_{\substack{B\ne C\\B,C\ne A}}m_B^2m_C\,K_{A;BC}^{\,i}
    \\[1mm]
    &\quad
    -G^3\sum_{\substack{B\ne C\\B,C\ne A}}m_Bm_C\left[
    \frac12 m_B\,H_{A;BC}^{\,i}
    +\frac12 m_C\,H_{A;CB}^{\,i}
    \right]
    \\[1mm]
    &\quad
    -G^3\sum_{B\ne A}\frac{n_{AB}^i}{r_{AB}^4}
      \left(\frac18 m_A^2m_B+2m_A m_B^2\right)
    +(I_i)_A
    +O(c^{-2}) .
\end{aligned}
\end{equation}

\begin{equation*}
\begin{aligned}
    \mathscr{C}_{AB}^{i}
    &=\frac{P_{AB}^{ij}(a_A^{\mathrm N})^j}{4r_{AB}^2}
    +\frac{(2n_{AB}^in_{AB}^j-\delta^{ij})(a_B^{\mathrm N})^j}{2r_{AB}^2}\\
    &\quad+\frac{1}{4r_{AB}^3}\Big[
    (-10\alpha_A+12\alpha_B)v_A^i+(5\alpha_A-10\alpha_B)v_B^i\\
    &\qquad\qquad+n_{AB}^i\big(7\mathbf v_A\cdot\mathbf v_B-4v_B^2+18\alpha_A^2-32\alpha_A\alpha_B+18\alpha_B^2\big)\Big].
\end{aligned}
\end{equation*}
\[A\ne B,C,\qquad B\ne C\qquad\text{in the explicit }K/H\text{ formulas below.}\]
\begin{align*}
    K_{A;BC}^{\,i}
    &\equiv
    n_{AB}^i\left(
    \frac{2}{r_{AB}^3 r_{AC}}
    -\frac{r_{AC}}{r_{AB}^3 r_{BC}^2}
    -\frac{r_{BC}^2}{r_{AB}^3 r_{AC}^3}
    +\frac{r_{AB}}{r_{AC}^3 r_{BC}^2}
    \right)
    \\
    &\quad
    +n_{AC}^i\left(
    \frac{3}{r_{AC}^4}
    -\frac{1}{r_{AC}^2 r_{BC}^2}
    +\frac{1}{r_{AB}^2 r_{AC}^2}
    +\frac{1}{2r_{AB}^2 r_{BC}^2}
    -\frac{3r_{BC}^2}{2r_{AB}^2 r_{AC}^4}
    -\frac{3r_{AB}^2}{2r_{AC}^4 r_{BC}^2}
    \right),
    \\[1mm]
    H_{A;BC}^{\,i}
    &\equiv
    n_{AB}^i\left(
    \frac{3}{2r_{AB}^4}
    +\frac{1}{2r_{AB}^3 r_{BC}}
    +\frac{r_{AC}}{r_{AB}^3 r_{BC}^2}
    -\frac{3r_{AC}}{2r_{AB}^4 r_{BC}}
    -\frac{3r_{AC}^2}{2r_{AB}^3 r_{BC}^3}
    -\frac{3r_{AC}^2}{2r_{AB}^4 r_{BC}^2}
    +\frac{3r_{AC}^3}{2r_{AB}^4 r_{BC}^3}
    \right)
    \\
    &\quad
    +n_{AC}^i\left(
    -\frac{1}{2r_{AB}^2 r_{BC}^2}
    +\frac{1}{2r_{AB}^3 r_{BC}}
    +\frac{3r_{AC}}{2r_{AB}^2 r_{BC}^3}
    +\frac{r_{AC}}{r_{AB}^3 r_{BC}^2}
    -\frac{3r_{AC}^2}{2r_{AB}^3 r_{BC}^3}
    \right),
    \\[1mm]
    (I_k)_A
    &=
    \frac{G^3}{4\pi}
    \sum_{B\ne C}
    \sum_{D\ne B,C}
    m_Bm_Cm_D\,
    \operatorname{Pf}_{\mathbf x'}\!\int
    d^3x'\,
    (\partial_A)_k\frac{1}{r'_A}\,
    \\
    &\qquad\qquad\times
    (\partial_B)_j(\partial_C)_l
    \ln(r'_B+r'_C+r_{BC})\,
    (\partial_D)_j(\partial_D)_l
    \frac{1}{r'_D}.
\end{align*}
\begin{align*}
    r'_P&=|\mathbf x'-\mathbf z_P|, & (\partial_P)_j&=\frac{\partial}{\partial z_P^j},\\
    g_{BC}(\mathbf x)&=\ln(r_B+r_C+r_{BC}), & r_P&=|\mathbf x-\mathbf z_P|,\qquad \partial_i=\frac{\partial}{\partial x^i}.
\end{align*}
\begin{align}
    \mathcal G_{A;BC}^{\,ij}
    &\equiv \bigl(\partial_i(\partial_B)_j g_{BC}\bigr)_A,
    &
    \widetilde{\mathcal G}_{A;BC}^{\,ij}
    &\equiv \bigl(\partial_i(\partial_C)_j g_{BC}\bigr)_A,
    \\
    \mathcal G_{A;BC}^{\,ijk}
    &\equiv \bigl(\partial_i(\partial_B)_j(\partial_B)_k g_{BC}\bigr)_A,
    &
    \widetilde{\mathcal G}_{A;BC}^{\,ijk}
    &\equiv \bigl(\partial_i(\partial_C)_j(\partial_C)_k g_{BC}\bigr)_A,
    \\
    \mathcal J_{A;BC}^{\,ijk}
    &\equiv \bigl(\partial_i(\partial_B)_j(\partial_C)_k g_{BC}\bigr)_A,
    &
    \mathcal J_{A;CB}^{\,ijk}
    &\equiv \bigl(\partial_i(\partial_C)_j(\partial_B)_k g_{BC}\bigr)_A.
\end{align}
\[ A\ne B,C,\qquad B\ne C. \]
\begin{align}
    \mathcal G_{A;BC}^{\,ij}
    &=
    -\frac{P_{AB}^{ij}}{r_{AB}\,S_{ABC}}
    -\frac{N_{ABC}^iL_{ABC}^j}{S_{ABC}^2},
    \\
    \widetilde{\mathcal G}_{A;BC}^{\,ij}
    &=
    -\frac{P_{AC}^{ij}}{r_{AC}\,S_{ABC}}
    +\frac{N_{ABC}^iM_{ABC}^j}{S_{ABC}^2},
\end{align}
\begin{align}
    \mathcal G_{A;BC}^{\,ijk}
    &=
    -\frac{\mathcal T_{AB}^{\,ijk}}{S_{ABC}}
    -\frac{N_{ABC}^i}{S_{ABC}^2}
    \left(
    \frac{P_{BC}^{jk}}{r_{BC}}
    +\frac{P_{AB}^{jk}}{r_{AB}}
    \right)
    +\frac{P_{AB}^{ij}L_{ABC}^k+P_{AB}^{ik}L_{ABC}^j}
    {r_{AB}\,S_{ABC}^2}
    +2\,\frac{N_{ABC}^iL_{ABC}^jL_{ABC}^k}{S_{ABC}^3},
    \\
    \widetilde{\mathcal G}_{A;BC}^{\,ijk}
    &=
    -\frac{\mathcal T_{AC}^{\,ijk}}{S_{ABC}}
    -\frac{N_{ABC}^i}{S_{ABC}^2}
    \left(
    \frac{P_{BC}^{jk}}{r_{BC}}
    +\frac{P_{AC}^{jk}}{r_{AC}}
    \right)
    -\frac{P_{AC}^{ij}M_{ABC}^k+P_{AC}^{ik}M_{ABC}^j}
    {r_{AC}\,S_{ABC}^2}
    +2\,\frac{N_{ABC}^iM_{ABC}^jM_{ABC}^k}{S_{ABC}^3},
\end{align}
\begin{align}
    \mathcal J_{A;BC}^{\,ijk}
    &=
    \frac{P_{BC}^{jk}N_{ABC}^i}{r_{BC}\,S_{ABC}^2}
    -\frac{P_{AB}^{ij}M_{ABC}^k}{r_{AB}\,S_{ABC}^2}
    +\frac{L_{ABC}^jP_{AC}^{ik}}{r_{AC}\,S_{ABC}^2}
    -2\,\frac{N_{ABC}^iL_{ABC}^jM_{ABC}^k}{S_{ABC}^3},
    \\
    \mathcal J_{A;CB}^{\,ijk}
    &=
    \frac{P_{BC}^{jk}N_{ABC}^i}{r_{BC}\,S_{ABC}^2}
    +\frac{P_{AC}^{ij}L_{ABC}^k}{r_{AC}\,S_{ABC}^2}
    -\frac{M_{ABC}^jP_{AB}^{ik}}{r_{AB}\,S_{ABC}^2}
    -2\,\frac{N_{ABC}^iM_{ABC}^jL_{ABC}^k}{S_{ABC}^3}.
\end{align}

\subsection{\((\partial_i \hat W_{jk})_A\)}
\begin{align}
    (\partial_i \hat W_{jk})_A
    &=
    -\,G\sum_{B\ne A}\frac{m_B\,n_{AB}^i}{r_{AB}^2}
    \Bigl(v_B^jv_B^k-\delta^{jk}v_B^2\Bigr)
    \nonumber\\[1mm]
    &\quad
    +\frac{G^2}{4}\sum_{B\ne A}m_B^2\,
    \frac{
    2n_{AB}^i\,P_{AB}^{jk}
    +P_{AB}^{ij}n_{AB}^k
    +P_{AB}^{ik}n_{AB}^j
    }{r_{AB}^3}
    \nonumber\\[1mm]
    &\quad
    -G^2\sum_{\substack{B<C\\ B,C\ne A}}
    m_Bm_C\,\mathcal D_{ABC}^{\,ijk}
    \nonumber\\[1mm]
    &\quad
    -G^2 m_A\sum_{B\ne A}m_B\,\mathcal E_{AB}^{\,ijk}
    +O(c^{-2}) .
    \label{eq:reg-dWhat-A}
\end{align}

\begin{align}
    \mathcal D_{ABC}^{\,ijk}
    &\equiv
    \frac{2N_{ABC}^i\,P_{BC}^{jk}}
    {r_{BC}\,S_{ABC}^2}
    -\frac{
    P_{AB}^{ij}M_{ABC}^k
    +P_{AB}^{ik}M_{ABC}^j
    }{r_{AB}\,S_{ABC}^2}
    +\frac{
    L_{ABC}^jP_{AC}^{ik}
    +L_{ABC}^kP_{AC}^{ij}
    }{r_{AC}\,S_{ABC}^2}
    \nonumber\\[1mm]
    &\quad
    -2\,\frac{
    N_{ABC}^iL_{ABC}^jM_{ABC}^k
    +N_{ABC}^iL_{ABC}^kM_{ABC}^j
    }{S_{ABC}^3},
    \\[2mm]
    \mathcal E_{AB}^{\,ijk}
    &\equiv
    \frac{
    2n_{AB}^i\,P_{AB}^{jk}
    +3P_{AB}^{ij}n_{AB}^k
    +3P_{AB}^{ik}n_{AB}^j
    }{4\,r_{AB}^3}.
\end{align}

\subsection{\((V_j\partial_i V_j)_A\)}
\begin{align}
    (V_j\partial_i V_j)_A
    &=
    -G^2\sum_{B\ne A}
    \frac{m_B^2v_B^2}{r_{AB}^3}n_{AB}^i
    \nonumber\\
    &\quad
    -G^2\sum_{\substack{B<C\\ B,C\ne A}}
    m_Bm_C(\mathbf v_B\!\cdot\!\mathbf v_C)
    \left(
    \frac{n_{AB}^i}{r_{AB}^2r_{AC}}
    +\frac{n_{AC}^i}{r_{AC}^2r_{AB}}
    \right)
    +O(c^{-2}) .
    \label{eq:reg-VjdVj-A}
\end{align}

\subsection{\((V^2\partial_i V)_A\)}
\begin{align}
    (V^2\partial_i V)_A
    &=
    -G^3
    \left[
    \sum_{B\ne A}\frac{m_B^2}{r_{AB}^2}
    +2\sum_{\substack{B<C\\ B,C\ne A}}\frac{m_Bm_C}{r_{AB}r_{AC}}
    \right]
    \left(
    \sum_{D\ne A}\frac{m_Dn_{AD}^i}{r_{AD}^2}
    \right)
    +O(c^{-2}) .
    \label{eq:reg-V2dV-A}
\end{align}

\subsection{\((V_j\partial_i V)_A\)}
\begin{align}
    (V_j\partial_i V)_A
    &=
    -G^2\sum_{B\ne A}\frac{m_B^2v_B^j}{r_{AB}^3}n_{AB}^i
    \nonumber\\
    &\quad
    -G^2\sum_{\substack{B<C\\ B,C\ne A}}m_Bm_C
    \left(
    \frac{v_B^j n_{AC}^i}{r_{AB}r_{AC}^2}
    +\frac{v_C^j n_{AB}^i}{r_{AC}r_{AB}^2}
    \right)
    +O(c^{-2}) .
    \label{eq:reg-VjdV-A}
\end{align}

\subsection{\((V\partial_i V_j)_A\)}
\begin{align}
    (V\partial_i V_j)_A
    &=
    -G^2\sum_{B\ne A}\frac{m_B^2v_B^j}{r_{AB}^3}n_{AB}^i
    \nonumber\\
    &\quad
    -G^2\sum_{\substack{B<C\\ B,C\ne A}}m_Bm_C
    \left(
    \frac{v_B^j n_{AB}^i}{r_{AB}^2r_{AC}}
    +\frac{v_C^j n_{AC}^i}{r_{AC}^2r_{AB}}
    \right)
    +O(c^{-2}) .
    \label{eq:reg-VdVj-A}
\end{align}

\endgroup

\section{Explicit Integrand Formulas}
\label{app:explicit-integrand-formulas}

This appendix collects the explicit algebraic formulas for the compact
quantities used in the final regularized expressions
\eqref{eq:DneqA-centered-shell-form} and
\eqref{eq:DA-Hadamard-finite-part}.  The small cross-reference tags in
those two formulas point to the definitions below.  The numerical implementation and
validation tests are described in
Secs.~\ref{sec:numerical-evaluation-nonclosed} and
\ref{sec:numerical-results-benchmarks}.

We factorize
\begin{align}
    (\mathcal Y_{ijk}^{(BC)})_A
    &=
    \mathcal K_{ij}^{(BC)}
    \mathcal U_A^k,
    \label{eq:FABC-factorized}
\end{align}
where
\begin{align}
    \mathcal K_{ij}^{(BC)}
    &:=
    (\partial_B)_i(\partial_C)_j g'_{BC},
    &
    \mathcal U_A^k
    &:=
    \frac{n_A'{}^k}{(r_A')^2}.
    \label{eq:K-U-def}
\end{align}

The remaining compact kernels and local coefficients appearing in the two
final sectoral results are
\begin{align}
    \bigl(\mathcal Z_k^{(BC)}(\mathbf x')\bigr)_A
    &=
    \mathcal K_{ij}^{(BC)}(\mathbf x')
    \frac{
        n_A'{}^k
        \left(3n_A'{}^i n_A'{}^j-\delta_{ij}\right)
    }{(r_A')^5},
    \label{eq:app-Z-explicit}
    \\
    \bigl(\mathcal P_{k,1}^{(BC)}(\mathbf y)\bigr)_A
    &=
    y^a\mathcal K_{ij,a}^{(BC,A)}
    \frac{
        n^k\left(3n^in^j-\delta_{ij}\right)
    }{r^5},
    \label{eq:app-P-A1-explicit}
    \\
    \bigl(\mathcal C_k^{(BC)}\bigr)_A
    &=
    \frac{4\pi}{15}
    \left(
        -2\mathcal K_{ii,k}^{(BC,A)}
        +3\mathcal K_{ik,i}^{(BC,A)}
        +3\mathcal K_{kj,j}^{(BC,A)}
    \right).
    \label{eq:app-C-explicit}
\end{align}
Here \(\mathbf y=\mathbf x'-\mathbf z_A\),
\(r=|\mathbf y|=r_A'\), and \(\mathbf n=\mathbf y/r\).  The factor
\(\mathcal K_{ij}^{(BC)}\) and the local derivative
\(\mathcal K_{ij,a}^{(BC,A)}\) are written explicitly below, so
Eqs.~\eqref{eq:FABC-factorized}--\eqref{eq:app-C-explicit}, together with
Eqs.~\eqref{eq:KBC-explicit} and
\eqref{eq:DA-K-first-coeff-explicit}, contain no additional unevaluated
potentials.

\subsection{\texorpdfstring{The \(\mathcal K_{ij}^{(BC)}\) Factor}
{The K Factor}}

Define
\begin{align}
    \mathcal S_{BC}
    &:=
    r_B'+r_C'+r_{BC},
    \label{eq:SBC-def}
    \\
    n_B'{}^i
    &:=
    \frac{x'{}^i-z_B^i}{r_B'},
    &
    n_C'{}^j
    &:=
    \frac{x'{}^j-z_C^j}{r_C'},
    &
    n_{BC}^i
    &:=
    \frac{z_B^i-z_C^i}{r_{BC}}.
\end{align}
Introduce
\begin{align}
    M_{ij}^{(BC)}
    &:=
    n_{BC}^i n_{BC}^j-\delta_{ij},
    \\
    p_i^{(B)}
    &:=
    n_{BC}^i-n_B'{}^i,
    \\
    q_j^{(C)}
    &:=
    n_{BC}^j+n_C'{}^j.
\end{align}
Then
\begin{align}
    \mathcal K_{ij}^{(BC)}
    &=
    \frac{
        M_{ij}^{(BC)}
    }{
        r_{BC}\mathcal S_{BC}
    }
    +
    \frac{
        p_i^{(B)}q_j^{(C)}
    }{
        \mathcal S_{BC}^2
    }.
    \label{eq:KBC-explicit}
\end{align}

\subsection{\texorpdfstring{The Coefficient
\(\mathcal K_{ij,a}^{(BC,A)}\)}
{The First Local Coefficient}}

For the \(D=A\) finite-part subtraction,
\begin{align}
    \mathcal K_{ij,a}^{(BC,A)}
    &=
    -\frac{
        (n_{AB}^a+n_{AC}^a)
        \left(
            n_{BC}^i n_{BC}^j-\delta_{ij}
        \right)
    }{
        r_{BC}S_{ABC}^2
    }
    \nonumber\\
    &\quad
    -
    \frac{
        (\delta_{ai}-n_{AB}^a n_{AB}^i)
        (n_{BC}^j+n_{AC}^j)
    }{
        r_{AB}S_{ABC}^2
    }
    \nonumber\\
    &\quad
    +
    \frac{
        (n_{BC}^i-n_{AB}^i)
        (\delta_{aj}-n_{AC}^a n_{AC}^j)
    }{
        r_{AC}S_{ABC}^2
    }
    \nonumber\\
    &\quad
    -
    \frac{
        2(n_{AB}^a+n_{AC}^a)
        (n_{BC}^i-n_{AB}^i)
        (n_{BC}^j+n_{AC}^j)
    }{
        S_{ABC}^3
    },
    \label{eq:DA-K-first-coeff-explicit}
\end{align}
where
\begin{align}
    S_{ABC}
    &:=
    r_{AB}+r_{AC}+r_{BC}.
\end{align}

\bibliographystyle{unsrtnat}
\bibliography{references}

@article{Fienga2024,
  author  = {Fienga, Agn{\`e}s and Minazzoli, Olivier},
  title   = {Testing theories of gravity with planetary ephemerides},
  journal = {Living Reviews in Relativity},
  volume  = {27},
  number  = {1},
  pages   = {1},
  year    = {2024},
  doi     = {10.1007/s41114-023-00047-0}
}

@article{doi:10.1142/S0218271816300032,
author = {Ni, Wei-Tou},
title = {Solar-system tests of the relativistic gravity},
journal = {International Journal of Modern Physics D},
volume = {25},
number = {14},
pages = {1630003},
year = {2016},
doi = {10.1142/S0218271816300032},

URL = { 
    
        https://doi.org/10.1142/S0218271816300032
    
    

},
eprint = { 
    
        https://doi.org/10.1142/S0218271816300032
    
    

}
}

@incollection{lorentz1937motion,
  title={The motion of a system of bodies under the influence of their mutual attraction, according to Einstein’s theory},
  author={Lorentz, Hendrik Antoon and Droste, Johannes},
  booktitle={Collected Papers: Volume V},
  pages={330--355},
  year={1937},
  publisher={Springer},
  address={Dordrecht},
  doi={10.1007/978-94-015-3445-1_11}
}

@article{7c44806c-bf75-3a8d-b622-7761a1e00cd9,
 ISSN = {0003486X, 19398980},
 URL = {http://www.jstor.org/stable/1968714},
 author = {A. Einstein and L. Infeld and B. Hoffmann},
 journal = {Annals of Mathematics},
 number = {1},
 pages = {65--100},
 publisher = {[Annals of Mathematics, Trustees of Princeton University on Behalf of the Annals of Mathematics, Mathematics Department, Princeton University]},
 title = {The Gravitational Equations and the Problem of Motion},
 urldate = {2026-06-10},
 volume = {39},
 year = {1938}
}

@article{9c298a78-4f65-3b62-b231-cb66fa9d86b7,
 ISSN = {0003486X, 19398980},
 URL = {http://www.jstor.org/stable/1969015},
 author = {A. Einstein and L. Infeld},
 journal = {Annals of Mathematics},
 number = {2},
 pages = {455--464},
 publisher = {[Annals of Mathematics, Trustees of Princeton University on Behalf of the Annals of Mathematics, Mathematics Department, Princeton University]},
 title = {The Gravitational Equations and the Problem of Motion. {II}},
 urldate = {2026-06-10},
 volume = {41},
 year = {1940}
}

@article{einstein1949motion,
  title={On the motion of particles in general relativity theory},
  author={Einstein, Albert and Infeld, Leopold},
  journal={Canadian Journal of Mathematics},
  volume={1},
  number={3},
  pages={209--241},
  year={1949},
  publisher={Cambridge University Press}
}

@ARTICLE{1965ApJ...142.1488C,
       author = {{Chandrasekhar}, S.},
        title = "{The Post-Newtonian Equations of Hydrodynamics in General Relativity.}",
      journal = {\apj},
         year = 1965,
        month = nov,
       volume = {142},
        pages = {1488},
          doi = {10.1086/148432},
       adsurl = {https://ui.adsabs.harvard.edu/abs/1965ApJ...142.1488C}
}

@ARTICLE{1969ApJ...158...55C,
       author = {{Chandrasekhar}, S. and {Nutku}, Yavuz},
        title = "{The Second Post-Newtonian Equations of Hydrodynamics in General Relativity}",
      journal = {\apj},
         year = 1969,
        month = oct,
       volume = {158},
        pages = {55},
          doi = {10.1086/150171},
       adsurl = {https://ui.adsabs.harvard.edu/abs/1969ApJ...158...55C}
}

@article{10.1143/PTP.50.492,
    author = {Ohta, Tadayuki and Okamura, Hiroshi and Kimura, Toshiei and Hiida, Kichiro},
    title = {Physically Acceptable Solution of Einstein’s Equation for Many-Body System},
    journal = {Progress of Theoretical Physics},
    volume = {50},
    number = {2},
    pages = {492-514},
    year = {1973},
    month = {08},
    issn = {0033-068X},
    doi = {10.1143/PTP.50.492},
    url = {https://doi.org/10.1143/PTP.50.492},
    eprint = {https://academic.oup.com/ptp/article-pdf/50/2/492/5228680/50-2-492.pdf},
}

@article{10.1143/PTP.51.1220,
    author = {Ohta, Tadayuki and Okamura, Hiroshi and Kimura, Toshiei and Hiida, Kichiro},
    title = {Higher Order Gravitational Potential for Many-Body System},
    journal = {Progress of Theoretical Physics},
    volume = {51},
    number = {4},
    pages = {1220-1238},
    year = {1974},
    month = {04},
    issn = {0033-068X},
    doi = {10.1143/PTP.51.1220},
    url = {https://doi.org/10.1143/PTP.51.1220},
    eprint = {https://academic.oup.com/ptp/article-pdf/51/4/1220/5221727/51-4-1220.pdf},
}

@article{PhysRevD.43.3273,
  title = {General-relativistic celestial mechanics. I. Method and definition of reference systems},
  author = {Damour, Thibault and Soffel, Michael and Xu, Chongming},
  journal = {Phys. Rev. D},
  volume = {43},
  issue = {10},
  pages = {3273--3307},
  numpages = {0},
  year = {1991},
  month = {May},
  publisher = {American Physical Society},
  doi = {10.1103/PhysRevD.43.3273},
  url = {https://link.aps.org/doi/10.1103/PhysRevD.43.3273}
}

@article{PhysRevD.45.1017,
  title = {General-relativistic celestial mechanics. {II}. Translational equations of motion},
  author = {Damour, Thibault and Soffel, Michael and Xu, Chongming},
  journal = {Phys. Rev. D},
  volume = {45},
  issue = {4},
  pages = {1017--1044},
  numpages = {0},
  year = {1992},
  month = {Feb},
  publisher = {American Physical Society},
  doi = {10.1103/PhysRevD.45.1017},
  url = {https://link.aps.org/doi/10.1103/PhysRevD.45.1017}
}

@article{PhysRevD.47.3124,
  title = {General-relativistic celestial mechanics. III. Rotational equations of motion},
  author = {Damour, Thibault and Soffel, Michael and Xu, Chongming},
  journal = {Phys. Rev. D},
  volume = {47},
  issue = {8},
  pages = {3124--3135},
  numpages = {0},
  year = {1993},
  month = {Apr},
  publisher = {American Physical Society},
  doi = {10.1103/PhysRevD.47.3124},
  url = {https://link.aps.org/doi/10.1103/PhysRevD.47.3124}
}

@article{PhysRevD.49.618,
  title = {General-relativistic celestial mechanics. {IV}. Theory of satellite motion},
  author = {Damour, Thibault and Soffel, Michael and Xu, Chongming},
  journal = {Phys. Rev. D},
  volume = {49},
  issue = {2},
  pages = {618--635},
  numpages = {0},
  year = {1994},
  month = {Jan},
  publisher = {American Physical Society},
  doi = {10.1103/PhysRevD.49.618},
  url = {https://link.aps.org/doi/10.1103/PhysRevD.49.618}
}

@article{1985Lagrangians,
  title={Lagrangians for n point masses at the second post-Newtonian approximation of general relativity},
  author={ Damour, Thibault  and  Sch{\"a}fer, Gerhard },
  journal={General Relativity \& Gravitation},
  volume={17},
  number={9},
  pages={879-905},
  year={1985},
}

@article{SCHAFER1987336,    
        title = {Three-body hamiltonian in general relativity}, journal = {Physics Letters A}, 
        volume = {123}, 
        number = {7}, 
        pages = {336-339}, 
        year = {1987}, 
        issn = {0375-9601}, 
        doi = {https://doi.org/10.1016/0375-9601(87)90389-6}, 
        url = {https://www.sciencedirect.com/science/article/pii/0375960187903896}, author = {Gerhard Schäfer}
}

@article{Transactions,
  author  = {{International Astronomical Union}},
  title   = {Chapter {IV}: Resolutions of the {General Assembly}},
  journal = {Trans. Int. Astron. Union},
  volume  = {24B},
  pages   = {33},
  year    = {2001},
  doi     = {10.1017/S0251107X00009007}
}

@article{heinze2026nbody2pnhamiltoniannumerical,
  author  = {Heinze, Felix M. and Sch{\"a}fer, Gerhard and Br{\"u}gmann, Bernd},
  title   = {The {N}-body {2PN} Hamiltonian and numerical integration of the equations of motion},
  journal = {Phys. Rev. D},
  volume  = {113},
  pages   = {104066},
  year    = {2026},
  doi     = {10.1103/9vhx-6wkk}
}

@misc{heinze2026fullyanalyticexpression2pn,
  title         = {A Fully Analytic Expression for the {2PN} {$N$}-Body Hamiltonian},
  author        = {Heinze, Felix M. and Sch{\"a}fer, Gerhard and Br{\"u}gmann, Bernd},
  year          = {2026},
  eprint        = {2609.17510},
  archivePrefix = {arXiv},
  primaryClass  = {gr-qc},
  url           = {https://arxiv.org/abs/2609.17510}
}

@article{blanchet1998gravitational,
  title={Gravitational field and equations of motion of compact binaries to 5/2 post-Newtonian order},
  author={Blanchet, Luc and Faye, Guillaume and Ponsot, Benedicte},
  journal={Physical Review D},
  volume={58},
  number={12},
  pages={124002},
  year={1998},
  publisher={APS}
}

@article{Blanchet_2000,
   title={Hadamard regularization},
   volume={41},
   ISSN={1089-7658},
   url={http://dx.doi.org/10.1063/1.1308506},
   DOI={10.1063/1.1308506},
   number={11},
   journal={Journal of Mathematical Physics},
   publisher={AIP Publishing},
   author={Blanchet, Luc and Faye, Guillaume},
   year={2000},
   month=nov, pages={7675–7714} 
}

@article{blanchet2024postnewtoniantheorygravitationalwaves,
  author        = {Blanchet, Luc},
  title         = {Post-Newtonian theory for gravitational waves},
  journal       = {Living Reviews in Relativity},
  volume        = {27},
  number        = {1},
  pages         = {4},
  year          = {2024},
  doi           = {10.1007/s41114-024-00050-z},
  url           = {https://doi.org/10.1007/s41114-024-00050-z},
  eprint        = {1310.1528},
  archivePrefix = {arXiv},
  primaryClass  = {gr-qc}
}

@article{SchaeferJaranowski2024,
  author  = {Sch{\"a}fer, Gerhard and Jaranowski, Piotr},
  title   = {Hamiltonian formulation of general relativity and post-Newtonian dynamics of compact binaries},
  journal = {Living Reviews in Relativity},
  volume  = {27},
  number  = {1},
  pages   = {2},
  year    = {2024},
  doi     = {10.1007/s41114-024-00048-7},
  url     = {https://doi.org/10.1007/s41114-024-00048-7},
  issn    = {1433-8351}
}

@incollection{BrumbergKopeikin1989Frames,
  author    = {Brumberg, V. A. and Kopeikin, S. M.},
  editor    = {Kovalevsky, Jean and Mueller, Ivan I. and Kolaczek, Barbara},
  title     = {Relativistic Theory of Celestial Reference Frames},
  booktitle = {Reference Frames: In Astronomy and Geophysics},
  year      = {1989},
  publisher = {Springer Netherlands},
  address   = {Dordrecht},
  pages     = {115--141},
  isbn      = {978-94-009-0933-5},
  doi       = {10.1007/978-94-009-0933-5_6},
  url       = {https://doi.org/10.1007/978-94-009-0933-5_6}
}

@article{BrumbergKopeikin1989Earth,
  author  = {Brumberg, V. A. and Kopeikin, S. M.},
  title   = {Relativistic Reference Systems and Motion of Test Bodies in the Vicinity of the Earth},
  journal = {Il Nuovo Cimento B (1971--1996)},
  year    = {1989},
  volume  = {103},
  number  = {1},
  pages   = {63--98},
  doi     = {10.1007/BF02888894},
  url     = {https://doi.org/10.1007/BF02888894},
  issn    = {1826-9877}
}

@article{Kopeikin2019Multipoles,
  author    = {Kopeikin, Sergei M.},
  title     = {Covariant Equations of Motion of Extended Bodies with Arbitrary Mass and Spin Multipoles},
  journal   = {Physical Review D},
  volume    = {99},
  number    = {8},
  pages     = {084008},
  year      = {2019},
  month     = apr,
  doi       = {10.1103/PhysRevD.99.084008},
  url       = {https://doi.org/10.1103/PhysRevD.99.084008}
}

@article{Kopeikin2020Lagrangian,
  author    = {Kopeikin, Sergei M.},
  title     = {Post-Newtonian Lagrangian of an {$N$}-Body System with Arbitrary Mass and Spin Multipoles},
  journal   = {Physical Review D},
  volume    = {102},
  number    = {2},
  pages     = {024053},
  year      = {2020},
  month     = jul,
  doi       = {10.1103/PhysRevD.102.024053},
  url       = {https://doi.org/10.1103/PhysRevD.102.024053}
}

@article{Soffel2003IAU2000,
  author  = {Soffel, M. and Klioner, S. A. and Petit, G. and Wolf, P. and Kopeikin, S. M. and Bretagnon, P. and Brumberg, V. A. and Capitaine, N. and Damour, T. and Fukushima, T. and Guinot, B. and Huang, T. and Lindegren, L. and Ma, C. and Nordtvedt, K. and Ries, J. and Seidelmann, P. K. and Vokrouhlick{\'y}, D. and Will, C. M. and Xu, C.},
  title   = {The {IAU} 2000 Resolutions for Astrometry, Celestial Mechanics and Metrology in the Relativistic Framework: Explanatory Supplement},
  journal = {The Astronomical Journal},
  volume  = {126},
  number  = {6},
  pages   = {2687--2706},
  year    = {2003},
  doi     = {10.1086/378162},
  eprint  = {astro-ph/0303376},
  archivePrefix = {arXiv}
}

@article{Chu2009NBody2PN,
  author  = {Chu, Yi-Zen},
  title   = {The n-body problem in General Relativity up to the second post-Newtonian order from perturbative field theory},
  journal = {Physical Review D},
  volume  = {79},
  number  = {4},
  pages   = {044031},
  year    = {2009},
  doi     = {10.1103/PhysRevD.79.044031},
  eprint  = {0812.0012},
  archivePrefix = {arXiv},
  primaryClass = {gr-qc}
}

@article{PhysRevD.62.124015,
  title     = {Post-Newtonian gravitational radiation and equations of motion via direct integration of the relaxed Einstein equations: Foundations},
  author    = {Pati, Michael E. and Will, Clifford M.},
  journal   = {Phys. Rev. D},
  volume    = {62},
  issue     = {12},
  pages     = {124015},
  numpages  = {28},
  year      = {2000},
  month     = nov,
  publisher = {American Physical Society},
  doi       = {10.1103/PhysRevD.62.124015},
  url       = {https://link.aps.org/doi/10.1103/PhysRevD.62.124015}
}

@article{PhysRevD.65.104008,
  title     = {Post-Newtonian gravitational radiation and equations of motion via direct integration of the relaxed Einstein equations. {II}. {Two-body} equations of motion to second post-Newtonian order, and radiation reaction to 3.5 post-Newtonian order},
  author    = {Pati, Michael E. and Will, Clifford M.},
  journal   = {Phys. Rev. D},
  volume    = {65},
  issue     = {10},
  pages     = {104008},
  numpages  = {21},
  year      = {2002},
  month     = apr,
  publisher = {American Physical Society},
  doi       = {10.1103/PhysRevD.65.104008},
  url       = {https://link.aps.org/doi/10.1103/PhysRevD.65.104008}
}

@article{PhysRevD.75.124025,
  title     = {Post-Newtonian gravitational radiation and equations of motion via direct integration of the relaxed Einstein equations. {V}. Evidence for the strong equivalence principle to second post-Newtonian order},
  author    = {Mitchell, Thomas and Will, Clifford M.},
  journal   = {Phys. Rev. D},
  volume    = {75},
  issue     = {12},
  pages     = {124025},
  numpages  = {12},
  year      = {2007},
  month     = jun,
  publisher = {American Physical Society},
  doi       = {10.1103/PhysRevD.75.124025},
  url       = {https://link.aps.org/doi/10.1103/PhysRevD.75.124025}
}

@article{PhysRevD.89.044043,
  title     = {Incorporating post-Newtonian effects in {$N$}-body dynamics},
  author    = {Will, Clifford M.},
  journal   = {Phys. Rev. D},
  volume    = {89},
  issue     = {4},
  pages     = {044043},
  numpages  = {14},
  year      = {2014},
  month     = feb,
  publisher = {American Physical Society},
  doi       = {10.1103/PhysRevD.89.044043},
  url       = {https://link.aps.org/doi/10.1103/PhysRevD.89.044043},
  note      = {Erratum: Phys. Rev. D 91, 029902(E) (2015)},
  eprint    = {1312.1289},
  archivePrefix = {arXiv}
}

@article{Will2014,
  author    = {Will, Clifford M.},
  title     = {The Confrontation between General Relativity and Experiment},
  journal   = {Living Reviews in Relativity},
  volume    = {17},
  number    = {1},
  pages     = {4},
  year      = {2014},
  month     = jun,
  doi       = {10.12942/lrr-2014-4},
  url       = {https://doi.org/10.12942/lrr-2014-4},
  eprint    = {1403.7377},
  archivePrefix = {arXiv},
  primaryClass = {gr-qc}
}

@article{Will2014Triple,
  title     = {Post-Newtonian effects in {$N$}-body dynamics: Relativistic precession and conserved quantities in hierarchical triple systems},
  author    = {Will, Clifford M.},
  journal   = {Classical and Quantum Gravity},
  volume    = {31},
  number    = {24},
  pages     = {244001},
  year      = {2014},
  month     = dec,
  doi       = {10.1088/0264-9381/31/24/244001},
  url       = {https://doi.org/10.1088/0264-9381/31/24/244001},
  eprint    = {1404.7724},
  archivePrefix = {arXiv}
}

@article{PhysRevD.83.084013,
  title     = {Characterization of the gravitational wave emission of three black holes},
  author    = {Galaviz, Pablo and Br{\"u}gmann, Bernd},
  journal   = {Phys. Rev. D},
  volume    = {83},
  issue     = {8},
  pages     = {084013},
  numpages  = {16},
  year      = {2011},
  month     = apr,
  publisher = {American Physical Society},
  doi       = {10.1103/PhysRevD.83.084013},
  url       = {https://link.aps.org/doi/10.1103/PhysRevD.83.084013}
}

@article{Lousto_2008,
  doi       = {10.1088/0264-9381/25/19/195019},
  url       = {https://doi.org/10.1088/0264-9381/25/19/195019},
  year      = {2008},
  month     = sep,
  volume    = {25},
  number    = {19},
  pages     = {195019},
  author    = {Lousto, Carlos O. and Nakano, Hiroyuki},
  title     = {Three-body equations of motion in successive post-Newtonian approximations},
  journal   = {Classical and Quantum Gravity}
}

\end{document}